\documentclass{report}
\begin{document}
\ifx\href\undefined\else\hypersetup{linktocpage=true}\fi 
\author{Anders K.H. Bengtsson \footnote{e-mail: anders.bengtsson@hb.se}}
\title{Quantum Computation:\\ A Computer Science Perspective
\thanks{Work supported by the Swedish KK-foundation under the PromoteIT program.}}
\date{February 22, 2005}   
\maketitle
\input{epsf}
\begin{abstract}
The theory of quantum computation is presented in a self contained way from a computer science perspective. The basics of classical computation and quantum mechanics is reviewed. The circuit model of quantum computation is presented in detail. Throughout there is an emphasis on the physical as well as the abstract aspects of computation and the interplay between them. 

This report is presented as a Master's thesis at the department of Computer Science and Engineering at G{\"o}teborg University, G{\"o}teborg, Sweden.

The text is part of a larger work that is planned to include chapters on quantum algorithms, the quantum Turing machine model and abstract approaches to quantum computation. 
\end{abstract}

\tableofcontents

\chapter*{Foreword}

The intended readership for this master's thesis in Computer Science is primarily the computer scientist wishing to get an idea of what quantum computing is about. But I also have physicists in mind. Therefore, the physicist will find material on physics that will appear to be obvious and the computer scientist will find material on computers that will likewise appear to be trivial. So perhaps the reader who will benefit the most from the text is the one who is unfamiliar with both subjects. The point is that I'm actually not writing for the lucky few who have expertise in both fields but rather for those who come from either field, or from none of them. The text is thus basically introductory, but not elementary.

There is also a further point. Since quantum computation straddles the borderline between physics and computing science, it is interesting to spell out the basic assumptions and facts of both fields in some detail. 

Obviously, this text can be seen as a review article. But I have no intention to treat every aspect of the subject which is simply to vast. The depth of the treatment will also vary considerably. Some basic definitions and some, in my opinion fundamental, results, will be spelt out in detail, whereas many topics that a comprehensive text would treat, will be passed over rapidly. The principle behind these choices is that I will attempt to be detailed on issues that has a bearing on the connections between physics and computation. What has been left out can be found in the textbook literature and original articles on the subject as well as in other review articles. 

The text is mostly written in theoretical physics style, introducing no more formalism than needed to make the arguments clear. The degree of formalization will vary. A high level of formalization throughout tends to make the text unreadable, whereas a low level of formalization might leave the reader unnecessarily confused. Definitions, derivations and results are presented and proved in the running text, but occasionally, due to the nature of subject, a more formal style will be adopted. I've chosen a level of formalization that I found appropriate and in the end it reflects my own taste. 

There are of course lots of review article on quantum computation. I have therefore decided not to repeat to much of the standard calculations and derivations, instead focusing on what I find interesting, trying to put forward a slightly different perspective, and instead being detailed on points that are often glossed over. In this respect I hope this text can be a complement to the many excellent books and reviews already in circulation, a few of which are \cite{Gruska1999,Steane1998,GalindoDelgado2001,WilliamsClearwater,NielsenChuang2000}.

One seldom learns a subject by reading just one book or just one review article. In writing chapter 4 on introduction to quantum mechanics, I realized how much is left implicit, even though you try to make the text self contained. If you haven't already mastered a subject, perhaps you cannot gain so much from just one review - you must read several articles and books to see the subject treated in different ways. 

\subsubsection{Outline of contents}
Chapter 1 is an introduction to text and a motivation for studying quantum computation. Some fundamental questions on the connection between physics and computation will be mentioned. They will be returned to in a planned part II of this work.

Chapter 2 is an overview of the central concepts of classical computation such as the notions of computational models, computability and complexity theory. Together with Chapter 5 on general quantum theory it serves as the foundation for a treatment of quantum computational models and quantum algorithms.

Chapter 3 is a brief introduction to quantum computation. It serves mainly as motivating the subsequent two chapters on quantum mechanics. 

Chapter 4 contains a quite extensive introduction to quantum mechanics written in a physics style. Three important models are treated in some detail; a particle trapped in a potential well, the harmonic oscillator and the theory of angular momentum. Apart from being important in quantum physics, these models are the standard ones employed when teaching introductory quantum mechanics. All concepts of quantum mechanics can be introduced while studying these simple models.

Chapter 5 then sets up the formal theory of quantum mechanics in terms of linear operators on Hilbert spaces. After that, the stage is set for treating quantum computation.

Chapter 6 describes in an abstract way the quantum circuit model.

As this text is mainly on the abstract and theoretical aspects of classical and quantum computational models, not very much will be said on practical realizations of quantum computing devices, or quantum computers for short. Presumably, the theoretical aspects of the subject matter will remain relevant, while the practical, implementational details are likely to undergo more dramatic change.

One last remark. My initial intentions was to treat also the Quantum Turing machine model and quantum algorithms. However, the scope of the project would then have gone beyond the boundaries of a masters project. For this reason, these topics will be left for a part II.

\subsubsection{Acknowledgment}
This work was done with support from the Swedish Knowledge Foundation under the Promote IT program.

I would like to thank professor Bengt Nordstr{\"o}m for supervising the project and for valuable discussions on computing science in general and the theory of computability in particular.

I also thank Ingemar Bengtsson for reading and commenting on the manuscript.

\chapter{Introduction}

Computer science, and in particular the theory of computation, can be studied without explicit regard to physics. The whole area of research into classical computability is phrased without any reference to physics or even real computing machines. The related areas of syntax and semantics of programming languages make no reference to anything more real than symbol shuffling by abstract machines. 

Classical computation is a discrete process. Whether viewed in terms of Turing machines, RAM\,-machines or operational semantics of programming languages in terms of abstract stack machines\footnote{An abstract stack machine is a notational system for giving step-by-step meaning to the primitives of a programming language.}, it really just amounts to string processing or symbol manipulation. The number of symbols is finite and the number of basic operations is finite. A program is a finite set of instructions in terms of the operations acting on the set of strings built out of the symbols. Seen in this way, computation seems to be detached from physical reality, and any 'system' that 'understands' the rules can perform the computation. 

From a practical point of view, the software/hardware division also stresses this apparent independence of physics. The software in the form of computer programs written in any of the many hundreds of invented programming languages are again just strings of symbols. They seem to have no more connection to physics than the ink with which they are recorded on paper. When they are compiled and stored electronically, the link with physics is somewhat more pronounced but still weak. It is upon actually running the program, which always entails the motion of some physical system, that the physical nature of computation comes into focus. This is obvious if the algorithm is carried out by hand or using some mechanical computing device. 

So there is a link, however weak, to some physical substratum, and it is not possible to severe this link completely. On the other hand, it is a fundamental property of reality that it is possible work and solve computational problems at abstract levels without having to check physical realizability at every step. This is analogous to the process of abstraction which is so characteristic of computer science. By abstraction, ever more powerful and complicated computational tools can be invented, which, once it has been ascertained that they can be implemented in terms of more primitive structures, can be used to solve more difficult computational problems without checking the implementation at every step. 

But if we work the other way, from abstract levels of programming structures to more concrete primitives, then eventually we will arrive at some physical system, a {\it computing machine}, that actually performs the physical motion needed in order to carry out the computations. In digital computers this is switching voltage levels in transistors, which in its turn involves the collective motion of large numbers of electrons.

Thus, as was pointed out and studied by Landauer \cite{Landauer61-82-96}, information is always carried by some physical medium, and likewise computation is a physical process constituted by some well defined motion of a physical system.

%*************************
%THE THEORY OF COMPUTATION
\section{The theory of computation}
The theory of computation arose in the nineteen thirties as a response to problems in the foundations of mathematics and logic \cite{Gandy1994}, in particular in connection to David Hilbert's {\it Entscheidungsproblem}. The Entscheidungsproblem is a problem within the formal or axiomatic approach to mathematics. Hilbert's program was to formalize mathematical theories into a set of axioms defining relations between the undefined primitive notions of the theory, and a set of rules of deduction. In this way one should be able secure the foundations of mathematics as well as mechanize the process of theorem proving. Good properties, like consistency and completeness, should be possible to ascertain within the axiomatic system.

The axiomatic approach itself has a long history dating back to antiquity. After the invention of the calculus by Newton and Leibniz in the mid seventeenth century, there was a very rapid progress in the fields of applied mathematics and physics. The new mathematics was phrased in an axiomatic language but the underlying concepts were intuitive and often vague. In the background history of Hilbert's approach we find attempts to secure the foundations of such concepts as infinitesimals, limits, real numbers, functions and derivatives to name a few. As an aside it is interesting to note the very close interplay between mathematics and physics during this period. Apart from being a theoretical subject of its own, mathematics is also the language of the physical sciences and of technology.

Hilbert's formalistic approach to mathematics made a distinction between the syntactic aspects of mathematics, i.e. the axioms and the rules of deduction, and the semantic aspects, i.e. what the mathematical concepts and theorems actually mean. 

Physicist, engineers and applied mathematicians are normally interested in the meaning of mathematics. Phenomena in the real world, and whole areas of science, are modeled using mathematics. On the other hand, once the modeling is done, the actual calculations can be performed without considering the interpretation. In  practice there is always an intricate interaction between modeling, calculation and interpretation. But the point is clear, the strength of mathematics derives from this division into syntactic rules of calculation and its semantic, or intuitive, interpretation in terms of objects in the physical world.

The same interplay between syntax and semantics is, of course, present in computer science itself. We write computer programs in order to solve scientific, engineering, economic, administrative, everyday and entertainment problems. But the programs run on computers that perform purely syntactic symbol shuffling. In the theory of programming languages there is also this division between the syntactic and the semantic aspects of programming and program execution.

%************************************************	
%THE INPUT/OUTPUT MODEL OF PYSICS AND COMPUTATION
\section{The input/output model of physics and computation}
The paradigm of computation is capable of encompassing widely different systems. On a certain level of abstraction, the description of a physical system and a computing machine is very similar, not to say identical. Almost everything conceivable can be described by an $input\rightarrow function\rightarrow output$ model. A {\it physical system} is defined as some well defined portion of space and time with a well defined interface and interaction with the environment. Then specifying the input using some labeling, the output can, in principle at least, be computed using the dynamical laws. 

In computer science, we always know the dynamics of the system, because this is the program, and we set up the system in order for it to compute previously unknown outputs from given inputs. Furthermore, essentially due to the discrete finite nature of input and output, there is a well agreed on paradigm for this input-processing-output model. As soon as the labeling (the alphabet) of the input/output states are defined, the computation is just a syntactically ruled shuffling of the labels. 

In physics the focus is different. First, we sometimes don't know the dynamical laws. The very object of fundamental physics is to investigate the dynamics through theory, experiments and observations. 

Is there a difference in the computational strength of different physical systems? What algorithms can be performed with what physical systems? These are questions not normally posed in theoretical computer science where the discussion is from the outset framed within the classical computational models, all of which are basically notational systems. 

However, it is generally believed that any physical system constrained to work in a discrete stepwise fashion, working to precisely and finitely stated rules according to the logical description of a Turing Machine or an electronic computer, or even a human computer as envisioned by Turing, are equivalent. 

The Church\,-Turing thesis identifies the set of effectively or intuitively calculable functions with the set of functions computable within any of the classical computational models. In a historical context, effective computability meant computability by an abstract human being working to precise rules. 

The thesis has however acquired connotations connecting it to machine computation, in particular electronic digital computing machines. In this latter sense the thesis is true; what can be computed by a general purpose digital computer can be computed by a Turing machine. This is due to the fact that a digital computer can be modeled as an abstract RAM machine. Whether any {\it conceivable} physical computing machine is constrained by the thesis is not known (but see \cite{Gandy1980}). This is a question about all of physics, and we don't know all of physics yet. 

On the other hand, the precise mapping between models of computation and human and digital machine computation and the consequent possibility to study computation in the abstract has lead to the view that the limits of computation are set by mathematics and logic. The development of quantum computation has, to a degree, challenged this view of computation. Since computations are basically physical processes when actually carried out, it can be argued that what can be computed is a question of physics, not a question of mathematics or logic \cite{Deutsch1985}.

%**********************************
%CLASSICAL PHYSICS AND THE COMPUTER
\section{Classical physics and the computer}
The CPU of a digital electronic computer as well as the main memory used for intermediate storage consist of huge amounts of transistors and other semiconducting devices working in an {\bf on/off} fashion corresponding to distinct voltage levels. These voltage levels constitutes a concrete realization of the abstract bit of information. The semiconductors in their turn are arranged into circuits implementing logical gates. The precise mapping between abstract logical gates working on bits and circuits working with voltage levels are the basis for the success of the electronic digital computer. But without the extreme fastness with which the switching between {\bf on} and {\bf off} can be performed (on the order of nanoseconds) the computer would not be so powerful. There is also an engineering aspect of this. The transistors must work in parallel, and in practice the CPU clock controls the working of the computer so that at each time tick, bit flips are performed in parallel. 

The electronic digital computer can therefore be seen as a special physical system constrained or engineered to work in a discrete way. All operations performed by the computer are discrete, but the underlying physical processes are continuous, or at least the description of these physical processes is continuous. The actually bit flips between zero and one, when studied at the physical level take a certain amount of time and the transition between voltage levels can actually be studied by employing a good enough oscilloscope. But once one has abstracted away from these physical considerations, the operations of the electronic computer could as well be performed by other physical systems, for example electric circuits working with magnetic relays. The performance would be much slower and other engineering problems would ensue.\footnote{Actually such computers, and computers based on vacuum tubes, preceded the solid state electronics computer.}

The workings of a transistor, as far as electronics goes, can be described by classical electrodynamics. But electronics is not far from quantum physics. Transistors are quantum mechanical devices that could not possibly be understood or built without a knowledge of quantum physics. However these transistors are wired to work in a discrete fashion as switches. Logically there is no difference between a transistor switch, a mechanical switch or an electromechanical switch. The only difference lies in performance measures like speed, reliability and energy consumption. The underlying physics of the transistor must be understood in terms of quantum mechanics, but once that is done, the transistor as a circuit element can be understood in classical terms. And furthermore, as pointed out above, these circuit elements can be abstractly modeled and reliably worked with without at every step consulting the underlying physics. When implementing the circuit, design issues like power supply, switching times, delays et cetera must of course be faced, but this does not, in principle, influence the logical design of the circuit.

In the end it all comes down to the fact that we can build fast electronic computers that can effectively carry out algorithms. These computers are based on quantum mechanical physical systems {\it constrained} to work in a discrete, classical  fashion. Furthermore, miniaturization notwithstanding, these systems involve the collective behavior of large numbers of particles (electrons), thus relying on statistical properties of the systems. 

In contrast, in the would be quantum computers, it is the individual behavior of the particles we have to rely on. This is at the same time the source of the power of quantum computation, as well as the source of the engineering problems of actually building devices capable of enacting quantum algorithms.

%*******************
%QUANTUM COMPUTATION
\section{Quantum computation}

In 1980, the physicist Richard Feynman pointed out that digital computers cannot simulate quantum systems without an exponential slowdown \cite{Feynman1980}. Feynman wasn't particularly interested in approximating quantum physics, what he discussed was exact simulation, the question of whether digital computers could do exactly the same as the quantum system would do. He came to the conclusion that present day physics does not allow this, essentially due to a mismatch between the discrete nature of digital computers and the exponentially large state space of quantum systems. Feynman's interest in the physics of computers sparked off the research into quantum computation in the 1980's. 

Other lines of research that contributed to the initial impetus of quantum computation was the work of Bennet \cite{Bennet1973} and Fredkin and Toffoli \cite{FredkinToffoli1982} on reversible computation, as well as the previously cited work by Landauer \cite{Landauer61-82-96}.

In quantum computing we are interested in the computational strength of physical systems that by their very nature must be analyzed or understood according to quantum mechanics. Quantum computation relies on the exact manipulation of individual quantum physical objects, in distinction from the classical computer, where averaged classical statistical properties of the objects suffices. This is the source of the strength of quantum computation as well the difficulties in actually building quantum computing devices. There is to be no transition to the classical regime during the computation, as that would destroy the very features that lends the quantum computer its strength.

Some researchers in the field remark that, as classical physics is fundamentally wrong, any correct theory of computability must be based on quantum mechanics. This point of view is very clearly stated in the paper by Deutsch \cite{Deutsch1985} where it is claimed that there is a physical assumption underlying the Church\,-Turing thesis. 

Deutsch argues that the Church\,-Turing thesis is actually at variance with classical physics, but that it can be rephrased in agreement with quantum physics. According to this point of view, quantum computers are not fundamentally more powerful than Turing Machines, though they might be faster. The basic line of the argument is that the continuous nature of classical physics makes it in principle impossible to simulate a classical physical system by a discrete computer. But quantum physics is fundamentally discrete, and therefore the Church\,-Turing thesis connects effective methods not to classical computing machines but to quantum computing machines. The argument is not entirely convincing, and will be returned to in part II.

And with this we plunge into the details! But first a disappointing remark is perhaps in order. There are no quantum computers as yet if one does not count experimental setups working on the equivalent of a few bits. Because of this, not much will be said here about realizations. Anything written about practical implementations of quantum computing devices will surely soon be outdated by new experimental developments, whereas the theoretical part of the topic is likely to be more stable.

\chapter{Classical computation}
Quantum computation relies on the ability for quantum mechanical physical systems to perform computations. In order to prepare some common ground for the discussion, we will review the theory of classical computation in terms of Turing machines, the Church-Turing thesis and the limitations of computation.

There are two fundamental questions in the theory of computation; what can be computed in principle, and what can be computed efficiently. The first question is addressed by the theory of computability and second by the theory of complexity.  In order to discuss these questions in general without reference to any particular computing machine or programming language, one must work within some abstract mathematical model of computation. Still, it must be possible understand precisely the relationship between the model and the properties of actual physical computing machines. 

The models of computation developed in the nineteen thirties were all attempts to capture in precise mathematical terms what is meant by a computation. In order to answer the Entscheidungsproblem\footnote{This was apparently not the only impetus to this work, see \cite{Gandy1994}.}, i.e. David Hilbert's question of whether there exist a mechanical method to determine if mathematical statements are true or not,\footnote{For Hilbert, truth of a statement was equivalent to it being a theorem, otherwise one would have to distinguish between truth and derivability when stating the Entscheidungsproblem. John von Neumann discussed the problem in terms of provability.} it was necessary to have a precise definition of a mechanical method in order to treat the question with mathematical tools. In this context, mechanical does not necessarily mean, and in fact did not mean, a procedure performed by a machine. Mechanical means algorithmic.

As it turned out, the different models put forward; Church's $\lambda$-calculus, Herbrand-G\" odel's recursive functions and Turing's {\it automatic computing \linebreak machines} \footnote{The term is Turing's own.}, were all shown to be equivalent in the sense that they all defined the same set of computable functions (see several papers in \cite {Herken1994}). All three models were meant to be abstract mathematical models of computation. Turing, however, phrased his concepts in terms of machines reading and writing symbols on a tape, and compared the process of computation to humans working with paper and pencil. And it is also with the Turing model that the connection to present day digital computers and to physics is most clearly seen. 

The formal models of computation must be connected to the intuitive notion of an algorithm. The models are meant to capture what it means to carry out a mechanical, or algorithmic, procedure. 

The Church-Turing thesis is often quoted in this context, but there seem to be some confusion as to what it actually says. After 50 years of dramatic evolution of digital computers, the thesis has perhaps not surprisingly, acquired connotations or meanings not present in the original formulation \cite{Copeland-SEP2002}. We will be concerned with what the thesis did say when it was first formulated, what is normally meant by it nowadays, and how it relates to quantum computation.

%SOME DEFINITIONS
%****************
\section{Some definitions}
The following definitions are for the benefit of the reader not familiar with computer science terminology, and for fixing the sense in which the terms will be used in this text.

%********************************
\subsection{Algorithm} 
It is not possible to give a formal definition of the concept of an algorithm but it can be characterized well enough so that no ambiguity remains as to the meaning of the term. This remark is true for all the concepts treated in this section. Perhaps characterization is a better term to use.
An algorithm consists of a set of instructions for carrying out a certain task. In computer science the task is a computation, a notion that will be defined below. The concept can, and must be, further elaborated by the following clauses.
\begin{itemize}
\item The set of instructions should be precise and unambiguous. The number of instructions should be finite and each instruction should be finite in length.
\item A machine or a human can execute it.
\item There should be no room for subjective decisions, appeal to human intelligence or creative intervention of the user.
\item It should solve some general problem.
\item It need not be phrased in any particular language, programming or natural.
\end{itemize}

The first three clauses imply that all creative or intelligent effort goes into the task of finding or constructing the algorithm. Once the algorithm is known, it should be possible to carry it out automatically or mechanically. The fourth clause has to do with the fact that we are not in general interested in particular cases, rather we want so solve sets of problems, often parameterized by a collection of variables. Therefore a general algorithm has a domain of definition, which is the set of meaningful, or allowed, input values or instances. The last point means that algorithms have an abstract existence independent of any particular language. In practice, an exact programming language or pseudo-code language is useful in order to satisfy the first three clauses. 
The term {\it mechanical method} or {\it effective method} can be considered to be synonymous with algorithm. The word {\it procedure}, can be used instead of method. Sometimes, the word {\it general} will be used to emphasize that we are considering methods applicable to a range or set of problems.

The computational models of the thirties identified this informal notion of an algorithm with precisely defined formal models of computation.

Note that we do not include {\it termination} among the characteristics of algorithms. That would be inappropriate for two different reasons. Firstly, many algorithms are not meant to terminate, at least not before we actively choose to terminate them. Examples are operating systems, web servers and lots of everyday applications like word processors. Secondly, termination is a non-trivial issue that has to do with executing, or running, the algorithm. This will be discussed in the next section.

%********************************
\subsection{Computation}
By a computation we mean the actual carrying out of an algorithm. From this follows that computations are processes taking place in time, that can be carried out by either machine, human or any other suitable physical system. The only requirement is that the computing system 'understands' the language used to write the algorithm in, and thus is able to carry out the instructions.
 
This distinction between an algorithm as a passive description of a computation and a computation as an actual enacting of an algorithm is not always upheld. The terms are often used interchangeably. In practical work with computers this does not lead to any confusion but when discussing fundamental questions of principle it is helpful to maintain this distinction.
 
When it comes to quantum computation and quantum algorithms the distinction is somewhat more acute. At the present time there are no quantum computers, so there is nothing to run the quantum algorithms on. Furthermore, it is not practical to simulate quantum computations on classical computers, as the time evolution of a quantum mechanical system that is inherent in quantum computation requires exponential resources!

If the above characterization of an algorithm is applied to a human performing a computation, the question can be asked as to what are the limitations of algorithms or computation. What can be calculated effectively, or mechanically, is precisely what can be done by following an algorithm with the additional clause that the algorithm should always produce the desired result in a finite number of steps. This question, whether the algorithm terminates or not, turns out to be a nontrivial issue as already noted.

%********************************
\subsection{Program}
A {\itshape program} is an algorithm written in a certain language. The term program is used in two slightly different, but related senses. 

In the first sense, we are referring to a program written in a general purpose programming language. Such a program should be possible to run on the appropriate computing machine without further work, except possible compilation. Hence the program must contain all circumstantial information like include or import statements for supporting files and software. The program furthermore should handle input and output of data, either in an interactive way through standard input and output devices or via a file system. A program is therefore a practical embodiment of an abstract algorithm. 

In the second sense, the term program is used for a collection of instructions for a computation in an abstract computational model like Turing machines. In this case, there need not be a physical machine to carry out the computation. But it should be possibly to carry it out (by a human) by adhering to the rules and specifications of the computational model. 

In some cases the instructions might be ordered in a {\it list}. In that case, we consider the execution order to be given by the ordering of the instructions in the list, possibly with branching of to labels in the list.\footnote {This is necessary in order to implement {\bf if} $<\!condition>\!$ {\bf then} $<\!statement\!>$ {\bf else} $<\!statement\!>$, and {\bf while} $<\!condition\!>$ {\bf do} $<\!statement\!>$ programming primitives.} 

When the program is actually a {\it set} of instructions, no execution ordering is presupposed. The machine looks up the proper instruction to execute depending on the state of the machine and the data. This is the way a Turing machine computes.

%********************************
\subsection{Process}
By a process we mean a program in execution. In some contexts, notably operating systems, the word process is reserved for executing programs that are not meant to terminate. In the present context we are primarily interested in terminating processes and I will use the term in both senses, letting the context determine which meaning is referred to.

Thus computation and process emphasizes the physical and dynamical side of our topic, whereas algorithm and program emphasizes the logical and mathematical side.

%********************************
\subsection{Alphabets, Strings and Numbers}
Numbers and strings of symbols are fundamental to computer science. Just as a computation can be thought of as the calculation of numerical values of a function, it can as well be regarded as the processing of strings of symbols. The equivalence follows from the fact that any finite set of finite strings of symbols drawn from a finite alphabet can be put in a one-to-one correspondence with a subset of the natural numbers. Any enumeration of the strings in some lexicographic order will do. We will make these concepts somewhat more precise.

%********************************
\subsubsection{Numbers}
By a number we mean, unless otherwise stated, a natural number, i.e. a member of the infinite set ${\bf N}=\{0,1,2,\ldots\}$. This set can be defined {\itshape inductively} starting from the number $0$ and adding, in a step by step fashion, the successors. Informally,
$$
\cases{
\mbox{0 is a natural number,} \cr
\mbox{If $n$ is a natural number, then the succesor $n+1$ is a natural number.}}
$$
This is not really a good definition, since in writing $n+1$ for the successor of $n$ we are in fact presupposing the numbers together with addition. But it captures the idea behind the following more formal definition.

The set ${\bf N}$ of natural numbers are defined by the clauses

\begin{equation}\label{eq: natnumdef}
\cases{0\in {\bf N} \cr n\in {\bf N} \Rightarrow S(n)\in {\bf N}}
\end{equation}
where $S(n)$ denotes the successor of $n$.

Arithmetical operations, like addition and multiplication can be defined on this basis \cite{Peano1889}.

By ${\bf N}^d$ we denote the set of all $d$-tuples $(n_1,n_2,\ldots,n_d)$ of numbers. 

%********************************
\subsubsection{Strings and languages}
The concept of a {\itshape symbol}, or a {\itshape token}, will be taken to be intuitively given and not further analyzed. An {\itshape alphabet} is non-empty a set of symbols, generically denoted by $\Sigma$. A {\itshape string} (over an  alphabet) results when the symbols taken from an alphabet are written consecutively.  The order in which the symbols are written within the string matters. There must be no blanks or commas or other separators between the symbols in a string, instead, if blanks or other separators are needed they should be included among the symbols. A {\itshape language} is a set of strings. The following notation is useful to have at hand.

The set of all strings over the alphabet $\Sigma$ is denoted by $\Sigma^*$ and it includes the empty string $\cal{E}$. An equivalent term for string is {\it word}. $\Sigma^+$ denotes the set of strings with the empty string excluded. In some contexts, strings will be enclosed by " ", as is prevalent in computer programming languages. Now, a {\it language} is an arbitrary subset of $\Sigma^*$. This is all we will need from the general theory of formal languages (cf. \cite{Revesz1983}).

%********************************
\subsubsection{Bit strings}
One particular important kind of strings are the bit strings. They are based on the alphabet $\{0,1\}$. A {\itshape bit string} is any combination of the symbols $0$ and $1$ written without any blank separators. By the notation $\{0,1\}^k$ we mean a bit string of length $k$ which will also be more explicitly denoted by $"b_1b_2\cdots b_k"$. Such a string is naturally interpreted as a k-bit binary number. Then the corresponding base-10 representation of the binary number can be used as a shorthand for the string, as in the example $"0101"={\bf 5}$.

%********************************
\subsection{Functions}
The concept of a function is supposed to be well-known, but we record the basic definition here for completeness and fixing our notation. A function can be regarded as a mapping between two sets $X$ and $Y$, taking an element $x$ in $X$ and mapping it into a unique element $y$ in $Y$. This is formally written as $f:X\rightarrow Y$. When we want to focus on generic, or particular elements that are connected by the mapping, we write $y=f(x)$ or $b=f(a)$. The element $b$ in $Y$ is called the  {\it image} (under the mapping $f$) of the element $x$ in $X$.

A more formal definition of functions is based on the concept of a relation.

A {\it relation} is a subset of all ordered pairs $(x,y)$ where $x\in X$ and $x\in Y$. A function $f$ is a special kind of relation for which we require: if $(x,y)\in f$ and $(x,z)\in f$, then $y=z$. This condition expresses the uniqueness requirement, that for each $x$ in $X$ there is a unique $y$ in $Y$. This is the only restriction on a function, and consequently, the concept of a function is very general indeed.\footnote{As we will see, when the sets $X$ and $Y$ are infinite, just a denumerable subset of all possible functions according to this definition are actually possible to compute by algorithmic methods.} 

If the function is defined for all elements in $X$, it is a {\it total} function, otherwise it is a {\it partial} function. The elements of $X$ for which the function is defined is called the {\itshape domain of definition}. The elements of $Y$ which are images of elements of $X$ are called the {\it range}.

We will almost exclusively consider functions from $N^d$ to $N$, from $\{0,1\}^k$ to $\{0,1\}^l$, or from $\Sigma^*$ to $\Sigma^*$.

\subsubsection{Enumerations}
A function is said to be {\it one-to-one} if the image of two different elements in the domain are different. If every element in the set $Y$ is the image of an element in the set $X$ (i.e. the range of the function is the whole set $Y$) then the function is said to be {\it onto}. 

Furthermore, a total function which is both one-to-one and onto is called a {\it one-to-one correspondence}. Such functions are useful for comparing the numbers of elements in two sets, as they as the name suggests, sets up a one-to-one correspondence between the sets. If the one-to-one correspondence is between a set $X$ and a subset of the natural numbers, then it can be used to count, or {\it enumerate}, the elements in $X$. Such an enumeration also provides an ordering of the elements in the set, as the following paragraph makes explicit in the case of strings.

The number of elements in a set is called its {\it cardinality}. A finite set has a cardinality which is a natural number. An infinite set, the elements of which can be put in a one-to-one correspondence with the set $N$ of natural numbers, is said to have cardinality $\aleph_0$.

%********************************
\subsubsection{Lexicographic ordering}
It is often useful to be able to order strings in a lexicographic order. The definition is mimicked on the ordering of words in the English language. An example makes the concept clear. Suppose $\Sigma=\{S_1,S_2,S_3\}$, then the lexicographic ordering of the strings is the infinite list 

$$
[{\cal E},S_1,S_2,S_3,S_1S_1,S_1S_2,S_1S_3,S_2S_1,S_2S_2,S_2S_3,\ldots ].
$$

We can define a function $lex$ from set of all strings $\Sigma^*$ to the set of integers $N$,

$$
lex: \Sigma^*\rightarrow N,
$$

where in particular $lex({\cal E})=0$. 

Clearly, the lexicographic ordering provides an enumeration of the strings in the language.

%********************************
\subsection{Decision procedures and Computation procedures}
It is sometimes useful to distinguish between algorithms for computing values of functions and algorithms for yes/no decisions. In the first case, the object of the algorithm is to compute values for functions $f:N^d\rightarrow N$. Such an algorithm can be called a {\itshape computation procedure}. Since it is possible to set up one-to-one correspondences between natural numbers and strings, computation procedures can also be viewed as {\itshape string processing}; an input string $w$ is processed into an output string $f(w)$. In this case we are considering functions $f:\Sigma^*\rightarrow\Sigma^*$. But it is often natural to think of computation procedures as computing values of numerical functions.

In the second case, the object is to decide questions like for example; Is the number $n$ prime? Does the number $a$ divide the number $b$? Such questions define properties $P(n)$, binary relations $R(a,b)$, or in the general case, n-ary relations $R(a_1,\ldots,a_n)$. Algorithms for such questions are called {\itshape decision procedures}. 

Decision procedures are often formulated in terms of languages. A language is a set of strings constructed in some way or satisfying certain properties. The problem is to determine for an arbitrary string over the alphabet whether it belongs to the language or not. This is a typical decision problem, having a {\it yes} or {\it no} answer. In fact, all decision problems can be formulated in terms of language membership. We will return to these notions in section 2.3.4.

%A NOTE ON THE CONNECTION TO EVERYDAY COMPUTING
%**********************************************
\section{A note on the connection to everyday computing}
The contents of this chapter might seem remote from everyday computers and their uses, and it is perhaps interesting to make a comment on the connection. As already remarked, many algorithms are not meant to terminate. We could take a word processor as an example. A word processor is a complicated piece of software, which apart from presenting the user with a graphical interface, also should respond to input from the keyboard as well as files read from storage media. Such programs are said to be event-driven. This means that they only perform actions when called for by request from the user, otherwise they are idle. (There might be actions like rewriting the screen invoked by other parallel running processes.) When the user hits a key, this triggers a computation that results in the text stored in the memory and displayed on the screen being altered. This can in fact be seen as a computation of a function, or better still, as string processing. The input is the present text stored (the coded text really) together with the code for the key. The output is the new text. Depending on what action is requested, different computations are performed on the text. Thus the concept of a computation is very strong and in fact incorporates all types of data processing made by digital computers. 

% THE CLASSICAL TURING MACHINE MODEL OF COMPUTATION
%**************************************************
\section{The classical Turing machine model of computation}
In a mathematical oriented approach to the theory of computability the distinction between passive algorithm and active computation can easily be overlooked. Just one example will suffice to make the point clear. Most popular or semi-popular accounts of Turing Machines abound with animistic phrases like reading, writing, moving etc. This conjures up the image of a magnetic read-and-write head moving along the tape, erasing and writing information. Exact mathematical notions can replace this suggestive but imprecise terminology, one classic example of which is \cite{Davis1958}. On the other hand, upon reading such a mathematical account of computability, it seems that the mathematical formulation has got rid of all reference to motion or time steps. Close scrutiny however reveals the distinction between algorithm and computation even in this case. The Turing machine program is the set of instructions for the machine together with specification how to present the machine with data and how to read off data. This is clearly passive. However in order to actually perform the computation inherent in the set of instructions, someone, machine or human, has to perform the computational steps.

%********************************
\subsection{Informal description of Turing machines}
A very good description of Turing machines can be found i Turing's original paper \cite{Turing1936}. In fact, most modern informal descriptions are just rephrasings of Turing's own words. This is true also for what follows here.

The machine consists of a memory, a read-and-write head and a processing unit.

The memory is a tape which is divided into distinct squares, also called cells. It is infinite to the left and to the right\footnote{For practical purposes, if one would like to write a computer program emulating a Turing machine, it might be easier to consider a one-way tape with a start square to left and infinite to the right.}. The memory tape is used for giving input to the machine, for storing intermediate data during computation and for writing output.

The read-and-write head can move along the tape. It can read symbols written on the tape (this is called scanning) and it can write symbols on the tape.

The symbols can be any symbols, but they must come from a finite alphabet $\Gamma=\{S_0,S_1,\ldots,S_n\}$.

The machine has a finite set of elementary operations that it can perform at each step in the computation. These are 
\begin{itemize}
\item move one step to the right
\item move one step to the left
\item write a symbol
\item erase a symbol
\item halt
\end{itemize}

This can symbolically be written as a set of operations 
$$
O=\{moveright, moveleft, write(S_i),erase,halt\}
$$

Note that reading a symbol need not be considered to be an operation. In fact, the machine always reads the symbol written on the scanned square. In some formulations, the operation erase is replaced by writing a special symbol called a {\itshape blank}, i.e. by the operation $write(blank)$.

The halt operation can be implemented in different ways.

The machine is controlled by a set of instructions. This is the program. In order to distinguish the instructions, the machine is considered to be in a set of different {\itshape machine states}. The states are numbered or given symbolic names from a set $Q=\{q_0,q_1,...,q_n\}$. Each instruction consists of four symbols (present state, scanned symbol on tape, operation, new state) or $(q_i,S ,op,q_j$) where $q_i\in Q, S \in \Sigma, op \in O, q_j\in Q$.

The program is executed by a control unit. Execution starts in a special initial machine state $q_0$ scanning the leftmost symbol on the tape. At the beginning of the computation, all but a finite number of tape squares are blank. This is true throughout the computation. At each step of the execution, the control unit checks through the list of instructions to find an instruction that matches the present state of the machine and the scanned symbol. Each cycle of the execution therefore consists of the following actions:

\begin{itemize}
\item get present state $q_{present}$
\item get scanned symbol $\alpha_{scanned}$
\item find matching instruction $(q_{present}, \alpha_{scanned}, op, q_{new})$
\item execute the instruction $op$
\item change to the new state $q_{new}$ as given in the matching instruction
\end{itemize}

%********************************
\subsubsection{Some comments}
Infinite memory in the form of an infinite tape is of course impossible in reality. But this is not a problem. At any stage of the computation, only a finite set of squares is needed. Should the machine ever run out of tape, a finite amount of new tape can always be added to the right or to the left. The number of squares on the tape thus need not be actually infinite, only potentially infinite.

As regards the symbols, the simplest choice is $\{0,1\}$ where '0' serves the purpose of a blank. Numbers are coded as strings of '1' separated by '0'. The number 0 itself must be coded as '1' in order to distinguish it from a blank, and consequently 1 is coded as "11" and so on. If one wants to use a more efficient binary coding, one can use an alphabet consisting of '0', '1' and a blank separator '$\#$'.

In some formulations of Turing machines, the operations of writing and moving are combined into a single operation. In that case an instruction consists of five symbols $q_iS_kS_lq_jM$, where $M$ denotes a move.

Furthermore, a special halt instruction is not needed. The machine stops or halts when the control unit cannot find any matching instruction. In practice, though, it is convenient to include an explicit halt instruction. In fact, when discussing decision problems in terms of Turing machines, it is natural to have two halting states, for example named by {\it yes} and {\it no}.

The names of the states are arbitrary, they can be named in any way that serves the purpose of clarity.

In the next section, a formal definition of a Turing machine is given. It does not entirely conform to the informal description given above. The reader unfamiliar with Turing machines might benefit from comparing the details.

%********************************
\subsection{Formal definition of a Turing Machine Model}
There are lots of variations of the basic definitions of a Turing machine in the literature, differing in details and level of formalization. I will choose an approach that is quite formal in order to remove as much of physics that is possible, following \cite{Davis1958}. See also \cite{Sipser1997} for a modern treatment. The parts of the definition are all motivated by the intended semantics of Turing machines.

%********************************
\subsubsection{Program}
Consider an alphabet $A_{M}$ consisting of the following {\it tokens}
\begin{enumerate}
\item a finite non-empty set of {\itshape symbols} $\Sigma=\{S_1,S_2,\ldots,S_n\}$, not containing the special {\it tape blank} $\#$ or any other tape markers. $\Sigma$ is called the {\it input alphabet}.
\item a finite non-empty set of {\itshape tape symbols} $\Gamma$, one of which is the tape blank $S_0=\sqcup$ used to mark tape ends. $\Gamma$ also contains any other tape markers such as $\#$ used to separate the input into tuples. Note that $\Sigma\subseteq\Gamma$.
\item a finite non-empty set of {\itshape internal configurations}  $Q=\{q_1,q_2,\ldots,q_n\}$, also called {\it machine states}.
\item a (small) finite set of {\it halting configurations} $Q_h$, where $Q_h\cap Q=\emptyset$.
\item a set of {\itshape moves} $M=\{L, R\}$ \footnote{Sometimes it is useful to include a "no move" S (Stay).}  
\end{enumerate}

One of the machine states $q_0$ is singled out as the {\it start} state. It is also convenient to single out halt states. If the machine is programmed for computation problems, one halt state, $q_h$, suffices. If the machine is used for decision problems, two halt states $\{q_y,q_n\}$ corresponding to the answers yes or no, are singled out. Note that the halting states are not in $Q$.

An {\itshape expression} is a finite sequence of tokens chosen from $A_{M}$.
An {\itshape instruction} is an expression having one of the following forms
\begin{itemize}

\item $q_iS_kS_lq_jR$ 
\item $q_iS_kS_lq_jL$
\end{itemize}
The intuition is that, if the machine is in the configuration $q_i$ scanning the symbol $S_k$, it prints the symbol $S_l$, changes configuration to $q_j$ and it makes a move $R$ or $L$.

A {\itshape Turing machine} $M$ has a {\itshape program} $P_{M}$  that is a finite non-empty set of instructions. The program can be thought of as defining a {\itshape transition function} 

\begin{equation}\label{eq: deltatransfundet}
\delta: Q\times \Gamma \rightarrow (Q\cup Q_h)\times\Gamma\times  M.
\end{equation}

This definition makes explicit that there are no transitions from the halting configurations.

As an example of the correspondence between instructions and the transition function, note that the instruction $q_iS_kS_lq_jR$ corresponds to $\delta(q_i,S_k)=(q_j,S_l,R)$. If the transition function is undefined for a certain $q_iS_k$ then there simply is no instruction in the program $P_{M}$ with the first two symbols equal to $q_iS_k$. In that case the machine gets stuck. This should be considered as a programming error, unless the configuration $q_i$ is not one of the halting configurations.

No two instructions have the first two symbols $q_iS_k$ the same. This means that at each step of the computation, the action of the machine is uniquely determined. Therefore, what we have defined so far are {\it deterministic} Turing machines. Removing this restriction leads to the classes of non-deterministic, probabilistic and quantum Turing machines respectively. These will be considered in sections 2.7, 2.8 and 6.2.

To prepare the way for this generalization, the transition function can be defined in a different way that will be useful when discussing generalizations of the Turing machine concept. Define the function $\Delta$

\begin{equation}\label{eq: Deltatransfundet1}
\Delta: Q\times \Gamma\times (Q\cup Q_h)\times\Gamma\times M\rightarrow \{0,1\}.
\end{equation}

Clearly, this is a function from instructions to the set $\{0,1\}$. For all instructions in the program $P_{M}$, $\Delta$ evaluates to 1. Otherwise it evaluates to 0 (i.e. the instruction is not contained in the program). 

This can be formalized somewhat further. Consider the set $I$ of all possible instructions 

\begin{equation}\label{eq: instrset}
I=Q\times \Gamma\times (Q\cup Q_h)\times\Gamma\times M,
\end{equation}

This is a finite set and a program $P_M$ is a subset of this set, or $P_M\subseteq I$.\footnote{The power set ${\cal P}(I)$ of $I$ is the set of all Turing machine programs for the given alphabet and configuration set. So we could also write $P_M\in{\cal P}(I)$.}

The function $\Delta $ can therefore be written as 

\begin{eqnarray}\label{eq: Deltatransfundet2}
\Delta : I\rightarrow \{0,1\}
\mbox{   where   }
\Delta(i) = \cases{1,&if $i\in P_{M}$\cr
	0,&if $i\notin P_{M}$\cr}.
\end{eqnarray}

Note that in this way $\Delta$ is naturally defined as a total function on the finite set of instructions $I$.

%********************************
\subsubsection{Data}
A {\itshape tape expression} is an expression consisting entirely of symbols from the set $\Gamma $. Denoting generic tape expressions by calligraphic letters, for example $\cal T$, we can consider tape expressions as split in left $\cal L$ and right $\cal R$ parts, and $\cal T=\cal L\cal R$. By convention, we always mark tape ends by the symbol $=\sqcup$, so that the tape actually looks as $\sqcup\cal L\cal R\sqcup$.

\subsubsection{Executing Turing Machine}
An {\itshape instantaneous description} $\alpha$ of $T\!M$ is an expression satisfying the following requirements
\begin{enumerate}

\item it contains exactly one $q_i$  
\item this $q_i$  is not the rightmost token
\item it does not contain R or L
\item for all the symbols $S_k$ in $\alpha$ , $S_k\in\Gamma$.
\end{enumerate}

The state $q_i$ is called the {\itshape instantaneous internal configuration} at $\alpha$. The symbol $S_k$ immediately to the right of $q_i$ in $\alpha$  is called the {\itshape scanned tape symbol}. In practice, an instantaneous description is a tape expression with exactly one configuration symbol $q_i$ inserted directly to the left of the scanned symbol. Since there must always be a scanned symbol, $q_i$ cannot be the rightmost token. Formally, a general instantaneous description can be written as ${\cal L}q_i{\cal R}$ where the right tape expression $\cal R$ must be non-empty.

So far everything is static. In order for the Turing machine to actually perform a computation we need a {\itshape computation relation} or a set of rewrite rules $\alpha\rightarrow\nolinebreak\beta$, allowing us to pass from one instantaneous description to another. In the following, $\cal X$ and $\cal Y$ denotes tape expressions, possibly just strings of blanks.

The computation relation $\alpha\rightarrow\beta$ is defined by

\begin{enumerate}
\item If $\alpha={\cal X}q_iS_k{\cal Y}$ and $q_iS_kS_lq_jR \in P_{M}$ then $\beta={\cal X}S_lq_j{\cal Y}$
\linebreak//print and move right

\item If $\alpha={\cal X}q_iS_0$ and $q_iS_0S_lq_jR \in P_{M}$ then $\beta={\cal X}S_lq_jS_0$
 \linebreak//print and right move at right end of tape, insert blank

\item If $\alpha={\cal X}S_mq_iS_k{\cal Y}$ and $q_iS_kS_lq_jL \in P_{M}$ then $\beta={\cal X}q_jS_mS_l{\cal Y}$
 \linebreak//print and move left

\item If $\alpha=q_iS_0{\cal Y}$ and $q_iS_0S_lq_jL \in P_{M}$ then $\beta=q_jS_0S_l{\cal Y}$.
\linebreak//print and left move at left end of tape, insert blank
\end{enumerate}

In practice, the rewrite rules are applied by searching for an instruction having the first two symbols matching the machine state and the scanned tape symbol. Then, at each step in the computation, the machine scans the symbol on the tape, prints a new symbol and performs a move according to the rules.

An instantaneous description is {\itshape terminal} if none of the rewrite rules apply.

A {\itshape computation} is a finite sequence $\alpha_1,\alpha_2,...,\alpha_p$ of instantaneous descriptions such that $\alpha_i\rightarrow\alpha_{i+1}$ for $1\leq i < p$ and such that $\alpha_p$ is terminal. The result of the computation is written $M(\alpha_1)$ which we define as $\alpha_p$. Using the notation $\rightarrow^*$ to denote a computation in several steps, we have

\begin{equation}\label{eq: CompRelManySteps}
\alpha_1\rightarrow^*\alpha_p=M(\alpha_1).
\end{equation}

This is the formal definition. Some comments are obviously in order. The question of whether the Turing machine halts or not is the same as whether there exists a computation or not. The tape is considered to be potentially infinite. This is taken care of by the computation rules (2) and (4) which have the effect of inserting blank squares at the ends of the tape when the machine is about to run off the tape. In any computation, only a finite amount of tape is ever used.

In order to get something done with this model, a few more choices must be made. We need a way to represent input data as tape expressions and a way to read off output from the terminal description. Some convention is needed for how to start the computation and what should be considered a proper terminal state. 

\subsubsection{A note on terminology}
What I have denoted by the term {\it internal configuration} is often denoted by the term {\it state} in the literature on classical Turing machines. My terminology instead follows that of \cite{Davis1958}, who uses the term {\it internal configuration}. It is more appropriate in the present context where we subsequently want to consider quantum Turing machines. There, we want to reserve the word {\it state} to denote the {\it quantum state} made up of the internal configuration of the machine together with the tape expression. This is what I (again following Davis) denote by {\it instantaneous description}. Thus, we define the {\it state} of a Turing machine to be synonymous to the instantaneous description. It seems reasonable in the present context to let the quantum physics usage of the word state to take precedence.

To summarize, the term state is equivalent to the term instantaneous description. When the set $Q$ is refereed to, I use the terms {\it(internal) configuration} and {\it machine state} as synonyms. So the qualifiers {\it internal} and {\it machine} are equivalent. 

Furthermore, to have some connection with intuition, we can think of the tape as the {\it memory}, the contents of which is the tape expression. Then it makes sense to think of the set of internal configurations as the {\it processor}.\footnote{It is actually a finite state machine.} The scanned tape symbol can likewise be marked by a {\it cursor}.

\subsubsection{Representing numeric input and output data}
Suppose we want to compute numerical functions $f:N^d\rightarrow N$. The simplest choice is to use the a one symbol alphabet with $S_1=1$ and a unary representation of numbers. Since we need to distinguish the number 0 from a blank, we let 0 be represented by 1, 1 by 11, 2 by 111 etc. Sets of numbers are represented as unary numbers separated by the separator \#. So a pair $(3, 5)$ is represented by the tape expression 1111\#111111. The generalization to n-tuples is obvious. The following notation is convenient.

Let $\overline{n}=\underbrace{11...1}_{n+1}$. Then the d-tuple $(n_1,n_2,...,n_d)$ is represented by the tape expression

\begin{equation}\label{eq: numericdata}
\overline{(n_1,n_2,...,n_d)}=\overline{n_1}\#\overline{n_2}\#\ldots\#\overline{n_d}.
\end{equation}

In order to start the computation according to the definition; an {\itshape initial instantaneous description} must be given. We set 
\begin{equation}\label{eq: initinstdescr}
\alpha_1=q_1\overline{(n_1,n_2,...,n_d)}.
\end{equation}

The numeric result of the computation should be read off from the terminal configuration. The only available way to do this is to remove the single $q_i$, upon which we get a tape expression, which must be interpreted as a number. A simple interpretation is to count the number of occurrences of 1, neglecting \#. Another choice, more restrictive, is to demand that the terminal state consists of a single consecutive stretch of 1's on an otherwise blank tape. 

The question of which terminal states should count as yielding acceptable output is really a question of how to code output data, but it affects the way programs for the machine are written. A choice often made is to demand that the machine should halt scanning the leftmost symbol on an otherwise blank tape.  Then one has to add instructions to clean up the tape after the computation proper is finished and then move left to the leftmost symbol. Whether this is worthwhile is a matter of taste. Formally, this choice of output coding corresponds to a terminal state of the form $\alpha_p=q_h\overline{n}$. 

This means that there is no instruction having the first two tokens $q_h1$ . Thus $q_h$ is the halting state (or one of the halting states). I will call this a  {\itshape standard terminal configuration}.\footnote{For practical programming purposes, one can note that three situations can be envisioned; (1) the computation does not terminate (halt) and no output data results, (2) the computation terminates in a standard configuration and (3) the computation terminates in a non-standard configuration. It seems that allowing this last case to define output data, though not wrong in principle, is a bit risky in practice as one has less control over the workings of the computation.}
  
Let us finally connect these, admittedly a bit heavy-handed notations, to functions by explicit identifying computations and functions. 

We associate a function $f_{M}:N^d\rightarrow N$ with a Turing machine $M$ in the following way.

For each d-tuple $(n_1,n_2,...,n_d)$ we set the initial state $\alpha_1=q_1\overline{(n_1,n_2,...,n_d)}$.

(a) If there exists a computation $\alpha_1,\alpha_2,...,\alpha_p$ such that 

$$
M\big(q_1\overline{(n_1,n_2,...,n_d)}\big)=q_h\overline{n}\] then \[f_{M}(n_1,n_2,...,n_d)=n
$$

(b) If no computation exists then $f_{M}(n_1,n_2,...,n_d)$ is undefined.

\subsubsection{More efficient numeric input/output conventions}
The unary description of numeric data is highly inefficient. It takes an exponential amount of tape space to represent a number as compared to a binary representation. Using $n$ bits, which can be written on $n$ tape cells, numbers ranging from 0 to $2^n-1$ can be represented, giving a logarithmic decrease of space requirements. It is convenient to use an alphabet $\{0,1,\#\}$ with an explicit blank symbol '$\#$' used to separate the numbers and write numbers in binary notation. 

Leaving the actual encoding of numbers open, (unary, binary or in any other base), we can use the same notation as in the preceeding section. Let $\overline n$ be the code for the number $n$ and $\overline{(n_1,n_2,...,n_d)}$ the code for the $d$-tuple of numbers $(n_1,n_2,...,n_d)$, then 

$$
f_{M}(n_1,n_2,...,n_d)=M(q_1\overline{(n_1,n_2,...,n_d)},
$$ 
if the computation exists, otherwise the function is undefined.

It is often convenient to simplify the notation somewhat and simply write $M(n_1,n_2,...,n_d)$ for $M(q_1\overline{(n_1,n_2,...,n_d)}$, so that $f_M(x)$ and $M(x)$ denotes the same object.

\subsubsection{The blank symbol confusion}
There seems to some confusion in the literature as to how to treat the blank symbol. In formal language theory, if one wants to have strings with blanks in them, the obvious way is simply to include a special blank symbol among the symbols. Since a string is always finite, there is no need to designate the beginning and end of a string in any special way. In particular, starting a string with blanks, or ending it with blanks, makes no sense. Such blanks would be trimmed away. 

However, in Turing machine theory, the tape is potentially infinite, and the machine needs some way to know where the right and left ends of the actually used part of the tape is. The obvious way would be to use the blank symbol as such a designator. This is often phrased as, '... a string written on an otherwise blank tape ...'. But then, if the blank is in the alphabet, and the string written on the tape contains blanks, the machine will not know whether a blank designates the end of the actually used tape, or if it is a blank within the string (as in the case where the blank separates the numbers $n_i$ in a $d$-tuple input). One way around this dilemma is to use two consecutive blanks, $\sqcup\sqcup$, to designate tape ends. In that case, the languages defined over the alphabet, needs to exclude strings with two consecutive blanks, otherwise, the confusion remains. Thus, when languages $L$ over the alphabet $\Sigma$ is mentioned, it is understood that no words in the language contain two consecutive blanks. 

Another way is to include two different 'blank' symbols, for example $\{\sqcup,\#\}$, one $\#$ denoting 'string blanks' or input separators, the other $\sqcup$ designating the 'left' and 'right' ends of the tape. This is the convention used in the present work. Any other 'language' blanks play no role in defining the general model of Turing machines and need only be defined in specific examples.

\subsubsection{String processing}
The model is, of course, not restricted to computing numeric functions. In general, a Turing machine, performs string processing, taking an input string $w_i$ from the set of strings $\Gamma^*$, producing an output string $w_o$ if it halts on the input. More formally, the Turing machine $M$ defines a partial function 

\begin{equation}\label{eq: StringProcessing1}
f_M:\Gamma^*\rightarrow\Gamma^*
\end{equation}
where
\begin{equation}\label{eq: StringProcessing2}
f_M(w_i)=w_o\in \Gamma^* 
\end{equation}
if  $M$ halts on input $w_i$, undefined otherwise.

\subsubsection{The state graph}
Since the set of tape expressions is infinite, the space of instantaneous descriptions, or states, is also infinite. A computation can be viewed as a directed graph in this space. This graph will be denoted with $G_c=(S_v,T_e)$ where $S_v$ ({\it state vertices}) denotes the set of vertices corresponding to states in the computation, and $T_e$ ({\it transition edges}) denotes the set of edges corresponding to transitions, i.e. computational steps. Two vertices $v_i$ and $v_j$ are connected by an edge if there is a corresponding instruction in the program, taking the machine from state $v_i$ to state $v_j$. 

Note that for a deterministic Turing machine, the state graph is simply a path in the state space. For a computation that halts, i.e. a computation that starts in a certain state and ends in another state, the path is non-intersecting. This can be understood as follows. If the path intersected itself, so that the machine returned to an 'earlier' state, then the machine would enter an infinite loop, and would not halt. Therefore, terminating computations corresponds to linear paths. 

\subsubsection{Concluding the formal definition of a Turing machine}
When defined in this way, everything looks static. Where does motion enter? Well, the computation, i.e. the series of instantaneous descriptions must be computed, at least once for each input $d$-tuple. Someone or something has to do this, human or machine. This is where motion enters. This is obvious if one considers doing the calculation with pen and paper. 

Also note that it is sometimes convenient to work with Turing machines with several tapes with concomitant read/write heads. 

Apart from the question of actually performing the computations, this is a formal theory of computation. There are many other models of computation. Off the formal ones, we have the (Herbrand-G{\"o}del) recursive functions, Church's $\lambda$-calculus, both contemporary with the Turing model. The RAM (Random Access Machine) model is close to an actual computer. Then there are lots of simplified programming languages, containing just the bare minimum of constructions. A survey of computational models can be found in \cite{Emdeboas1990}.

%********************************
\subsection{Syntax and semantics}
It is interesting in this context to digress slightly and discuss the question of syntax vs semantics for this model. The Turing machine model reduces computation to syntax. Everything written above could easily be phrased in terms of a specification of a formal language. No meaning is conferred to the elements of the model. One does not need to understand the tape expressions or the instructions in order to carry them out. An executing Turing machine just performs meaningless string processing.

The grammar of the language is the specification of what is well-formed programs (sets of instructions) and what constitutes well-formed instantaneous descriptions, in particular the initial description. The computational rules, also syntactically defined, then tells us how to perform computations within the model. The data itself has no meaning, it is just a string of symbols.

The semantics of the model enters through the interpretation of the initial and final tape expressions as defining sets of natural numbers, or objects from some other set of mathematical objects, functionally related via the computation. Thus, we can say that we understand a Turing machine if we understand what it computes.

%********************************
\subsection{Decision procedures and Computation procedures revisited}
We will now refine our notions about algorithms for decisions and computations respectively.

\subsubsection{Deciding recursive languages}
Let ${\bf L}\subset\Gamma^*$ be a language, i.e. a set of strings defined over the alphabet $\Gamma$. Next, let $M$ be a Turing machine and $x\in\Gamma^*$ be an input string. We say that $M$ {\it decides} the language $\bf L$ if the following conditions hold

\begin{equation}\label{eq: DecideLanguage}
\cases{M(x)\succ q_{y} &if $x\in {\bf L}$ \cr M(x)\succ q_{n} &if $x\notin {\bf L}$}.
\end{equation}

Here we use the notation $M(x)\succ q_{y}$ to denote the sentence: "The machine $M$ on input string $x$ halts in the configuration $q_y$", and similarly in the other case.

If the language $\bf L$ is decided by some Turing machine $M$, then $\bf L$ is a {\it recursive language}.

There is also a weaker form of decision procedures.

\subsubsection{Recognizing recursively enumerable languages}
We say that $M$ {\it recognizes} the language $\bf L$ if the following conditions hold

\begin{equation}\label{eq: AcceptLanguage}
\cases{M(x)\succ q_{y} &if $x\in {\bf L}$ \cr M(x)\succ\triangleright &if $x\notin {\bf L}$},
\end{equation}
where by $M(x)\succ\triangleright$ we denote the sentence: "The machine $M$ on input string $x$ does not halt".

If the language $\bf L$ is accepted by some Turing machine $M$, then $L$ is a \nobreak {\it recursively enumerable language}. It is obvious that a recursive language is also recursively enumerable. The weaker form is still strong enough to enumerate the strings in the language. By judiciously employing the machine $M$ that accepts the language, the strings of the language can be enumerated. The intuition is that, if a string belongs to the language, it will be found eventually. But for a string not yet accepted, there is no way of ascertaining that it does not belong to the language.

When a machine is used for decision problems, the output is really encoded in the halting states $\{q_y,q_n\}$ and the tape contents at halting have no special significance.

\subsubsection{Computing recursive functions}
Suppose that $f$ is a function from $\Gamma^*$ to $\Gamma^*$. If there is a Turing machine that computes $f$ as in (\ref{eq: StringProcessing1}) and (\ref{eq: StringProcessing2}), f is called a {\it recursive function}.

%********************************
\subsection{The Church-Turing Thesis}
The Church-Turing thesis identifies the set of effectively (intuitively) computable functions with the set of functions computable within any of the classical computational models; Turing machines, $\lambda$-definable functions or general recursive functions. It was originally formulated by Church in terms of general recursive functions, but Turing made similar remarks in reference to his model, hence the name Church-Turing thesis (see several articles in \cite{Herken1994}). Historically, effective computability meant computability by a human comput{\itshape or} who works to precise rules. Later the thesis has acquired connotations connecting it to machine computation, in particular electronic digital computing machines. In this sense the thesis is certainly true; what can be computed by a general purpose digital computer can be computed by a Turing machine.\footnote{The role of the Turing machine model of computation for the development of the modern digital computer is discussed in \cite{Goldstine1972}.}

The literature contains stronger statements to the effect that anything that can be computed by a machine can be computed by a Turing machine (for a discussion, see \cite{Copeland-SEP2002}). This is a much stronger statement. It is a statement about every conceivable physical system that can be harnessed to perform computations. Whether it is true or not is unknown. To determine if this statement might be true or not, we would have to analyze the general computational characteristics of physical systems. Such an investigation seems to require a complete theory, or set of theories, covering all of physics. Even though many physicist are pursuing research into finding a "Theory of Everything", it is far from clear whether such a theory exists, or if it can be found in any near future. And should such a theory exist, we know nothing of its implications for computability.

It is not really primarily a question of theory, but rather of phenomena. New physical phenomena might very well be discovered in the future that require new theoretical concepts for their explication. 
%********************************
\subsection{Computability}
A classic result in the theory of computability is that there are non-computable functions. This follows, almost trivially, once one has accepted the following three propositions; 
\begin{itemize}
\item{(i)} the set of Turing machine programs is enumerable,
\item{(ii)} there are non-denumerably many functions $f:N\rightarrow N$,
\item{(iii)} the Church-Turing thesis.
\end{itemize}

The first two propositions has the status of mathematical theorems as they can be formulated within precisely defined formalisms. The third, cannot be proved as it relates the intuitive notion of an effective method to the formal models of computation.  

The enumerability of the Turing machines follows from the requirement that every Turing machine program must be stated as a finite set of instructions, each instruction being built from a finite number of tokens. Each program therefore is a finite string of tokens from a finite alphabet, and since such strings can be enumerated, the set of Turing machine programs is enumerable. For a complete proof, an explicit numbering of the programs must be given, but that can be done based on G\" odel numbering. Also, by explicitly giving an enumeration, a particular non-computable function can be exhibited. 

The non-enumerability of the functions $f:N\rightarrow N$ can be proved using a diagonal argument. The proof, though standard and well-known, will be given here since it illustrates the {\it diagonal method}\footnote{The method was invented by G. Cantor.} in a simple setting. First note that we are considering all such functions, both partial and total. Suppose that we are given an enumeration of all functions $F=\{f_n\}_{n=0}^{\infty}$. We can then define a new function $u$, called the anti-diagonal function \cite{BoolosJeffrey1974}, where
$$
u(n) = \cases{1,&if $f_n(n)$ is undefined\cr
	f_n(n)+1,&otherwise.\cr}
$$
This is a well-defined total function. Note that questions of computability do not enter at this stage. If the list $F$ is complete, then the function $u$ must be one of the functions in the list, say $f_m$ and thus $u(x)=f_m(x)$ for every number $x$. In particular $u(m)=f_m(m)$ Using the definition of $u$ we get
$$
u(m) = f_m(m) = \cases{1,&if $f_m(m)$ is undefined\cr
	f_m(m)+1,&otherwise.\cr}
$$
This contradiction proves that the list $F$ cannot be complete and the set of all functions cannot be enumerated in any way. 

Computability enters when we ask the question whether the anti-diagonal function can be computed or not. If the list $F$ was compiled using Turing machines, i.e. if the list is a list of all Turing computable functions, then the proof shows that there are Turing non-computable functions. This argument can be used on any computational model. If $F$ is a list of all functions computable within a certain named model, then there are functions that are not computable within this model. A priori, different models of computation could give rise to different sets of computable functions. It is an empirical fact that this is not the case.

It turns out that all computational models, claiming to capture the idea of effective computability, that has been considered so far, can be proved to yield the same set of computable functions. 

The next question is, can the function $u$ be computed effectively at all, using some intuitive model? If that is the case, then the classic computational models are to narrow. On the other hand, if the Church-Turing thesis is true, then the function $u$ is absolutely uncomputable.

It is clear what the key point is. If we consider total functions, i.e. functions defined for all numbers, then the diagonal argument shows that any computational model that computes total functions is too narrow. In this case, the clause taking care of the cases when the function is undefined, is not needed. The anti-diagonal function can be defined, and it is easily proved that it cannot occur in the list of functions. Therefore the model is incomplete. But in this case, the anti-diagonal function is intuitively computable. This is often phrased as saying that we can diagonalize out of any computational model for total functions. This intuitive computation of the anti-diagonal function relies on examining the list of functions and computing its values based on this list, so it could also be seen as a meta-computation. 

The question of whether it is possible to diagonalize out of the model or not when partial functions are allowed, depends on whether there is any general procedure to determine if a function is  defined for a certain number or not. If there is such a procedure, it could be used to compute the anti-diagonal function.

Now, for the Turing machine model, a function is left undefined for a certain argument in two cases. Either the machine stops in a non-standard configuration, a case which can be taken care of by proper programming. Or the machine never stops. If there is a general effective method which is capable of determining (in a finite amount of time) whether or not Turing computations halt, then this method could be used to diagonalize out of the Turing model. This is the {\it halting problem}. It can, in fact, be shown that the halting problem is unsolvable within the Turing model.

If one could devise a computational model, formal or intuitive, which were able to solve the Turing halting problem, then that model would in some sense be stronger than the Turing model. To date, there is no such model.

\subsubsection{The halting problem}
The general halting problem is the problem of designing an effective method, intuitive or within the Turing machine model, to determine whether a particular Turing machine $M_n$ will ever halt when started to compute with input data $m$. If a certain computation does not halt, this means that the corresponding function is undefined. Therefore, the halting problem is closely related to the question of computability. 

The algorithm has access only to the Turing machine programs and the input data on the tape. This makes sense, because it is of no use just to set the machines running and wait to see if they will stop in a standard terminal configuration. The machines, 'destined' not to stop, will run forever, and the answer cannot be obtained by waiting.  

It can be proved that the halting problem is unsolvable within the Turing machine model. The proof is non-trivial and technical, and we will just outline it in the next section.

If the Church-Turing thesis is correct, the general halting problem is therefore unsolvable.

%********************************
\subsection{Universal Turing machines}
The Turing machines considered so far are special purpose machines. Each and every machine is constructed to solve a particular algorithmic problem, the program being encoded in the list of instructions. We will now argue that there exist universal Turing machines $U$, which act like general purpose computers. They are programmable in the sense that, given a description of a certain Turing machine $M$, and its input $x$, it mimics the computation of $M$. Leaving open the details for the moment, by a {\it description} of $M$, we mean a symbolic representation of the set of instructions for $M$ in the alphabet of the universal machine. In order not to clutter the notation, the description of $M$ will also be denoted by $M$ since any machine is essentially defined by its set of instructions anyway. So, if the result of running $M$ with input $x$ is $M(x)$, i.e. the function $f_M(x)$, then we write $M(x)=U({ M};x)$ to denote that the universal machine computes the same result when given as input, the description $M$ as well as the 'data' $x$. 

As a preliminary step, note that the Turing machines can be enumerated and collected into an infinite list $[M_i]_{i=1}^\infty$. The alphabets are fixed and the programs can written as strings by concatenating the instructions. Thus, the enumeration can be performed using a lexicographic ordering starting by first ordering all one-state machine programs, then all two-state machine programs and then continuing in this way.

The actual construction of universal machines is quite complicated if it is to be carried out in full detail. One complication is that the different machines $M$ could very well have different alphabets $\Gamma$ and $\Sigma$, and consequently, the universal machine must be able to accommodate a potentially infinite set of symbols. However, since for any particular machine, the set of symbols is finite, it is possible to map this set of symbols one-to-one onto a standard set, say $\Gamma=\{0,1\}$ and $\Sigma=\{0,1,\#,\sqcup\}$ using some binary coding. This will be our strategy. 

Furthermore, $U$ must be able to accommodate a potentially infinite set of labels for internal configurations of the simulated machines. This we also standardize by encoding the configuration labels using the very same alphabet $\Sigma$. In this way, both the input data and the program for the simulated machines are encoded using the same alphabet. This is useful, since it then makes sense to provide the program of a Turing machine as input to the universal machine. 

The internal configurations of $U$ itself may be labeled by any suitable set.

The construction of $U$ is simplified if it is built as a two-tape machine. The first tape can then be dedicated to storing the program for the machine being simulated. The second tape of $U$ is used to store the instantaneous descriptions of the simulated machine $M_i$. The specific set of instructions for $U$ itself, which in accordance to the Turing machine model, is not stored on any tape, but instead is part of its finite state control, can be thought of as an operating system.

We can now informally describe the workings of the universal machine. Upon being set in motion, it scans the leftmost symbol on the second tape (this is the starting configuration of $M_i$), then it scans the next symbol to the right (the symbol that $M_i$ itself would have scanned). Having done this it knows the both the internal configuration and scanned symbol of $M_i$. Then it scans the first tape, looking for a matching instruction. If such an instruction is found, it is performed on the second tape. Thus the first step in simulating $M_i$ is performed. Next it scans the second tape looking for a symbol corresponding to a configuration of $M_i$, then it scans the symbol to right (which again is the symbol scanned by $M_i$). Then it scans the first tape again looking for a matching instruction. Having found it, it is performed. Continuing in this way it is clear that the workings of $M_i$ is simulated. What remains to be done if the construction is to be carried out in detail is to code these operations in terms the primitives of $U$. 

\subsection{The halting problem is undecidable}
We are now in a position to state the halting problem and prove that it is undecidable. In order to use the formalism set up so far we will phrase the problem in terms of decision problem. 

Let $\bf H$ be a language defined by

\begin{equation}\label{eq: DefinitionHaltingLanguage}
{\bf H}=\{M;x:M(x)\not\succ\triangleright\},
\end{equation}

which is read out as "The language consisting of all strings that encode a Turing machine $M$ and an input $x$ such that the machine halts on the input.". 

\subsubsection{Theorem}
{\bf H} is recursively enumerable.
\subsubsection{Proof}
What is needed is a Turing machine $H$ that accepts the language $\bf H$. According to the definition recursively enumerable languages (\ref{eq: AcceptLanguage})

\begin{equation}\label{eq: AcceptHalting}
\cases{H(M;x)\succ q_y &if $M;x\in {\bf H}$ \cr H(M;x)\succ\triangleright &if $M;x\notin {\bf H}$}.
\end{equation}

But then $H$ is precisely a universal machine programmed so that it halts in the accepting configuration $q_y$ whenever the machine $M$ halts on input $x$.

\subsubsection{Theorem}
{\bf H} is not recursive.
\subsubsection{Proof}
Suppose contrary to the proposition that there exist a Turing machine $H$ that decides $\bf H$. This means, according to (\ref{eq: DecideLanguage}), that we have

\begin{equation}\label{eq: DecideHaltingAssumption}
\cases{H(M;x)\succ q_y &if $M;x\in {\bf H}$ \cr H(M;x)\succ q_n &if $M;x\notin {\bf H}$}.
\end{equation}

Now since both the input data and the program are encoded using the same alphabet, it is possible to write programs for the universal machine that takes programs for other Turing machines as input. Consider therefore a new machine $D$ which is a modification of the machine $H$ and which takes the description $M$ as input. It is defined by 

\begin{equation}\label{eq: DiagonalMachine}
D(M): \cases {D(M)\succ\triangleright \mbox{    \bf{   if}   } H(M;M)\succ q_y \cr 
D(M)\succ q_y \mbox{    \bf{if}   } H(M;M)\succ q_n}
\end{equation}
 
Under the assumption that $H$ exists, this is a perfectly well defined computation. $D$ can be explicitly defined by just adding a small set of instructions to $H(M;x)$ directing it to move right forever in the case that $H(M;M)$ accepts (which it does eventually by assumption), or directing it to accept if $H(M;M)$ accepts (which it also does eventually by assumption). 

Let us spell out the definition of $D$ explicitly. On given input $M$, $D$ first simulates the universal machine $H$ on input $M;M$. Then, in the case where $M$ should have entered the accepting configuration $q_y$, $D$, which is reprogrammed (as compared to $H$) to 'loop', just continues to move forever right along the tape. In the case where $M$ should have entered the rejecting configuration $q_n$, $D$ is reprogrammed to accept.

But now we can ask how $D$ behaves when it is run with a description of itself as input, i.e. how does $D(D)$ behave? The definition of $D(D)$ immediately gives

\begin{equation}\label{eq: DiagonalMachineSelfApplied}
D(D): \cases {D(D)\succ\triangleright \mbox{    \bf{   if}   } H(D;D)\succ q_y \cr 
D(D)\succ q_y \mbox{    \bf{if}   } H(D;D)\succ q_n}.
\end{equation}

Let us analyze this. Does $D(D)$ halt or not? 

Suppose it does not halt, i.e. $D(D)\succ\triangleright$. That case occurs when $H$ accepts the input $D;D$. But then it follows from the assumption (\ref{eq: DecideHaltingAssumption}) about $H$ that $D;D\in {\bf H}$. This, in its turn implies $D(D)\not\succ\triangleright$ using the definition (\ref{eq: DefinitionHaltingLanguage}) of the language $\bf H$.

On the other hand, suppose $D(D)$ does halt. That case occurs when $H$ does not accept the input $D;D$. But then it follows from the assumption (\ref{eq: DecideHaltingAssumption}) about $H$ that $D;D\notin {\bf H}$. Consequently, $D(D)$ does not halt according to the definition (\ref{eq: DefinitionHaltingLanguage}) of the language $\bf H$.

Both ways, we get a contradiction. The conclusion is that the universal machine $H$ deciding $\bf H$ does not exist.

Note, that this proof hinges on a delicate interplay between the definition (\ref{eq: DefinitionHaltingLanguage}) of the halting language  $\bf H$, the assumed properties of the universal machine $H$ purported to decide $\bf H$ and the derived properties of the 'diagonal' reprogrammed machine $D$.

%BOOLEAN LOGIC
%*************
%\section{Boolean logic}

%CLASSICAL CIRCUIT MODEL OF COMPUTATION
%**************************************
\section{The classical circuit model of computation}
The circuit model of computation is based on the classical logical gates like AND, OR and NOT. Since the 70\rq s these are implemented as physical devices in the form of TTL or CMOS technology. In a microprocessor there are millions of gates, but they are also packaged in components containing a few gates, which can be wired together on circuit boards using traditional soldering techniques. The abstract logical values $\{true, false\}$ are represented by voltage levels. In this section we will review the circuit model as a theoretical model of computation, but everything in this model have a physical realization in terms of real world circuits and wires. 

A circuit is made up of wires and gates. The inputs and outputs of the gates are bits, either represented by $\{true, false\}$ or more conveniently by $\{0,1\}$. A single gate might have any number of inputs and outputs, though in practice the basic building blocks have just a few inputs and outputs. 

A circuit with k inputs and l outputs corresponds to a function \linebreak $f:\{0,1\}^k\rightarrow\{0,1\}^l$. 

\begin{figure}[h]
\epsfbox{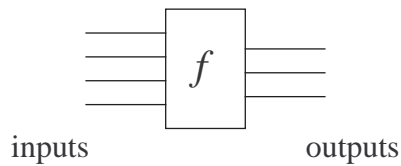}
\caption{A general gate.}
\label{fig: GeneralGate}
\end{figure}

Bits are carried from one gate to another through wires. By connecting gates with wires a circuit is built. No loops or feedback are allowed in the circuit as that generally leads to instabilities. Wires can be split into two or more wires, thus duplicating the bit they are carrying.\footnote{This is sometimes coded in terms of a FANOUT gate, see below.} All circuits can be built using just one type of logical gate, often chosen to be the 2-input gate NAND. NAND is an AND gate followed by a NOT gate. The NAND gate is therefore said to be {\it universal} for classical (non-reversible) computation. Circuits are often easier to construct and understand if one allows oneself to use a larger set of gates: NOT, AND, OR, NAND, XOR. 

%\clearpage
\subsubsection{The basic circuit elements}

\begin{figure}[h]
\epsfbox{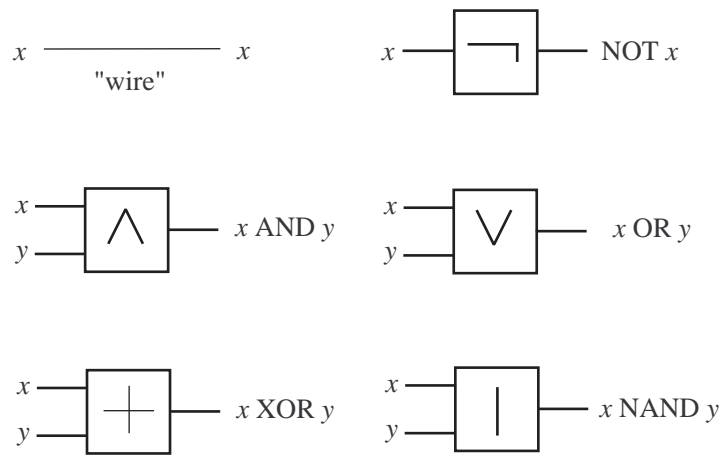}
\caption{The basic circuit elements.}
\label{fig: BasicCircuitElements}
\end{figure}

\subsubsection{Input/output relations for the basic circuit elements}

The outputs of these gates for different combinations of inputs are given by the following table.

\vskip 1cm
\begin{tabular}{|c|c|c|c|c|c|}
\hline
x & y & x AND y & x OR y & x XOR y & x NAND y\\ \hline
0 & 0 & 0 & 0 & 0 & 1 \\ 
0 & 1 & 0 & 1 & 1 & 0 \\ 
1 & 0 & 0 & 1 & 1 & 0 \\ 
1 & 1 & 1 & 1 & 0 & 0 \\ \hline
%\caption{The basic circuit elements.}
%\label{tab: BasicCircuitElements}
\end{tabular}
\vskip 1cm

The NOT gate simply flips 0 to 1 and 1 to 0.

\vskip 0.5cm
\begin{tabular}{|c|c|}
\hline
x & NOT x\\ \hline
0 & 1 \\ 
1 & 0 \\ \hline
\end{tabular}
\vskip 0.5cm

Apart from the logical gates one also needs the FANOUT gate. It is not really a logic gate but just a splitting off of a wire into several wires, all carrying the same bit as the original wire. Electrically this is in fact just a splitting off of a wire, although in practice, since in real wires there are small currents flowing, one might get problems with the voltage levels at the outputs if a gate is drained to heavily. There are therefore limits on maximum fanout for real electronic gates.

\begin{figure}[h]
\epsfbox{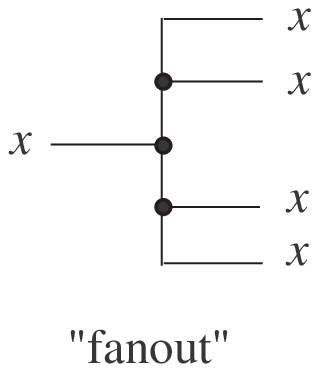}
\caption{The fanout gate.}
\label{fig: Fanout}
\end{figure}

Below, we give just on example of a simple circuit, the half-adder, which can be used as a building block in a circuit for binary addition.

\begin{figure}[h]
\epsfbox{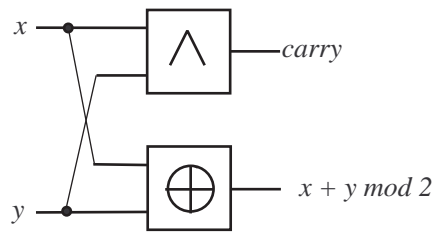}
\caption{A half-adder circuit.}
\label{fig: HalfAdder}
\end{figure}

One further concept is the {\it ancilla} or auxiliary (work) bit. It is a fixed bit, set to 0 or 1 once and for all. Physically this is realized by a fixed voltage.

As is well known, there is a complete isomorphism between the circuit model and Boolean algebra and the functions computed by a circuit are often called Boolean functions. In fact the easiest way to analyze a circuit is using Boolean algebra. There is also a very close correspondence with propositional logic, although in logic the focus is different.\footnote{In propositional logic the focus is on the logically true propositions and the notion of completeness, i.e. the question of whether the logically true, and only the logically true propositions can be derived within the system. This question has since long been settled.}  

Note that there is no computational steps involved here apart from the time it takes for the basic circuit elements to compute the outputs from the inputs. For real physical devices this time is of the order of nanoseconds. Apart from this time lag, the outputs appears as soon as the inputs are applied. 

%********************************
\subsection{The circuit model and non-computable functions}

We will now prove that there are circuits to compute any function \linebreak $f:\{0,1\}^k\rightarrow\{0,1\}^l$. The proof uses induction over the number of input bits.

\subsubsection{Theorem}
Every function $f:\{0,1\}^k\rightarrow\{0,1\}^l$ has a circuit that computes it.
\subsubsection{Proof}
First note that it suffices to prove the assertion for functions $f:\{0,1\}^k\rightarrow\{0,1\}$ as the $l$-bit output case is easily put together from $l$ 1-bit output functions. For $k=0$ there is nothing to prove. For $k=1$ there are four possible functions:
\begin{enumerate}
\item The identity. A circuit consisting of a single wire computes this function.
\item The bit flip. This function is computed by a {\bf NOT} gate.
\item The constant function with output 0. This is computed by an {\bf AND} gate with one input bit taken to be an ancilla bit equal to 0.
\item The constant function with output 1. This is computed by an {\bf OR} gate with one input bit taken to be an ancilla bit equal to 1.
\end{enumerate}

For the induction step, assume the assertion true for $k=n$ . Now let $f$ be a function of $n+1$ bits. Define $n$ bit functions $f_0$ and $f_1$
\[f_0(x_1,...,x_n)=f(0,x_1,...,x_n)\]
\[f_1(x_1,...,x_n)=f(1,x_1,...,x_n)\]
These are both  $n$-bit functions and are therefore computed by circuits. The function $f$ is now computed by the circuit implementing the formula
\[f(x_0,x_1,...,x_n)=(x_0{\bf AND}f_0(x_1,...,x_n)){\bf XOR}(({\bf NOT}x_0){\bf AND}f_1(x_1,...,x_n))\]
or using the more convenient Boolean algebra notation
\[f(x_0,x_1,...,x_n)=(x_0\wedge f_0(x_1,...,x_n))\oplus (\neg x_0\wedge f_1(x_1,...,x_n))\].
The result now follows by induction.

So, using gates and wires, any function $f:\{0,1\}^k\rightarrow\{0,1\}^l$ whatsoever can be computed. This does not, however, mean that any function $f:N\rightarrow N$ can be effectively computed using circuits. That would run counter to the well-established fact that there are non-computable functions. This is an interesting point that we will examine in some detail.
Suppose we want to compute a simple function like $f(x)=x^2$  for all values of the argument. No single circuit can do this, as it is immediately clear that the input must be represented in the form of a bit string, or a binary number, and each bit must be carried by a single wire. So a 2-bit input circuit can calculate the function for at most the numbers 0, 1, 2, 3. A 3-bit input circuit manages the numbers 0 through 7. So what we really need in order to compute the square function is a enumerable infinite family of circuits. Let us make this notion precise.

\subsubsection{Consistent circuit families}
A consistent circuit family consists of denumerably infinite set of circuits $\{C_n\}_{n=0}^{\infty}$ with the properties
\begin{enumerate}
\item The circuit $C_n$ has $n$ input bits and a finite number of extra ancilla bits as well as a finite number of output bits.
\item The output from $C_n$ is denoted by $C_n(x)$ and is defined for all binary numbers $x$ of at most $n$  bits of length.
\item If $m < n$  and $x$ is at most $m$ bits in length then $C_m(x)=C_n(x)$. This is the consistency requirement.
\end{enumerate}

What prevents the circuit model from yielding an effective method for computing every function whatsoever is the fact that there does not exist an effective method to construct the circuits in the family for every number $n$. 

By a {\itshape uniform circuit family} we mean a consistent circuit family for which there does exist an algorithm, for example running on a Turing machine, which computes a description of the circuit for every number $n$. In this way, the uniform circuit model is by definition equivalent to the other models of computation.

From this we see a fundamental difference between the circuit model and the Turing machine model. Once a Turing machine is programmed, it will compute the values of the function for every input number   for which it halts. A given circuit with a given number of input bits only computes the function for a finite range of values. Beyond this range of values, a new circuit (in the family) is demanded.

Another way of looking at the fact that non-uniform circuits compute all functions is that the non-uniform circuit model cannot be finitely described. The list of circuits is infinite and we have no finite way to generate the list of circuits. It therefore falls outside the characterization of a finitely defined algorithm.
It is clear that circuits can be built to compute any computable function, but in this way one gets specialized circuit families for each computational task. In order to get a universal model of computation, the circuit must be wired to perform a standard set of instructions on input data, the computational task itself being supplied as a program. This is the way an ordinary von Neumann architecture digital computer works in a (fetch instruction, fetch data, execute instruction, save data) cycle. The circuit must thus be clocked and in this way computational steps are introduced.

The circuit model is useful to describe quantum computation, but in itself it is rather awkward. What makes an algorithm for an infinite set of instances of a problem useful is the fact that once the algorithm is known, it permits us to obtain {\itshape new} knowledge. If we do not know the value of a computable function for a certain argument, run the algorithm to find that value out! Using circuits, new members of the circuit family must be computed in order to get new values of the function. 

In a sense, this is a reflection of the fact that the circuits really just furnish formulas for the function values, not algorithms. If there is an explicit formula for a function, no algorithm is needed to compute the values.
%*******************************
\subsection{Reversible gates}
The classical logical gates are all irreversible except for the {\bf NOT} gate. This means that the values of the input bits cannot be inferred from the values of the output bits. Just one example illustrates the point. If an {\bf AND} gate outputs 0, there is no way to know which of the possible input combinations 00, 01 or 10 resulted in the output.

In \cite{Landauer61-82-96} and \cite{Bennet1973} it was shown that an irreversible logical operation has to dissipate a certain minimum amount of energy. On the other hand, a reversible logical operation does not have to dissipate any energy. This lead to an interest in reversible computations, and this was also one of the initial motivations behind research into quantum computation, since the evolution of closed quantum systems are inherently reversible.

The first requirement for a gate to be reversible is that the number of output gates equals the number of input gates. A simple example is the {\bf CNOT} gate. It has two input bits and two output bits. 

\begin{figure}[h]
\epsfbox{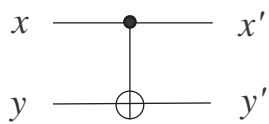}
\caption{The CNOT gate.}
\label{fig: CNOT}
\end{figure}

The function computed by this gate is shown in the 'truth' table below.
\vskip 1cm
\begin{tabular}{|c|c||c|c|}
\hline
x & y & x'& y' \\ \hline
0 & 0 & 0 & 0 \\ 
0 & 1 & 0 & 1 \\ 
1 & 0 & 1 & 1 \\ 
1 & 1 & 1 & 0 \\ \hline
\end{tabular}
\vskip 1cm

The name {\bf CNOT} stands for {\bf controlled-NOT}. The first input bit $x$ controls the second input bit in such a way that when $x=0$ , the second output bit $y'=y$, and when $x=1$, then $y'= {\bf NOT}\,y$. The first output bit $x'$ is always equal to $x$. Note that the gate can also be regarded as a generalization of {\bf XOR} since $y'=x\,{\bf XOR}\,y$.

The fact that the {\bf CNOT}-gate is reversible can be seen in two ways. First, by simply inspecting the truth table, it is clear that knowing $x'$ and $y'$, $x$ and $y$ can be uniquely derived. Secondly, if a second {\bf CNOT} is connected after the first {\bf CNOT}, the total effect will be same as just to unit wires, i.e. {\bf CNOT} is its own inverse.

The functional relations between inputs and outputs can thus be written

$$
\cases{x'=x\cr
y'=x\oplus y}.
$$

The CNOT gate can be used to make a copy of an input bit. If $y$ in the truth table is fixed to 0, both $x'$ and $y'$ are equal to $x$.

\vskip 1cm
\begin{tabular}{|c|c||c|c|}
\hline
x & y & x'& y' \\ \hline
0 & 0 & 0 & 0 \\ 
1 & 0 & 1 & 1 \\  \hline
\end{tabular}
\vskip 1cm
This can also be seen as a FANOUT.

Another reversible gate is the Toffoli gate. It has three input wires and three output wires. 

\begin{figure}[h]
\epsfbox{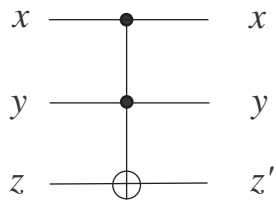}
\caption{The Toffoli gate.}
\label{fig: ToffoliGate}
\end{figure}

The function computed by this gate is shown in the 'truth' table below.

\vskip 1cm
\begin{tabular}{|c|c|c||c|c|c|}
\hline
x & y & z & x'& y'& z'\\ \hline
0 & 0 & 0 & 0 & 0 & 0\\ 
0 & 0 & 1 & 0 & 0 & 1\\ 
0 & 1 & 0 & 0 & 1 & 0\\
0 & 1 & 1 & 0 & 1 & 1\\
1 & 0 & 0 & 1 & 0 & 0\\
1 & 0 & 1 & 1 & 0 & 1\\
1 & 1 & 0 & 1 & 1 & 1\\ 
1 & 1 & 1 & 1 & 1 & 0\\ \hline
\end{tabular}
\vskip 1cm

The functional relations between inputs and outputs can be written

$$
\cases{x'=x\cr
y'=y\cr
y'=(x\wedge y) \oplus z}
$$

The first two bits, $x$ and $y$, can be regarded as {\it control} bits, they are not changed by the gate. Instead the AND of $x$ and $y$ determines whether the third bit $z$ is flipped or not. The third bit can therefore be regarded as a {\it target} bit. This terminology is used in quantum computation.

The reversibility of the Toffoli gate can be seen in exactly the same way as for the {\bf CNOT} gate.

The Toffoli gate turns out to be universal for reversible computation. This is easily seen as it can be wired as to mimic a NAND gate. Fixing the $z$-input wire to be 1, we get $z'=(x\wedge y) \oplus 1= \neg(x\wedge y)= x/y$. This is seen by restricting the truth table to the rows where $z=1$. 

\vskip 1cm
\begin{tabular}{|c|c|c||c|c|c|}
\hline
x & y & z & x'& y'& z'\\ \hline
0 & 0 & 1 & 0 & 0 & 1\\ 
0 & 1 & 1 & 0 & 1 & 1\\
1 & 0 & 1 & 1 & 0 & 1\\
1 & 1 & 1 & 1 & 1 & 0\\ \hline
\end{tabular}
\vskip 1cm

It can also be wired to mimic a two-wire FANOUT. Fixing the first input to 1 and the third to 0, the bit on the second input appears on the second and third output. This is seen by restricting the truth table to the rows where $x=1$ and $z=0$. 

\vskip 1cm
\begin{tabular}{|c|c|c||c|c|c|}
\hline
x & y & z & x'& y'& z'\\ \hline
1 & 0 & 0 & 1 & 0 & 0\\
1 & 1 & 0 & 1 & 1 & 1\\ \hline
\end{tabular}
\vskip 1cm
Clearly $y'=y$ and $z'=y$.

The bits that are fixed to constant values in these constructions are called {\it ancilla} bits.

We should also mention the Fredkin gate in this context. It is a universal reversible gate with three inputs and three outputs. It has one control bit $x$ and two target bits $y$ and $z$.\footnote{To be consistent with terminology, we ought to speak about input and output wires instead of input and output bits. However this is common abuse of language.} If the control bit is 0, the target bits goes through unchanged, whereas if the control bit is 1, the target bits are swapped, i.e. $y'=z$ and $z'=y$.

The Toffoli gate is more useful in quantum computation.

%**********
\subsection{Reversible circuits and un-computation}
Relying on the universality of Toffoli gates, a circuit wired with NAND and FANOUT gates can be rewired into a reversible circuit. In order to do that, extra "ancilla" bits are needed. Furthermore, the Toffoli gates outputs one or two extra bits (depending on whether they mimic FANOUT or NAND) not needed in the computation. These bits only serve the purpose of making the computation reversible. The extra output bits from each Toffoli gate add up to what essentially amounts to "garbage". It would be nice to be able to have the ancilla bits in a standard state and to get rid off the garbage bits. Simply erasing them will not do, as that would spoil the reversibility. However, there is a procedure to clean up the garbage using precisely this reversibility!

Suppose we have a non-reversible circuit computing a function $f$ on some $n$\,-bit input $x$. We want to do this computation reversibly while cleaning up the garbage. If the non-reversible computation is represented as 

\begin{equation}
x\rightarrow f(x),
\end{equation}

we can represent the reversible computation as

\begin{equation}
(x,a)\rightarrow (f(x),g(x)),
\end{equation}

where $a$ denotes the ancilla bits needed to wire the Toffoli gates and $g(x)$ denotes the resulting garbage bits. The ancilla bits, as well as $x$ can be thought of as bit strings stored in appropriately sized registers. 

In order to put the ancilla bits in a standard state, we allow the use of NOT gates. These are reversible. Then all ancilla bits can be 0's, using NOT gates where 1's are needed. So now we have $(x,{\bar 0})\rightarrow (f(x),g(x))$ with ${\bar 0}$ denoting a bit string with just 0's.

Furthermore, allowing the use of the CNOT gates (also reversible), we can do two things. First, a copy of the input bit string $x$ can be made, so that the computation now reads

\begin{equation}
(x,{\bar 0},{\bar 0})\rightarrow (x,x,{\bar 0})\rightarrow(x,f(x),g(x))
\end{equation}
where the first arrow corresponds to the copying action of the initial CNOT gates.

Now we can introduce the idea of {\it uncomputation}. This is a procedure that allows us to get rid of the garbage bits by inverting the circuit, and so to speak, uncompute the garbage back to ${\bar 0}$. But of course, the result of the computation $f(x)$ must be saved before. This can be done by introducing a fourth input register $y$ with the same size as the register needed to store the result $f(x)$. The register $y$ is not used until the computation of $f(x)$ is finished. Then CNOT gates are used to add the result bitwise to the bits in $y$. Now the computation reads (suppressing the initial $x$-copying),

\begin{equation}
(x,{\bar 0},{\bar 0},y)\rightarrow(x,f(x),g(x),y)\rightarrow(x,f(x),g(x),y\oplus f(x)).
\end{equation}

All steps performed in this computation up to the last ones involving $y$ are reversible and none of them affects the fourth register, so reversing this part of the computation, yields

\begin{equation}
(x,f(x),g(x),y\oplus f(x))\rightarrow(x,{\bar 0},{\bar 0},y\oplus f(x)).
\end{equation}

The complete computation now reads
\begin{equation}
(x,{\bar 0},{\bar 0},y)\rightarrow(x,{\bar 0},{\bar 0},y\oplus f(x)).
\end{equation}

Note that this is still a reversible computation, as the $\oplus$ part could also be reversed, giving the bit string $y$ back. But, of course, we don't want to do that. So, suppressing the ancilla bit strings, we have simply

\begin{equation}
(x,y)\rightarrow(x,y\oplus f(x)).
\end{equation}

%**********
\subsection{Reversible computation and physics}
Up to this point we have only discussed "logical" reversibility. The computation is reversible in the sense that the input can be recovered from the output by carefully keeping track of every bit during the computation. This is very close to "physical" reversibility by which is meant that the time evolution of a physical system can be reversed so that an initial state of the system can be recovered from a final state. When a computational process is carried out by a physical system\footnote{There must always be some underlying physical system performing the computation, even when computation is viewed abstractly as mere symbol shuffling. Someone or something must shuffle the symbols.} there is precisely such a time evolution involved, so the two concepts of reversibility must be closely connected.

The microscopical laws of dynamics are all reversible, whether classical or quantum. Reversible dynamics does not dissipate any energy. In order for this to make sense one must really talk about closed physical systems, i.e. systems that do not in any way interact with the environment. One must also have full control over all degrees of freedom, i.e. the dynamics of every degree of freedom must be governed by fundamental (time-reversible) equations of motion. Such systems are conservative, meaning that the time evolution is reversible and no energy is dissipated into the environment. 

Comparing this to a computational process we can guess that it is the loss of control of some of the individual bits, inadvertently or deliberately, that leads to energy dissipation in a computational process. 

Indeed, as has been studied by Landauer \cite{Landauer61-82-96}, erasure of one bit of information leads to an energy dissipation given by $k_BT\ln 2$. Here $k_B$ is Boltzmann's constant, a fundamental constant of physics relating  mechanical quantities to thermodynamical quantities like energy and entropy. $T$ is the temperature of the environment into which the energy is dissipated.

Physically, this possibility of performing computations reversibly is of more theoretical interest than practical. The solid state hardware of today dissipate energy far above the $k_bT\ln 2$ limit. Even if solid state circuits can be manufactured that performs reversible logical operations like the Toffoli gate, these devices must be powered by some voltage source like any other electronic gate. Tiny electric current will flow and there will be heat dissipation from electric resistance. Even if this effect can be minimized, perhaps by exploiting superconductivity, the inevitable weak interaction with the environment will generate noise that will have to be corrected. The error correction registers employed must eventually be erased, since memory is always finite, leading to energy dissipation.

For now, classical reversible computation serves just as a backdrop to quantum computation, which is inherently reversible. We will return to the subject of reversible computation in chapter 8 on physics of computation, and in particular to the question of the thermodynamics of computation.

%COMPARISON TO REAL COMPUTERS
%****************************
\section{Comparison to real computers}
Neither the Turing machine model nor the circuit model is very close to the actual workings of a modern digital computer. In what sense then are they models of real world computers? It is generally agreed that all present day general purpose digital computers are {\itshape von Neumann machines}, machines that store both data and program in a memory and which works in a cyclic way of (fetching instructions and data, executing instructions, storing data). This is a rather vague description of the basic workings of a computer and cannot by itself serve as a model of computation. There is however a computational model that very closely captures the workings of a modern computer - the Random Access Machine-model. It has a CPU with temporary storage registers, a program counter and an ALU - much as in a real world processor. The CPU is connected to a
(random access) memory which stores both data and program. In order to read and write in arbitrary memory locations, every memory cell have an address. The model can be programmed in an assembly-like language. It is clear from the close analogy to real computers that the model has expressive strength enough to support compilers and higher level languages.

The main difference between the Turing machine and real computers is that its memory is not accessible immediately. In order to read a square away from the present position of the read/write head, all intermediate squares must be traversed and read. 

The Random Access Machine (RAM) can reach an arbitrary memory cell in a single step. It can be considered a simplified model of real world computers.

The same functions are computable on Turing machines and on the RAM. 

%NON-DETERMINISTIC TURING MACHINES
%*********************************
\section{Non-deterministic Turing Machines}

The models of computation considered so far are deterministic, i.e. at each step in the computation, the next step is exactly determined by the program and the data. However, non-deterministic models of computation are theoretically very important and often lead to simplification of analysis, though they cannot in general be efficiently implemented. Referring back to the definition of a Turing machine, we see that in the set of instructions defining the program, there is at most one instruction with a certain combination of scanned tape symbol and machine state. This makes the computation deterministic, i.e. the action of the machine is uniquely determined. Removing this restriction leads to non-determinism. 

Informally, for any combination of scanned tape symbol and machine configuration, we allow a set of possible instructions. 

Formally this is easiest to formulate in terms of the transition function. Remember the transition function

$$
\delta: Q\times \Gamma \rightarrow (Q\cup Q_h)\times\Gamma\times M,
$$
which maps combinations of scanned symbol and machine configuration into the set of combinations of configurations, symbols and moves. 

Consider the set of all subsets of the set $(Q\cup Q_h)\times\Sigma\times M$, this is the power set ${\cal P} ((Q\cup Q_h)\times\Sigma\times M)$. Allowing several different instructions having the same tape symbol and configuration can be formulated in terms of a transition function that maps into this set of subsets

$$
\delta: Q\times \Gamma \rightarrow {\cal P}((Q\cup Q_h)\times\Gamma\times M).
$$ 

How does a non-deterministic Turing machine compute? Suppose that during the computation the machine finds a matching pair of scanned symbol and machine state for which there are several instructions. The computation then branches off into parallel computations, one for each possible way to proceed. The state graph for a non-deterministic computation is therefore a directed tree, whereas for a deterministic computation it is a list.

In order to actually carry out such a computation in parallel, one would have to assign new computational resources at each branch in the graph, in the form of  new processors or new Turing machines. In practice, this is not possible in the general case where the maximal number of processors are limited. 

The alternative would be to traverse the tree, breadth-first, using just one processor.\footnote{Depth-first traversal runs the risk of going down a non-terminating branch, so breadth-first is the best option in an actual simulation.} Exponential resources are needed in the generic case, in the form of increasing time and space requirements. 

It is quite easy to argue that the set of computable functions are the same. Suppose a partial function is computable by a non-deterministic Turing machine. This means that the function values are found at the ends of terminating branches of the computation graph. Performing the computation breadth-first on a deterministic Turing machine, we are guaranteed to eventually reach the halting states, possible after consuming an exponential amount of time and space. The computation might take exponential time to systematically work through the ever increasing number of branches, and use an exponential amount of tape to record information about the state at not yet processed branch points. Tape can be reclaimed but not time. Still, the function is computable on a deterministic Turing machine.

\subsection{A note on classical parallelism}
Non-determinism offers a kind of parallelism. Parallel computation and non-deterministic computation are overlapping concepts but they are not the same. Parallel computation does not involve an unbounded number of parallel processes, as there is always a maximum number processors available in any real machine. On the other hand, parallel processes can communicate, by shared data or by passing data (messages), and that need not the case for non-deterministic algorithms. Parallel algorithms and parallel computation is a huge subject and there are several different models for parallel computation but no generally agreed on paradigm. 

One might wonder if classical parallelism is a threat to the Church-Turing thesis? That is, is it possible to compute non-computable functions using parallel computation? The answer is no, and the argument is similar to the argument in the case of non-determinism.

%********************************
\section{Probabilistic Turing machines}
There is a close connection between probabilistic Turing machines and non-deterministic Turing machines. In non-deterministic machines, the computation can branch of into different sub-computations, in principle at every node in the computation. Consequently, the computation graph becomes tree. Now, if the computational graph edges leading out from a node allowing branching are assigned probabilities, and a probabilistic choice as to which edge to follow is made, we get a probabilistic Turing machine. In this case, the computation becomes a directed path through the computation graph. Of course, different runs of the same machine, with the same input, will give different paths depending on the random choices made at each branch node.

Given a perfect random number generator {\it rnd}, a probabilistic machine can be easily simulated on a deterministic machine by performing calls to the {\it rnd} at each step allowing for probabilistic choices.

Formalizing this concept will yield a first step towards an understanding of quantum Turing machines as well a providing a background for discussing where quantum computation departs from classical computation. Quantum Turing machines are treated in chapter 5.

The point of departure is the transition function $\Delta$ of section 2.3.2. There, the values 0 and 1 of the function determined whether the transition was present in the program or not. Thinking of the numbers 0 and 1 as probabilities one is lead to extend the range of $\Delta$ to numbers $p$ in the interval $[0,1]$, and interpreting $p$ the probability for the transition, i.e. defining

$$
\Delta: Q\times \Gamma\times (Q\cup Q_h)\times\Gamma\times M\rightarrow [0,1].
$$

Referring back to the computational tree of a non-deterministic machine, we can turn this into a computation tree for a probabilistic machine by marking up each branch node with probabilities. The probability to reach a certain node in the tree is the total calculated probability to reach that node from the initial starting node of the computation.

For theoretical purposes we could then consider the probabilistic machine as being, at each computational step, in a (classical) superposition of all the reachable states (from the start). Denoting the states ${\cal L} q_i{\cal R}$ with $|s)$,\footnote {The notation $|s>$ will subsequently be used for true quantum states.} we can formally write this superposition as a sum $\sum p_s|s)$ where the summation runs over all states reachable at the given stage in the computation. 

This superposition is an entirely theoretical construct. Physically there is no such superposition for a classical probabilistic Turing machine. Each separate execution of the machine simply traces out a path in the computation graph. Theoretically, however, we can speak of the machine as being in a superposed state. Observing (or measuring) the machine after a certain number of time steps, we will find the it in a certain state with a certain probability. This probability is the same as the probability to reach that state from the initial starting state. The classical computation of a probabilistic machine can be observed at each state, thus tracing out the particular execution path. This observation, or measurement is of no consequence for the future execution of the machine. 

Here we have two major differences as compared to quantum machines. Firstly, for quantum machines, the corresponding superpositions (somewhat differently defined though) do actually exist. Secondly, and as consequence of the reality of the superposition and the nature of quantum dynamics, the measuring a quantum Turing machine {\it\underline {will}} affect the future execution of the machine.

Returning to the classical probabilistic machine, it intuitively appears that restricting the branching probabilities to $\{0,1/2,1\}$, 1/2 corresponding to fair coin tossing, ought to be sufficient. This can in fact be proved [??].

%SOME COMPLEXITY THEORY
%**********************
\section{Some Complexity Theory}
Computability theory discusses what can be computed in principle by investigating the boundary between computable and uncomputable functions. Complexity theory discusses what can be computed in practice by analyzing the amount of resources needed in a computation.

The complexity is calculated by analyzing the algorithm, not by running the computation, so clearly, in order for complexity to make sense, it must be defined for decidable problems or computable functions.

Complexity is most conveniently discussed in terms of the computational resources required to decide recursive languages, i.e. in terms of decision problems. A computational resource can be {\it time}, roughly measured as the number of steps required by an algorithm. Another resource is {\it space}, corresponding to the amount of memory required. It could also be some other physical resource like energy, but time and space are the measures are the most important from the point of view of difficulty of algorithmic problems. In the circuit model, complexity is naturally measured in terms of the number of gates in the circuit.

A distinction is made between tractable problems and untractable problems. A problem is tractable if it can be solved on a computer using a reasonable amount of CPU-time and/or memory. It is well known that there is a dramatic difference in the growth rate of polynomial functions and exponential functions. Reasonable amount of resources are those that grows at most as a polynom in the size of the problem. 

Algorithms are not in general intended to solve particular problems, but rather sets of problems, parameterized in some way. A particular problem in the set is called an instance. In general the instances are increasing in size in terms of the parameters. When analyzing a certain algorithm for a certain problem (for example, insertion sorting for sorting) it is in general the worst-case behavior that is interesting. In that case we are interested in upper bounds on the amount of resources required by the algorithm.

When analyzing classes of algorithms for a certain problem (for example, the class of sorting algorithms) it is rather lower bounds that are in focus. We want to know the performance of the best possible algorithm.

We will now make these notions exact. First of all we need a model of computation. In general, different models of computation can have different strengths.\footnote{This is in contrast to the situation as regards computability.} However, the so called slowdown between different reasonable models is polynomial, and therefore not important in theoretically. Of, course, in practical computing, even small increases in speed can be important.

In order to treat the complexity of algorithmic problems in a uniform way, the problems are formulated in terms of formal language theory. Problems are coded using some alphabet and problem instances then corresponds to strings in the set of all strings $\Sigma^*$. A decision problem then amounts to deciding whether a given string belongs to the language (which is a subset $\Sigma^*$, see section X.X.X) or not.

This section on the theory of complexity will be very brief and just record the basic definitions and results of the topic without proofs or detailed explanations. A good modern reference is \cite{Sipser1997}, see also \cite{DavisSigalWeyuker1994} and \cite{Papadimitriou1994}.

%**********
\subsection{Measures of complexity}
Any 'reasonable' model of computation can be used to set up the theory of complexity. What is needed for measuring the time complexity is some consistent way of counting computational steps in terms of a unit of time for performing some elementary step. There is a good deal of arbitrariness here both regarding what is a step and what is an elementary unit. This arbitrariness is however inherent to the problem, and in the end does not matter much, as good measures of complexity only differs polynomially.

It is anyway not the exact number of steps that is important. Rather, we need a concept of complexity that is robust to incremental improvements in hardware and software, and yet sensitive to more dramatic developments, of which quantum computation is an example.\footnote{The extent to which quantum computation is stronger than classical is not yet fully understood.}

The Turing machine model will be used here. Time complexity will be defined in terms of the number of steps taken by the machine during the computation. Space complexity will be defined in terms of the maximal number of tape squares needed by the computation. These complexity measures will be functions from the size of the input to the number of computational steps and the number of tape squares respectively. Input size is defined as the number of written tape squares.

Let ${\mathbf L}$ be a recursive language decided by a Turing machine $M$. Referring back to section 2.3.4 we recall that this means that $M$ halts on all inputs $x$ in either the 'yes' or the 'no' configuration depending on whether the string $x$ belongs to the language or not. Computational resources is measured in terms of the size of the input $x$ and will be taken as the number of non-blank tape squares in the start configuration of the machine. This is called the {\it length} of the input, i.e. it is just the length of the string written on the tape in the start configuration.

This is a simplification, since the number of steps required by an algorithm may depend on several of the parameters defining the instance of the problem. For example, in a graph problem, both the number of nodes and the number of edges may have an effect on the running time of an algorithm. On the other hand, the maximum number of edges in a graph with $n$ nodes is $n(n-1)/2$, so using the square of the number of nodes we get a reasonable measure of the instance size. The idea is here that the instances are coded in some way on the machine tape, and coding both node data and edge data, we take this into account when we measure the instance size as the length of the string written on the tape.

\subsubsection{Asymptotic upper bounds}
The "big O" notation ${\mathcal O}$ is used to set {\it upper bounds} on the asymptotic behavior of functions. We want to capture the notion that the function $f$ is asymptotically bounded by the function $g$. 

Let $f$ and $g$ be two functions from the natural numbers to the positive real numbers.  $f(n)$ is in the class of functions ${\mathcal O}(g(n))$, or simply $f(n)={\mathcal O}(g(n))$ if there exist positive integers $c$ and $n_0$ such that $f(n)\leq cg(n)$ for every integer $n\geq n_0$. This simply says that for sufficiently large $n$, the function $f$ is bounded from above by the function $g$ apart from a constant factor.

\subsubsection{Asymptotic lower bounds}
For {\it lower bonds}, the "big Omega" $\Omega$ is used. 

Again let $f$ and $g$ be two functions from the natural numbers to the positive real numbers.  $f(n)$ is in the class of functions $\Omega(g(n))$, or simply $f(n)=\Omega(g(n))$ if there exist positive integers $c$ and $n_0$ such that $cg(n)\leq f(n)$ for every integer $n\geq n_0$. This simply says that for sufficiently large $n$, the function $f$ is bounded from below by the function $g$ apart from a constant factor.

\subsubsection{Asymptotic behavior}
If a function $f$ is in both ${\mathcal O}(g)$ and $\Omega(g)$, i.e. if it, apart from constant factors, is bounded both from above and below by the same function $g$, then it behaves asymptotically as $g$. The "big $\Theta$" notation is used to indicate this. 

Thus, $f(n)$ is in $\Theta(g(n))$ if it is in both ${\mathcal O}(g(n))$ and $\Omega(g(n))$.

\subsubsection{Time complexity}
The {\it time complexity} of a deterministic Turing machine $M$ is a function \break $f: N\rightarrow N$, where $f$ is the maximum number of steps performed by $M$ during any computation with input length $n$. 

This is also phrased in any of the following ways: $f$ is the running time of $M$, $M$ runs in time $f$, $M$ is a time $f$ machine.

The {\it time complexity class} $TIME(f(n))$, is defined as

\begin{equation}\label{TimeComplexityClass}
TIME(f(n))=\{\mathbf L\,|\,\mathbf L \mbox{ is decided by an}\;{\mathcal O}(f(n))\mbox{ time machine}\}.
\end{equation}

There is a corresponding notion of time complexity for non-deterministic computations. The {\it time complexity} of a non-deterministic Turing machine $M$ is a function $f: N\rightarrow N$, where $f$ is the maximum number of steps performed by $M$ on any branch of the computation with input length $n$. 

\subsubsection{Space complexity}
The {\it space complexity} of a deterministic Turing machine $M$ is a function \break $f: N\rightarrow N$, where $f$ is the maximum number of tape cells scanned by $M$ during any computation with input length $n$. 

This is also phrased in any of the following ways: $M$ runs in space $f$, $M$ is a space $f$ machine.

The {\it space complexity class} $SPACE(f(n))$, is defined as

\begin{equation}\label{SpaceComplexityClass}
SPACE(f(n))=\{\mathbf L\,|\,\mathbf L \mbox{ is decided by an}\;{\mathcal O}(f(n))\mbox{ space machine}\}
\end{equation}

\subsubsection{Analysis of algorithms}
Algorithms are analyzed by roughly estimating the number of steps required to perform the parts of the algorithm. Textbooks on complexity theory normally goes through the techniques of doing this. The details differs from model to model depending on the programming primitive available. We will bypass this topic here, and just rely on our intuition in the simple models we are concerned with, Turing machines and circuits.

\subsection{Complexity classes}
For easy reference, we will briefly review the definitions of the basic complexity classes.
\subsubsection{The class P}
The time complexity class {\bf P} is the collection of all languages that are in $TIME(n^k)$ for some constant k. That is, a language is in {\bf P} if it can be decided by a deterministic Turing machine whose running time is bounded from above by a polynomial in the number of steps.

\subsubsection{The class NP}
{\bf NP} is an extremely important time complexity class. Its name is an abbreviation of {\bf N}on-deterministic {\bf P}olynomial. It is defined as the collection of all languages that are in $NTIME(n^k)$ for some constant k. That is, a language is in {\bf NP} if it can be decided by a non-deterministic Turing machine whose running time is bounded from above by a polynomial in the number of steps. This class is potentially larger than {\bf P}, indeed ${\bf P}\subseteq{\bf NP}$.

There is another characterization of {\bf NP} that do not refer to non-deterministic computations. It is based on the easy (polynomial) verification of a "yes" instance by a so called {\it witness}. However, there need not be any such witnesses for "no" instances. 

A good example is factoring of integers. Suppose the language under consideration is the set of composite natural numbers, i.e. non-prime numbers. A "yes" instance, let's say a number $n$, can always be ascertained by exhibiting a factor, say $wy$. By simply dividing $n$ by $wy$ it can be verified (in polynomial time) that $n$ is indeed composite. On the other hand, supplying a "no" witness $wn$ is quite useless, since even if $wn$ does not divide $n$, there might very well be another "yes" witness not yet exhibited. So, without deeper insight into the problem of determining whether a number is prime or composite,\footnote{Recently, such insight has indeed been gained, showing that primality testing is in {\bf P} after all \cite{AgarwalSaxenaKayal2002}.} all tentative factors must be checked before the verdict prime can be passed.

A language $L$ is in {\bf NP} if there is a Turing machine such that
\begin{itemize}
\item If $x\in L$, there exist a witness $w$ such that when the machine is started with $x$ and $w$ as inputs, its halts in the "yes" state after a time polynomial in the size $|x|$ of $x$.
\item If $x\notin L$, then for all purported witnesses $w$, the machine halts in the "no" state after a time polynomial in the size $|x|$ of $x$, when started with $x$ and $w$ as inputs.
\end{itemize}

It is not known whether {\bf P} is a strict subset of {\bf NP}. The conjecture ${\bf P}\neq{\bf NP}$ is one of the main unsolved problems in complexity theory.

\subsubsection{The class PSPACE}
The class {\bf PSPACE} is the space analogue to {\bf P}. It is defined as follows.

The space complexity class {\bf PSPACE} is the collection of all languages that are in $SPACE(n^k)$ for some constant k. That is, a language is in {\bf PSPACE} if it can be decided by a deterministic Turing machine using a number of working bits polynomial in the input size. There is no limit to the amount of time used.

It is clear that {\bf P} is included in {\bf PSPACE} simply because a machine that halts after a polynomial number of steps can only traverse a polynomial number of tape squares. Thus, ${\bf P}\subseteq{\bf PSPACE}$, but it is not know whether the inclusion is strict, i.e. if ${\bf P}\neq{\bf PSPACE}$ or not.

\subsubsection{The class BPP}
If probabilistic algorithms are considered, then corresponding probabilistic complexity classes can be defined. The {\it bounded error probabilistic} class, {\bf BBP}, is defined to contain all languages $L$ that can be decided by a probabilistic Turing machine M, such that
 
\begin{itemize}
\item If $x\in L$, then M accepts $x$ with a probability at least 3/4.
\item If $x\notin L$, then M rejects $x$ with a probability at least 3/4.
\end{itemize}

The probability 3/4 is arbitrary, any probability strictly greater than 1/2 would suffice in the definition.

There are many more complexity classes, as well as lots of inclusion relations between them. The reader is referred to the literature for a thorough discussion. We will briefly return to the topic in chapter 6 on the complexity of quantum computation.

\chapter{Algebra of quantum bits}

There are a few different models of quantum computation in the literature. The most popular, and most thoroughly worked out, is the quantum circuit model \cite{Deutsch1989}. Quantum circuits are the quantum analogue of classical circuits built out of logic gates. Another model, the quantum Turing machine \cite{Deutsch1985} is the quantum analogue of the classical Turing machine. But just as general-purpose digital computers are not really built as Turing machines, it does not seem practical to build real quantum computers as quantum Turing machines. In this respect, quantum circuits seem to be closer to actual implementation as physical devices. 

This chapter is an introduction to the subject of quantum computation. The circuit model of computation introduced in chapter 2 will be elaborated and realized in terms of vectors and matrices, thus offering what could be called an algebra of bits and quantum bits. In this way we will be able to see precisely where the quantum paradigm of computation breaks away from the classical. Precise and general definitions of concepts, as well as a detailed treatment will follow in subsequent chapters.

%CLASSICAL AND QUANTUM PHYSICAL SYSTEMS
%**************************************
\section{Classical and quantum physical systems}
In contrast to the case of classical theory of computation, the physical substratum of the computer is more focused in the research on quantum computation. In part this is due to the very real problems of actually building devices capable of performing quantum computations. It is appropriate therefore to begin with a short discussion of the concept of a physical system.

A simple example of a classical system is a gas in container. Pressure, temperature, and volume give the macroscopic state of the gas. In classical physics, these variables can range over a continuous set of values corresponding to a continuous state space. The microscopic state is given by the values of all positions and all momenta of all the particles in the gas. This forms a huge continuous state space.

In quantum physics, state spaces can be discrete or continuous or both. Continuous state spaces occur for free particles or particles scattered off a potential, whereas discrete state spaces occur for bound state systems, notably particles bound by a potential field. The standard example of a bound state system is the hydrogen atom which can be in a set of discrete states, each state being characterized by values of energy and a couple of other variables. In this case the energy can only range over a discrete set of values. If the atom absorbs enough energy, it will become ionized and the electron will no longer be bound by the potential of the nucleus. This corresponds to the continuous part of the state space.

A gas of quantum particles in a container will have a huge discrete microscopic state space. There is no practical way to distinguish different states of such a system and it is useless for computational purposes. In order for a physical system to be useful as a computer, it must be possible to exercise precise control over the states of the system. Typically, it must be possible to prepare the system in an input state, and then let the system evolve according to dynamical laws (this corresponds to the program) and subsequently to measure an output state after the computation is completed.

To conclude, in both classical and quantum physics one speaks of the state of a physical system, and the states are characterized by the values of a number of variables. The state spaces in classical physics are continuous. This is true also for systems like the bit. In solid state devices, voltage levels represent the two states of the bit and there is a certain range within which these voltage levels are allowed to vary. However, the levels must be well separated so that no overlap of the ranges occur. This ensures the discrete digital nature of the device. 

We will return to the physics of computing system in the two last chapters.

%TWO STATE QUANTUM SYSTEMS AND THE QUANTUM BIT
%*********************************************
\section{Two-state quantum systems and the quantum bit}
The basic building block in most quantum computation models is the qubit. The qubit is a quantum generalization of the classical bit. A bit can be in any of the two well defined states {\bf 0} and {\bf 1}, and a classical memory register can be modeled by a string of bits. There is no interaction between the separate bits in the register. Information processing, or computation, can be regarded as bit flips performed on the register. Reading and writing single bits are the most primitive acts of computation. 

A qubit is a quantum system having two states. These states are denoted by $|0\rangle$  and  $| 1\rangle$.\footnote{The notation will be explained later on.} These states are the quantum versions of two states of the bit, {\bf 0} and {\bf 1}. The fundamental difference between classical and quantum physics is that whereas a classical system must be in a definite state, a quantum system can be in a superposition of a set of states. The bit must be either {\bf 0} or {\bf 1}. But the qubit can be in a complex linear combination of $|0\rangle$  and  $|1\rangle$, namely

\begin{equation}\label{eq: LinCombQubit}
|\psi\rangle = \alpha|0\rangle + \beta|1\rangle. 
\end{equation}

Here $\alpha$ and $\beta$ are complex numbers and $|\psi\rangle$ is used to denote the general state. Precise definitions of the quantum mechanical notations will be given in the next two chapters. Suffice it here to note that the proper framework for this is complex linear vector spaces, and consequently, the states $|0\rangle$  and  $|1\rangle$ can be thought of as basis states in such a space.

The typical example of a two-state quantum system is the spin states of spin\,-$1\over 2$ particle like the electron, but the precise physical nature of the system will not concern us at the moment. We will instead develop the theory of quantum computation based on generic two-state quantum systems.  Questions of practical implementations will be returned to in chapter 9.

There is a certain restriction on the complex numbers $\alpha$ and $\beta$ having to do with the interpretation of quantum mechanics. Classically, one can determine which state the bit is in, and one will get 0 or 1 according to the actual state of the bit. Quantum mechanically, the situation is different. 

The process of obtaining information out of the qubit is called a {\it measurement}. If the qubit is in (or is known to be in) either the state $| 0\rangle$ or the state $| 1\rangle$, a measurement performed on it will give the result 0 or 1 respectively. If however the qubit is in the general state $|\psi\rangle$ , the measurement will give 0 with probability ${|\alpha|}^2$ and 1 with probability ${|\beta|}^2$. There is no way, for a single qubit, to determine its precise state, i.e. there is no way to determine the values of $\alpha$ and $\beta$. If however we have a large collection of identically prepared qubits, repeated measurements on the qubits will yield statistical values for ${|\alpha|}^2$ and ${|\beta|}^2$. No single measurement can ever determine the values of $\alpha$ and $\beta$. 

However, since a measurement must yield either 0 or 1, this probabilistic interpretation gives the restriction

\begin{equation}\label{eq: QubitNormalization}
{|\alpha|}^2+{|\beta|}^2=1
\end{equation}
on the numbers $\alpha$ and $\beta$. 

A quantum measurement will have an effect on the state of the system after the measurement. If a measurement is performed on the general state $|\psi\rangle$ and the result is 1, the state will be $|1\rangle$ after the measurement. Likewise, if the result is 0, the state after the measurement will be  $|0\rangle$. This is general property of measurements.

There is a further difference with regard to classical physics. Classically it does not make sense to consider measuring the bit in any other state than {\bf 0} or {\bf 1} because there are no other states. However, quantum mechanically we can consider, for example, the special states

\begin{eqnarray}
|+\rangle = {1\over\sqrt 2}|0\rangle + {1\over\sqrt 2}|1\rangle\label{eq: PlusState}
\\
|-\rangle = {1\over\sqrt 2}|0\rangle - {1\over\sqrt 2}|1\rangle\label{eq: MinusState}.
\end{eqnarray}

Just as one can make a measurement on a general state $|\psi\rangle$ with respect to the states $| 0\rangle$ and $| 1\rangle$, one can measure with respect to the states $|+\rangle$ and $|-\rangle$. A qubit in the state $|0\rangle$, say, will when measured with respect to the states $|+\rangle$ and  $|-\rangle$, yield the result '+' with probability 1/2 and '$-$' with probability 1/2. 

These non-classical features of the theory will be elaborated in the next two chapters on quantum mechanics.

%MULTIPLE QUBIT STATES
%*********************
\section{Multiple qubit states}
Multiple qubit states are modeled on their classical analogue, the bit strings. A classical two-bit register can store any of the bit strings {\bf 00}, {\bf 01}, {\bf 10} or {\bf 11}. The quantum analogue of these strings are $| 00\rangle$, $| 01\rangle$, $| 10\rangle$ and $| 11\rangle$ respectively, and they can be regarded as a basis for a four-dimensional vector space. 

A general 2-qubit state can now be written as a linear combination of the basis states,
$$
|\psi\rangle = \alpha_{00}| 00\rangle + \alpha_{01}| 01\rangle + \alpha_{10}| 10\rangle +\alpha_{11}| 11\rangle
$$
Two facts can be noted at this stage. Firstly, as already noted for the single qubit, whereas a classical memory register must be in a definite state corresponding to the actual values of the stored bits, the quantum memory register can be in linear combination of all the basis states. This is referred to as superposition of states. Secondly, there are quantum states of the memory register that cannot be expressed as direct products of the basis states. One example is the state ${1\over\surd 2}(| 00\rangle + | 11\rangle)$  which in no way can be written as a product of single qubit states $| 0\rangle$ and $| 1\rangle$. This property of quantum mechanics is called {\it entanglement}. 

These features of quantum mechanics, superposition of states and entanglement, are crucial to the theory of quantum computation. 

If we denote a single bit by ${b}$, a classical n-bit string can be written as $b_1b_2\ldots b_n$. The quantum analogue is 
$| b_1b_2\ldots b_n\rangle$. A general state is a linear combination of these $2^n$ basis states. These states are called {\itshape computational basis states}.

%COMPUTATION
%***********
\section{Computation}
Computation can be seen as a transformation of an input state to an output state. If both input and output are represented by $n$ -bit strings, then the computation can be performed by applying an $2^n\times 2^n$ matrix to the input. 

At first sight one might be tempted to use $n\times n$ matrices, representing the states by $n$ -dimensional vectors, the components of which are taken to be the bits of the bit strings. However, that cannot work, as can be seen even in the simplest case of just one bit. Representing the input bit by the (one-dimensional) vector $i$ and the output by $o$, the computational relation connecting output and input is

$$
i\rightarrow o=Ci
$$
for some matrix $C$ which in this case is just a number. But then the bit flip $0\rightarrow 1$, $1\rightarrow 0$ cannot be represented with one and the same number $C$. Thus one bit of information must actually be represented by a two dimensional space. 

%*************
\subsubsection{Representations of classical bits and quantum bits}
We will introduce a convenient representation for both classical och quantum bits. The constructions are actually the same, but different notation will be used in order to highlight the differences between classical and quantum computation. The values for a classical bit will be denoted in boldface as {\bf 0} and {\bf 1} and they will be represented as two-dimensional vectors as

\begin{equation}\label{eq: BitRepresentation}
{\bf 0}=\pmatrix{1 \cr 0 \cr},\quad {\bf 1}=\pmatrix{0 \cr 1 \cr}.
\end{equation}

Correspondingly, the quantum bits will be represented in terms of the same vectors as

\begin{equation}\label{eq: QubitRepresentation}
|0\rangle=\pmatrix{1 \cr 0 \cr},\quad |1\rangle=\pmatrix{0 \cr 1 \cr}.
\end{equation}

Note one difference in interpretation in this context. As in the preceding section, the quantum states $|0\rangle$ and $|1\rangle$ are basis vectors in a complex vector space and consequently it makes sense to consider linear combinations as in equation (\ref{eq: LinCombQubit}). For the classical bit values {\bf 0} and {\bf 1} we introduce no such structure.\footnote{It could be somewhat artificially introduced in order to represent probabilistic computation. Still, even so there are fundamental differences as compared to quantum computation.}

Next, bit strings and multi-qubit states will be represented by direct products of these two-dimensional vectors. The rules are very simple, and we will write them out in the case of products of two and three vectors.

\begin{equation}\label{eq: DirectProductRule2}
\pmatrix{a_0 \cr a_1}\otimes\pmatrix{b_0 \cr b_1}=\pmatrix{a_0b_0\cr a_0b_1\cr a_1b_0\cr a_1b_1},
\end{equation}
\begin{equation}\label{eq: DirectProductRule3}
\pmatrix{a_0 \cr a_1}\otimes\pmatrix{b_0 \cr b_1}\otimes\pmatrix{b_0 \cr c_1}=\pmatrix{a_0b_0c_0\cr a_0b_0c_1\cr a_0b_1c_0\cr a_0b_1c_1\cr a_1b_0c_0\cr a_1b_0c_1\cr a_1b_1c_0\cr a_1b_1c_1\cr }.
\end{equation}

From these two cases, the principle should be clear. As an example, the four different two-bit strings can be represented by the vectors

\begin{equation}\label{eq: TwoBitStates}
{\bf 00}=\pmatrix{1 \cr 0 \cr 0 \cr 0 \cr},\quad {\bf 01}=\pmatrix{0 \cr 1 \cr 0 \cr 0 \cr},\quad {\bf 10}=\pmatrix{0 \cr 0 \cr 1 \cr 0 \cr },\quad {\bf 11}=\pmatrix{ 0 \cr 0 \cr 0 \cr 1 \cr}.
\end{equation}

The corresponding two-qubit states $|00\rangle,|01\rangle,|10\rangle$ and $|11\rangle$, have exactly the same vector representation. Following \cite{Mermin2002} I call these vectors the {\it classical basis}.

When the numbers of bits or qubits are large it is convenient to use a shorthand notation using the base-10 representation of the bit strings interpreted as binary numbers. For example, the bit string {\bf 0101} will be denoted ${\bf 5_4}$ and similarly for $|0101\rangle$ which is denoted by $|5\rangle_4$. The suffix is needed in order to remove any ambiguity as to the numbers of bits or qubits that the number represents.

%*************
\subsubsection{One-bit classical computations}
The bit-flip program, $flip$, can now be represented by the matrix

$$
FLIP=\pmatrix{0 & 1\cr 1 & 0},
$$
which, of course, corresponds to the logical operation $NOT$. In the context of quantum computation, this matrix is also called $X$ for reason that will become clear subsequently.

Furthermore, there are three more distinct programs, namely for a one-bit state, namely

$$
id: \matrix{0\rightarrow 0\cr 1\rightarrow 1},\quad set: \matrix{0\rightarrow 1\cr 1\rightarrow 1},\quad reset: \matrix{0\rightarrow 0\cr 1\rightarrow 0}.
$$
All these can be represented by $2\times 2$-matrices. Not that the two last computations, {\it set} and {\it reset}, are not reversible, whereas the first two, {\it not} and {\it id} are reversible. 

Reversibility in this context means that the input can be deduced from the output.

%*************
\subsubsection{Two-bit classical computations}
On the 2-bit states (\ref{eq: TwoBitStates}), certain $4\times 4$-matrices, represents computations. To take just one example, consider the operation of exchanging the values of the two bits.

\begin {equation}\label{eq: Swap}
swap: \matrix{00\rightarrow 00\cr 01\rightarrow 10\cr 10\rightarrow 01\cr 11\rightarrow 11}.
\end{equation}

A matrix effecting this transformation is

\begin {equation}\label{eq: SwapMatrix)}
SWAP=\pmatrix{1&0&0&0\cr 0&0&1&0\cr 0&1&0&0\cr 0&0&0&1}.
\end{equation}

%*************
\subsubsection{Multi-bit classical computations}
In this view of computation, $n$ -bit strings are represented by $2^n$ dimensional vectors and computations are represented by $2^n\times 2^n$ square matrices. The number of input and output bits are the same. 

Some restrictions on the allowed matrices can be derived. The input bit string is represented by a column vector with just one 1 and the rest entries 0. The output bit string must also be represented in the same way. This puts a severe restriction on the possible matrices. Consider applying a certain matrix $C$ to an $n$ -bit state

$$
\pmatrix{c_{11}&c_{12}&\cdots&c_{1n}\cr
	c_{21}&c_{22}&\cdots&c_{2n}\cr
	\vdots\vdots&\vdots\vdots& &\vdots\vdots\cr
	c_{n1}&c_{n2}&\cdots&c_{nn}\cr}\pmatrix{0\cr\vdots\cr\delta_j\cr\vdots\cr 0},
$$
where the $n$-bit state is represented by a column vector with just on entry $\delta_j$ different from 0 and equal to 1. Carrying out the matrix multiplication yields

$$
\pmatrix{c_{1j}\cr c_{2j}\cr\vdots\cr c_{nj}},
$$
i.e. it pulls out the $j$-th column from the matrix $C$. If this column vector is to represent an $n$-bit state, only one of the coefficients $c_{ij}$ can be equal to 1. 

This shows that the computation matrix $C$ is a matrix of zeros, except for precisely one entry in each column which is equal to 1. As an aside, note that for $n$-bit computations, represented by $2^n\times 2^n$ -matrices, there are $(2^n)^{2^n}=2^{n\cdot 2^n}$ different possible matrices. 

In the classical case working with this model of computation is very uneconomical as it requires exponential sized vectors and matrices, most of which entries are zero anyway. It corresponds to a unary notation for numbers.

Note, however, the very close correspondence with the circuit model. In fact, this model is a realization of the circuit model, hence the name {\it computational} (or better {\it classical} ) basis for the vectors (\ref{eq: TwoBitStates}
).

In  the classical case, we don't really need this expansion out of the bit strings into exponential sized vectors. There are much more efficient ways to process bit strings. 

\subsubsection{Transition to quantum computations}
The possibility to have linear combinations, called {\it superpositions}, of the basis vectors, marks the point of departure into quantum computations.\footnote{At least in this approach to the theory. There are other ways to look at it.}
A complex vector space is built on the basis vectors (which by the way, forms a normalized and orthogonal set). As an example, consider the case of 3-qubit states. Linear combinations can now be written nicely, 
$$
|\psi\rangle = \sum_{i=0}^7\alpha_i|i\rangle_3=
\alpha_0|0\rangle_3+\alpha_1|1\rangle_3+\ldots+\alpha_7|7\rangle_3=
$$
$$
\alpha_0|000\rangle+\alpha_1|001\rangle+\ldots+\alpha_7|111\rangle.
$$

The general case is 

\begin{equation}\label{nQubitLinComb}
|\psi\rangle = \sum_{i=0}^{2^n-1}\alpha_i|i\rangle_n.
\end{equation}

The coefficients are normalized

\begin{equation}\label{nQubitNormalization}
\sum_{i=0}^{2^n-1}|\alpha_i|^2=1.
\end{equation}

Having introduced the complex numbers $\alpha$ into the theory, there is no reason to work with the very restrictive set of matrices used in classical computation. A priori, any matric $C$ with complex entries could be contemplated as a candidate for a computation. There are however restrictions even in the quantum case that we will come to. But first, let us consider a few examples.

\subsubsection{One-bit quantum computations}
Consider first a one-qubit space. Define two new matrices $Y$ and $Z$ by

\begin{eqnarray}
Y=\pmatrix{0&-i\cr i&0}\label{eq: Ymatrix}
\\
Z=\pmatrix{1&0\cr 0&-1}\label{eq: Zmatrix}.
\end{eqnarray}

Acting with $Z$ on the 1-qubit basis vectors yields

$$
Z|0\rangle=Z\pmatrix{1\cr 0}=\pmatrix{1\cr 0}=|0\rangle,
$$
$$
Z|1\rangle=Z\pmatrix{0\cr 1}=-\pmatrix{0\cr 1}=-|1\rangle.
$$

Such a computation has no meaning classically. But quantum mechanically, the state (call it $|\phi\rangle$) resulting from acting with $Z$ on the state $|\psi\rangle$ of (\ref{eq: LinCombQubit})

$$
Z|\psi\rangle=Z\big(\alpha| 0\rangle + \beta| 1\rangle\big)=\alpha| 0\rangle - \beta| 1\rangle=|\phi\rangle,
$$
is, so to speak, no worse than $|\psi\rangle$ itself.

If one hides the intermediate steps, this simple calculation $Z|\psi\rangle=|\phi\rangle$, shows that $Z$ transforms the state $|\psi\rangle$ in one computational step. 

Let us also introduce one further matrix, of outmost importance in quantum computation, This is the so called {\it Hadamard} matrix

\begin{equation}\label{eq: Hadamard}
H={1\over\sqrt 2}\pmatrix{1&1\cr 1&-1}.
\end{equation}

This matrix is used build the linear superpositions of (\ref{eq: PlusState}) and (\ref{eq: MinusState}) out of the basis states, or as can be checked by a simple calculation

\begin{eqnarray}
H|0\rangle=|+\rangle\label{eq: HadamardOn0}
\\
H|1\rangle=|-\rangle\label{eq: HadamardOn1}.
\end{eqnarray}

%*************
\subsubsection{Multi qubit computations}
Generalizing to $n$-qubit states and $2^n\times2^n$ dimensional computational matrices, we have $C|\psi\rangle=|\phi\rangle$. Thus an exponential number of classical computational steps are performed in parallel.

If this is to be simulated on a classical computer, then of course, an exponential number of operations have to be performed anyway, and nothing is gained as compared to performing the classical computations

The situation is drastically changed if quantum devices can be built that actually performs the operation $C$ on the state $|\psi\rangle$.

%*************
\subsubsection{Restriction on quantum computation matrices}
A general quantum computation can now be written as

\begin{equation}\label{eq: GenQuanComp}
|out\rangle=C|in\rangle.
\end{equation}

In a real quantum computer, this process of transforming the input state into the output state is actually a dynamical process that occurs in time, we speak of {\it time evolution}. Now it is a fundamental property of quantum mechanics that, if no measurements are made, the time evolution is reversible. This means that the input can be inferred from the output. This, and other requirements, puts a restriction on the allowed computational matrices $C$. Deriving this restriction demands tools that will be developed in the following two chapters. Here we just state the result.

First, if we can find an inverse $C^{-1}$ to $C$ with the property $C^{-1}C=1$, then equation (\ref{eq: GenQuanComp}) can be inverted by multiplying through by $C^{-1}$,

$$
C^{-1}|out\rangle=C^{-1}C|in\rangle=|in\rangle,
$$
so that

$$
|in\rangle=C^{-1}|out\rangle.
$$

The requirement on the matrices $C$ that makes it straight forward to invert them is that they must be unitary. This means that $C$ is invertible and its inverse is equal to its conjugate transpose $(C^*)^T=C^{\dagger}$, or

$$
C^{-1}=C^{\dagger}.
$$

Taking the conjugate transpose is a computationally cheap process of rearranging and complex conjugating the elements of the matrix.

\chapter{Introduction to quantum mechanics}

Quantum mechanics is not a physical theory in itself, it is rather a framework in which physical theories must be formulated. If one takes a more fundamental, or philosophical point of view, quantum mechanics is a basic characteristic of reality which transcends all descriptions or theories of physical systems. It sets certain limits on what can known in principle about physical systems. The bare bones of quantum mechanics can be formulated as a few postulates which every quantum mechanical description of a physical system must conform to.

In the community of physicists, opinions differs as to the proper philosophical status of quantum mechanics. The majority view seems to be to take it as a fact of life, and since physical theories based on quantum mechanics in general agree very well with experiment, the only sensible thing to do is to go on and use it. There are features to quantum mechanics (for example the uncertainty principle and entanglement) that are considered to be counter intuitive from an everyday or classical physics perspective, but there is not a single experimental fact contradicting quantum mechanics. Quite to the contrary, the theory is verified every day in physics laboratories around the world. Quantum mechanics has furthermore been corroborated during the last twenty years by special experiments testing the very foundations of the theory \cite{AspectEtAl1982}. 

There are, however, and has always been, a strand of physicists uncomfortable with quantum mechanics. For them the theory is, though in practice successful, in principle tentative, and eventually due to be replaced by a more satisfactory theory. The discussion goes back to the very beginning of quantum mechanics and in particular to the famous Bohr-Einstein debate. 

Some physicists maintain that (as Niels Bohr is reported to have said) that if you're not confused by quantum mechanics, then you haven't understood it, while others, especially younger physicists, can't understand what all the fuss is about. Clearly, this has more to do with ones own philosophical outlook than with the theory itself, and we will leave this discussion here. As this is not a work on fundamental principles of natural philosophy, I will adopt the standard view that quantum mechanics is the proper framework for describing and understanding physical systems, and that classical theories offers at best very good approximations. It should be kept in mind though, that there are fundamental problems having to do with the relation between quantum mechanics and relativity, especially general relativity and the theory of gravitation. This is not, at least not yet, of any importance to the theory of quantum computation.

The term {\itshape quantum physics} thus refers to any physical system, or theory, formulated according to the postulates of quantum mechanics. The term {\itshape classical physics} on the other hand, refers to physics not formulated using quantum mechanics. Examples of classical theories are classical mechanics (or Newtonian mechanics), relativity (both special and general) and classical electrodynamics. These classical theories are, as already noted, excellent approximations to physical phenomena that takes place on a macroscopic scale, and often even to microscopic phenomena. But in principle, physics is 'quantum'. 

Many physical theories come in both a classical and a quantum version and there are well defined procedures to pass between them. The procedure of going from classical to quantum is called {\it quantization}. Very often, quantum theories are formulated by first writing down a classical theory which is then quantized according to a set of heuristic rules. 

Obviously, in order to work on quantum computation you need some grasp on quantum mechanics. It is in fact not difficult to rapidly gather together the basic elements of quantum mechanics on a couple of pages, and this seems to be what most review articles does. Such a brief expose through the quantum mechanical toolbox tends however to rather dull, and I think, fairly incomprehensible if you haven't already studied the subject.

I will adopt another strategy. Quantum mechanics will be introduced through a set of simple physical toy models. These will be the standard models that have traditionally proved their worth in physics education. In the course of working through the models, all relevant quantum mechanical concepts can be abstracted from these concrete models. We will recklessly assume that whats true in the particular case is true in the general case unless otherwise stated. Of course, such an approach is only useful in a first general introduction to a subject, and is not a substitute to proper study. I also think that this approach will be helpful when implementations of quantum computation in terms of physical devices are discussed briefly in chapter 9.

The simple model systems we will consider are:
 
\begin{itemize}
\item Particle in a potential box
\item Harmonic oscillator
\item Spin
\end{itemize}

There are a few popular formulations of quantum theory. One of them uses configuration space wave functions and their conjugate momentum space Fourier transforms. This is one of the traditional formulations, originating with Erwin Schr\"odinger, and it is traditionally called {\itshape wave mechanics}. Another formulation, contemporary with Scr\" odinger's, is Heisenberg's {\itshape matrix mechanics}. These are the original formulations of quantum mechanics and they were almost immediately shown to be equivalent. In the context of a theoretical study of quantum computation another formulation, somewhat more abstract, and which can be considered as a generalization of the other formulations, due to P.A.M Dirac, is more appropriate. A very readable account of this formulation is Dirac's own classic book \cite{Dirac1930}.

A fourth formulation, the {\itshape path integral formulation} , developed by R.P.N Feynman in the 1950's \cite{Feynman1948}, will not be mentioned here. It is of extreme usefulness in modern theoretical physics, but its methods does not seem to be needed in quantum computation.

%QUANTUM MECHANICS IN ONE SPACE DIMENSION
%****************************************
\section{Quantum mechanics in one space dimension}
As our introduction to quantum mechanics we will study a particle moving in one dimension of space under the influence of a potential. The state of the particle is described by the wave-function $\psi(x,t)$, where $x$ is the space coordinate and $t$ is the time. The states of a system can be described in different ways. This particular representation is called the {\itshape configuration space representation}, where {\itshape configuration} refers to using space to parameterize the state. The dynamics of the state is governed by the Schr\"odinger equation \cite{Schr1926}

\begin{equation}\label{eq: SchrodingerEquationGeneral}
i\hbar{\partial\over\partial t}\psi = H\psi.
\end{equation}

In this equation, $\hbar$ is a physical constant which sets the scale of quantum phenomena.\footnote{It's numerical value is  $1.054\cdot10^{-34}$ Js.}$H$ is the {\it Hamiltonian operator}. The equation equates the time rate of change of the wave function with the action of the Hamiltonian, thus the dynamics of the state is encoded in the form of $H$. In quantum computation, the 'program' of the quantum computer can be regarded as encoded in the Hamiltonian. But more on this later on.

The Hamiltonian is related to the classical energy of the system. In classical physics, a particle has a mechanical energy consisting of {\it kinetic} energy $K$ and {\it potential} energy $V$, and the total energy is $E=K+V$. The kinetic energy is given by

\begin{equation}\label{eq: KineticEnergy}
K={p^{2} \over 2m},
\end{equation}

where $p$ is the particle {\it momentum}, classically related to the velocity $v$ through $p=mv$ where $m$ is the particle mass. Thus, the kinetic energy can also be written as
 
$$
K={mv^{2} \over 2},
$$ 
 
a formula perhaps more readily recognized by non-physicists. However, the first form is the fundamental one.

The potential energy depends on the forces acting on the particle. Forces are not further analyzed in this context, and a formula is simply given for $V$. In general, it is a function of space and time, but we will only consider time-independent potentials.

Quantization is performed via the heuristic rules

$$
\mbox {\bf replace    } x \mbox {    \bf by    }x\cdot
$$
$$
\mbox {\bf replace    } p \mbox {    \bf by    }-i\hbar {\partial\over \partial x},
$$

or more concisely

\begin{eqnarray}
x\longrightarrow x\cdot\label{eq: QuantizationRulex}
\\
p\longrightarrow-i\hbar {\partial\over \partial x}\label{eq: QuantizationRulep}
\end{eqnarray}

In these rules, the left hand sides should be thought of as classical physics entities, whereas the right hand sides stands for the corresponding quantum mechanical operators. An operator can be either multiplication by a function $f\cdot$ or a differential operator $D$ (as in the second rule) acting on the state.\footnote{Other representations of quantum mechanical operators will appear subsequently.} If this sounds confusing, this is not the proper time for worry. It is best just to carry on in order to get a little bit more used to the quantum mechanical machinery.

If one applies these rules to the classical energy, one gets the Hamiltonian, or in formulas

\begin{equation}\label{eq: HamiltonianGen2}
E\longrightarrow H={1\over 2m}\Big(-i\hbar{\partial\over \partial x}\Big)^2 + V(x)=-{\hbar^2\over 2m}{\partial^2\over\partial x^2}+V(x).
\end{equation}

The Schr\"odinger equation now becomes

\begin{equation}\label{eq: SchrodingerEqGen2}
i\hbar{\partial \over \partial t}\psi=-{\hbar^2\over 2m}{\partial^2\over\partial x^2}\psi+V(x)\psi.
\end{equation}

This is a partial differential equation governing the time development of the system. This simple example captures most of the main features of quantum mechanics in this formulation. 

A more realistic system would be in three spatial dimensions. The force acting on the particle is given by the potential, examples of which could be the coulomb field from an atomic nucleus on an electron, forces from other electrons and perhaps time-dependent electromagnetic fields. But we will stick to this simple one-dimensional system and solve the equation in two cases; the {\it square well potential} and the {\it harmonic oscillator}. 

The first steps in the solution are general and does not depend on the form of the potential, except that it is assumed to be independent of time. The method is the standard separation of variables method used in solving partial differential equations. We review it here in order to highlight the aspects that will be abstracted later on.

%**********************************************
\subsection{Separation of space and time}
If the potential is time-independent, i.e. if $V(x,t)=V(x)$, the Schr\"odinger equation can be simplified by separating the variables
 
\begin{equation}\label{eq: SeparationAnsatz}
\psi(x,t)=\sum_n u_n(x)f_n(t).
\end{equation}

This is an ansatz for the solution, which can be justified referring to general theorems on partial differential equations. \cite{MorseFeshbach1953}. Here, $\{u_n\}$ and $\{f_n\}$ are enumerable infinite sets of functions. 

If the ansatz is inserted into the Schr\"odinger equation one gets

$$
i\hbar u(x){\partial f(t)\over \partial t}=-{\hbar^2\over 2m}f(t){\partial^2 u(x)\over\partial x^2}+V(x)u(x)f(t),
$$

which upon division by $u(x)f(t)$ yields
 
$$
i\hbar{1\over f}{\partial f\over \partial t}={1\over u}\lbrack -{\hbar^2\over 2m}{\partial^2 u\over\partial x^2}+Vu\rbrack.
$$

Now, since the left hand side is independent of $x$ and the right hand side is independent of $t$, both sides must be equal to the same constant $E$. This constant is called a {\it separation} constant. We thus get two ordinary differential equations, one for the time-dependent function $f$ and one for the space-dependent function $u$,

\begin{eqnarray}
i\hbar{\partial f\over \partial t}=Ef\label{eq: SeparatedEquationTime}
\\
-{\hbar^2\over 2m}{\partial^2 u\over\partial x^2}+Vu = Eu\label{eq: SeparatedEquationSpace}.
\end{eqnarray}

The first equation is easy to solve

\begin{equation}\label{eq: TimeEqSolution}
f(t)=C\exp({iEt\over \hbar}),
\end{equation}
where $C$ is a constant.

The second equation is an eigenvalue equation of the Sturm-Liouville type, and its general form can be written abstractly as

$$
Hu=Eu.
$$

The solutions to this equation will be precisely the functions $u_n(x)$ in the expansion (\ref{eq: SeparationAnsatz}). They are referred to as {\it eigenfunctions} and the constants $E_n$ as {\it eigenvalues}.

For special forms of the potential $V$ (and appropriate boundary conditions), the equation defines well-known systems of orthonormal functions $\{u_m(x)\}$ with the index $m$ running over some infinite subset of $Z$. The general solution to the wave equation can then be written by inserting (\ref{eq: TimeEqSolution}) into (\ref{eq: SeparationAnsatz})

$$
\psi (x,t)=\sum_n u_n(x)\exp({iE_nt\slash \hbar)},
$$

where the constant $C$ has been hidden in the as yet undetermined functions $u_n$.

%**********************************************
\subsection{Particle in a potential well}
With the general groundwork done, we will now turn to our first example, a quantum particle trapped within a container with impenetrable walls. In, reality there is no such thing, but it can be mimicked by choosing a potential of the form

\begin{equation}\label{eq: SquareWell}
V(x)=\cases{+\infty,&if $|x|>a$\cr 
0, &if $|x|<a$}
\end{equation}

with the walls at the locations $x=-a$ and $x=a$. The impenetrability of the walls is modeled by the infinite value for the potential outside the well. This potential is often called a {\it square well potential}.

Since the potential has three distinct regions, being discontinuous at $x=\pm a$, the equation must be solved in the three regions separately. However, since in the two regions $x<-a$ and $x>a$, the potential is infinite, the function $u$ must be equal to zero here. Furthermore, $u$ itself must be continuous at the potential walls. This translates into {\it boundary conditions} for the solution in the region $|x|<a$. Therefore, we get

\begin{equation}\label{eq: SquareWellEq}
-{\hbar^2\over 2m}{\partial^2 \over\partial x^2}u=Eu
\end{equation}
with boundary conditions
\begin{equation}\label{eq: BvSquareWellEq}
u(a)=u(-a)=0.
\end{equation}

In this form, the boundary conditions are quite easy to understand. In classical physics, no particle can pass from as region with finite potential energy into a region with infinite potential energy, as that would require an infinite kinetic energy. This is true in quantum physics also. And since the $u$ in some sense corresponds to the presence of the particle (in a way that will explained later), those boundary conditions corresponds to the impenetrable walls.

This differential equation has the well known solution 

$$
u(x)=A\sin kx+B\cos kx
$$

with $k=\sqrt{2mE\slash\hbar^2}$. 

Inserting the boundary conditions, we get the two equations

$$
\cases{A\sin ka+B\cos ka=0&\cr
-A\sin ka+B\cos ka=0},
$$

which reduce to

$$
\cases{A\sin ka=0&\cr
B\cos ka=0}.
$$

Since these sine and cosine expressions cannot be simultaneously zero, we get the solutions

$$
\cases{A=0&and $\cos ka=0 \Rightarrow ka={n\pi\slash 2}$  with $n$ odd\cr
B=0&and $\sin ka=0 \Rightarrow ka={n\pi\slash 2}$ with  $n$ even}.
$$

In fact it suffices to restrict the solutions to non-negative integers. The reason is that the solutions for negative integers are not linearly independent of the solutions with positive integers. This is apparent from the explicit form of the solutions

\begin{equation}\label{eq: SquareWellSolutions}
u_n(x)=\cases{B\cos ({n\pi x\slash 2a}),& $n$ odd\cr
A\sin({n\pi x\slash 2a}), & $n$ even}.
\end{equation}
Changing $n$ to $-n$ in these formulas have no effect for the cosine solutions, and for the sine solutions, the ensuing change of sign for sine can be absorbed into the constant $A$. 

The constants $A$ and $B$ are normalization constants to be determined by the normalization condition

\begin{equation}\label{eq: Normalization}
\int_{-a}^au_n(x)^*u_n(x)dx=1
\end {equation}

where '*' denotes complex conjugation. Clearly, some normalization of the solutions is needed, and this particular one is related to the interpretation of quantum mechanics where the wave functions $\psi(x)$ are interpreted as {\itshape probability amplitudes}. To say that $\psi$ is a probability amplitude is to say that the integral

$$
\int_c^d\psi(x)^*\psi(x)dx
$$

is the probability of detecting the particle in the interval $(c,d)$.

In fact, an even stronger property of our solutions can be inferred:

\begin{equation}\label{eq: OrtonormGen}
\int_{-a}^au_n(x)^*u_m(x)dx=\delta_{nm}=\cases{1, &if $n=m$\cr 0, &if $n\not=m$}.
\end{equation}

This equation expresses the {\itshape orthonormality} of the solutions, i.e the solutions are normalized and solutions with different index are orthogonal. This is a general property of solutions to eigenvalue problems. The general theory will be spelt out in chapter 5. 

The solutions $u_n(x)$ can now be considered to form a basis of a linear vector space. Any solution to the wave equation can be expressed as a linear combination (or superposition) of the basis functions

\begin{equation}\label{eq: BasisExpansion}
\psi(x)=\sum_{n=0}^\infty \alpha_nu_n(x),
\end{equation}

where $\alpha_n$ are complex numbers. These numbers are arbitrary apart from a global normalization. Since the total probability of detecting the particle in the box must be 1, we get

\begin{equation}\label{eq: OrtonormSpec}
\int_{-a}^a\psi(x)^*\psi(x)dx=\sum_0^\infty \alpha_n^*\alpha_n=1
\end{equation}

This follows from a nice calculation involving several of the formulas given in this section and some trigonometry. Performing this calculation gives quite a lot of insight into the mathematics of quantum mechanics.

Since we now know $k$, we can get a formula for the separation constant E, which is to be interpreted as the energy of the system

\begin{equation}\label{eq: EnergySquareWell}
E_n={\hbar^2k^2\over 2m}={\hbar^2\over 2m}{n\pi\overwithdelims () 2a}^2={\pi^2\hbar^2\over 8ma^2}n^2.
\end{equation}

In this way we get quantization of the energy. The energy can only take values determined by the integer $n$. This is symbolized by indexing the $E$ with $n$. 

But why is $E_n$ energy? This can be understood by going back to the classical equation for energy in terms of kinetic and potential energy $E=K+V$. In our case, the potential energy is zero inside the box, and we have simply $E=K$. Then using the quantization rules we got 

$$
E\longrightarrow H={1\over 2m}(-i\hbar{\partial\over \partial x})^2.
$$

If this Hamiltonian operator is applied to any of the solutions, the result is

$$
Hu_n(x)={1\over 2m}(-i\hbar{\partial\over \partial x})^2B\cos ({n\pi x\over 2a})=
$$
$$
{\hbar^2\over 2m}({n\pi\over 2a})^2B\cos ({n\pi x\over 2a})=({\hbar^2\pi^2\over 8ma}n^2)B\cos ({n\pi x\over 2a})=E_nu_n(x)
$$

In the next to last expression, we recognize the separation constant (eigenvalue) $E_n$. Thus, the physical interpretation of the eigenvalue equation 

$$
Hu_n(x)=E_nu_n(x),
$$

is that the eigenvalues corresponding to the Hamiltonian operator $H$ are the energies of the accessible states for the system. This makes sense, since the Hamiltonian itself is the quantum operator corresponding to the classical energy.

We have thus seen that the states of the system form an enumerable infinite set. The first stage of abstraction in quantum mechanics is to note that, although we might need special properties of these solutions, it is not be necessary to work with the explicit representation in terms of sine and cosine functions. Our next example system, the harmonic oscillator, will illustrate this.

%LINEAR HARMONIC OSCILLATOR
%**************************
\section{Linear harmonic oscillator}
The harmonic oscillator is a simple and extremely useful model of physical systems both in classical physics and quantum physics. Classically, it can be used to model mechanical vibrations. In quantum physics it is model for the modes of electromagnetic waves. Its usefulness stems from the fact that even complicated many-particle systems or continuous media can often be analyzed in terms of normal modes of vibration, perhaps after linearization, and furthermore, that it is a completely solvable model.

The model is also very useful in that it has enough features in order to develop large portions of elementary classical and quantum mechanics within it. This is precisely what will be done in this section. 

The simple classical harmonic oscillator consists of a particle connected with a spring to rigid wall. The force from the spring is proportional to the displacement of the spring from its natural length. If other forces like air-resistance and friction are neglected, the particle will oscillate forever once it is set in motion. During this oscillation, there will be an oscillation of energy between kinetic energy of motion and potential energy in the spring. The total energy is constant during the motion. The total energy is said to be a {\itshape constant of the motion}.  

When setting up a model for this system, physicists normally abstract away from the wall, instead considering a particle of mass $m$ attracted to a fixed center by a force that is proportional to the displacement from the center. This gives a more symmetrical formulation, and the displacement can take negative values. Letting $x$ denote the displacement from the center, the force acting on the particle is $F=-kx$. The negative sign makes the force attracting. The equilibrium position is in the center where $x=0$. The so called {\it conservative} forces, i.e forces that conserves total energy, can always be derived from a potential by the relation 

$$
F=-{d\over dx}V(x).
$$

It is thus easy to see that the potential energy for a harmonic oscillator is 

$$
V(x)=\frac{\rm 1}{\rm 2}kx^2.
$$

The total classical energy of the oscillator is 

\begin{equation}\label{eq: EnergyHarmOsc}
E=K+V={p^2\over 2m}+{1\over 2}kx^2.
\end{equation}

Since, as noted, $E$ is a constant during the motion, we can read off the oscillation of the energy between kinetic energy and potential energy. When $x=0$, corresponding to the particle passing through the center of motion, all energy is kinetic and the momentum $p$ takes its maximum value. On the other hand, when $p=0$, the kinetic energy is zero and all energy is potential. This corresponds to the turning points at the maximum distance from the center.

We will not analyze the classical model further, but turn directly to the quantum harmonic oscillator.

%**********************************************
\subsection{Quantization of the oscillator}
Referring back to the quantization rules of the previous section we can now easily write down the Hamiltonian for the harmonic oscillator. Applying the quantization rules to the energy (\ref{eq: EnergyHarmOsc}) we get

$$
E\longrightarrow H=-{\hbar^2\over 2m}{\partial^2\over\partial x^2}+{1\over 2}kx^2.
$$

The Schr\"odinger equation becomes

\begin{equation}\label{eq: HarmOscSchrEq}
i\hbar{\partial \over \partial t}\psi=-{\hbar^2\over 2m}{\partial^2\over\partial x^2}\psi+{1\over 2}kx^2\psi.
\end {equation}

The separation of variables and the solution for the time-dependence, i.e. the steps recorded in section 4.2.1, are exactly the same for the harmonic oscillator. We need only concentrate on the space-dependence. In fact, this is true for all systems for which the forces are time-independent. Thus, after separation of the variables we always end up with equation (\ref{eq: SeparatedEquationSpace}) of the previous section, with the appropriate potential.\footnote{In three dimensions of space, and in other coordinate systems than rectangular, the equation is more complicated - but in principle it is always the same equation with the relevant potential.} Therefore, the eigenvalue equation to solve is

\begin{equation}\label{eq: HarmOscEigvalEq}
-{\hbar^2\over 2m}{\partial^2\over\partial x^2}u+{1\over 2}kx^2u=Eu.
\end{equation}

At this stage, one can proceed as in the previous section, and solve this differential equation to obtain an infinite set of basis functions. The set of orthonormal basis functions can be written as

$$
u_n(x)=N_nH_n(\alpha x)exp({-\frac{\rm 1}{\rm 2}\alpha^2 x^2}).
$$

Here, $\alpha$ is a constant and $N_n$ is a normalization constant

$$
\alpha^4={mk\over\hbar^2}\qquad N_n=({\alpha\over \sqrt{\pi 2^n n!}})^\frac{\rm 1}{\rm 2}
$$

and $H_n$ are Hermite polynomials, the first few of which are

$$
H_0(x)=1,\qquad H_1(x)=2x,\qquad H_2(x)=4x^2-2.
$$

We will not derive this solution to the harmonic oscillator.\footnote{Any standard textbook on quantum mechanics contains the calculations\cite{Schiff, LandauLifschitz]}.} Instead we will introduce a more abstract, and more powerful formalism, which is also the standard formalism used in almost all applications of harmonic oscillators. This is the method of annihilation and creation operators and in the process of introducing them we will also introduce the very useful notation on {\it bra} and {\it ket} vectors invented by Dirac.

%**********************************************
\subsection{Operators for momentum and position}
One step in the quantization of a classical system is to replace classical dynamical variables with operators. Momentum $p$ and position $x$ are replaced by the momentum operator and the position operator respectively, often denoted by $\hat p$ and $\hat x$.\footnote{Pronounced p-hat and x-hat.} Explicit representations of these operators are

\begin{eqnarray}
\hat p=-i\hbar{\partial\over \partial x}\label{eq: MomentumOperator}
\\
\hat x=x\label{eq: PositionOperator},
\end{eqnarray}
which is the same representation of the operators as in equations (\ref{eq: QuantizationRulex}) and (\ref{eq: QuantizationRulep}). This particular representation is valid in the configuration space representation of the states using $x$-space wave-functions. Note that momentum is represented by a differential operator, and position by a multiplication. A consequence of this is that the order of application of the operators matters. A simple calculation illustrates this. Consider applying first $\hat x$ and then $\hat p$ to a state $\psi$, and then applying these operators in the reverse order, i.e. first $\hat p$ and then $\hat x$:

$$
\hat p\hat x\psi=-i\hbar{\partial\over \partial x}\big(x\psi\big)=-i\hbar\psi-i\hbar x{\partial\over \partial x}\psi=-i\hbar\big(1+x{\partial\over \partial x}\big)\psi,
$$
$$
\hat x\hat p\psi=-i\hbar x{\partial\over \partial x}\psi.
$$
Then subtract these expressions to get
$$
(\hat x\hat p-\hat p\hat x)\psi=i\hbar\psi.
$$

The state $\psi$ is completely arbitrary here, and can be removed, yielding the operator equation

\begin{equation}\label{eq: xpCommutator1}
\hat x\hat p-\hat p\hat x=i\hbar.
\end{equation}

This combination of operators is so important in quantum mechanics and so frequently occuring that a special notation is introduced. A {\itshape commutator bracket}, or simply a {\itshape commutator}, between two operators $\hat A$ and $\hat B$ is defined by 

\begin{equation}\label{eq: GenCommutator}
\lbrack \hat A , \hat B\rbrack=\hat A\hat B - \hat B\hat A.
\end{equation}

Using this notation, equation (\ref{eq: xpCommutator1}) can be written

\begin{equation}\label{eq: xpCommutator2}
\lbrack \hat x , \hat p\rbrack=i\hbar.
\end{equation}

This is a fundamental equation relating the operators $\hat x$ and $\hat p$ and it holds whatever representation is used. Therefore it can be used as a quantization condition.

For completeness, we also record the trivial commutators

\begin{eqnarray}
\lbrack \hat x , \hat x\rbrack=0\label{eq: xxCommutator}
\\
\lbrack \hat p , \hat p\rbrack=0\label{eq: ppCommutator}.
\end{eqnarray}

In general, an operator always commutes with itself.

%**********************************************
\subsection{Commutators}
If the commutator between two operators is non-zero, i.e if $\lbrack \hat A , \hat B\rbrack\not=0$, the operators are said to be non-commuting. In that case, as we have seen, the order in which they are applied to a quantum state matters. Sometimes, when there is no particular ordering to prefer, or when the ordering is ambiguous, a {\it symmetrical ordering} is chosen as a kind of default ordering 

\begin{equation}\label{eq: SymOrdering}
(\hat A \hat B)_{sym}={1\over 2}(\hat A \hat B + \hat B \hat A).
\end{equation}

%**********************************************
\subsection{A note on classical dynamics}
Classical dynamics for a particle in one dimension of space is governed by\break Newton's equation 

\begin{equation}\label{eq: NewtonEq}
ma=F,
\end{equation}

where $a$ denotes the acceleration. Acceleration is the time derivative of velocity, 

$$
a={dv\over dt},
$$

which in its turn is related to the particle momentum through the equation 

\begin{equation}\label{eq: VelocityMomentum}
v={p\over m}. 
\end{equation}

The force $F$ is determined by the potential $V$

$$
F=-{dV\over dx}.
$$

Combining these four equations while assuming that the mass is constant (the normal case), yields

\begin{equation}
{dp\over dt}=-{dV\over dx}.
\end{equation}

Since the kinetic energy $K$ does not depend on $x$, it is possible to replace the potential energy $V$ in this formula with the total energy E. It is practical to change notation slightly and use $H$ for the total energy also in the classical case, reserving $E$ for the quantum mechanical energy eigenvalues. Thus we write for the total energy $H=K+V$. This yields the dynamical equation

\begin{equation}\label{eq: HamEqdpdt}
{dp\over dt}=-{dH\over dx}.
\end{equation}

At this stage one might worry that the original Newton equation (\ref{eq: NewtonEq}) is a second order differential equation (remember, the acceleration $a$ is a second order derivative with respect to time), and (\ref{eq: HamEqdpdt}) is a first order differential equation. Something is missing. 

The missing ingredient is precisely the equation (\ref{eq: VelocityMomentum}) relating velocity and momentum, or rather this equation rewritten so that it relates velocity to the momentum derivative of the energy. Differentiating $H$ with respect to $p$ yields

$$
{dH\over dp}={dK\over dp}={d\over dp}{p^{2} \over 2m}={p\over m}.
$$

But this is precisely the velocity $v={dx\slash dt}$, so we get

\begin{equation}\label{eq: HamEqdxdt}
{dH\over dp}={dx\over dt}.
\end{equation}

These two equations, (\ref{eq: HamEqdpdt}) and (\ref{eq: HamEqdxdt}), are the fundamental dynamical equations of classical mechanics. This reformulation of Newtonian mechanics was performed during the $18^{\rm th}$ and $19^{\rm th}$ centuries by Euler, Lagrange, Hamilton and Poisson. It is quite general and it is the formulation of classical mechanics in which the translation to quantum mechanics is most easily performed. In the general theory, where there might be more than one particle, one considers a dynamical system described by a set of dynamical variables $\{(x_i,p_i)\}$.\footnote{These variables can be more general than ordinary position and momentum, including for example angular variables.} The energy, or Hamiltonian is function of these variables. Then the Hamiltonian equations of motion become

\begin{eqnarray}
{dx_i\over dt}={\partial H\over\partial p_i}\label{eq: HamEqGenx}
\\
{dp_i\over dt}=-{\partial H\over\partial x_i}.\label{eq: HamEqGenp}
\end{eqnarray}

Note how beautiful and symmetrical these equations are! However, they can still be rewritten in an even more compact way in which the transition to quantum mechanics is most easily performed. In order to do that, consider a dynamical system with coordinates and momenta $\{(x_i,p_i)\}$ where the indices ranges from 1 to $n$, $n$ being the number of particles of the system. Let $A$ and $B$ be two dynamical variables, both differentiable functions of the $x's$ and $p's$. The {\itshape Poisson bracket} is defined by

\begin{equation}\label{eq: PoissonBracket}
\{A,B\}=\sum_1^n\;\Big({\partial A\over\partial x_i}{\partial B\over\partial p_i}-{\partial B\over\partial x_i}{\partial A\over\partial p_i}\Big).
\end{equation}

As an example, calculate the brackets $\{x,H\}$ $\{p,H\}$ for a one-dimensional system, i.e $n=1$.

$$
\{x,H\}={\partial x\over\partial x}{\partial H\over\partial p}-{\partial H\over\partial x}{\partial x\over\partial p}={\partial H\over\partial p}
$$
$$
\{p,H\}={\partial p\over\partial x}{\partial H\over\partial p}-{\partial H\over\partial x}{\partial p\over\partial p}=-{\partial H\over\partial x}
$$

Combining these equations with (\ref {eq: HamEqGenx}), (\ref {eq: HamEqGenp}) and generalizing to $n$ particles, the Hamiltonian equations of motion can be written compactly

\begin{eqnarray}
{dx_i\over dt}=\{x_i,H\}\label{eq: ParticlesDynEqx}
\\
{dp_i\over dt}=\{p_i,H\}\label{eq: ParticlesDynEqp}.
\end{eqnarray}

In these equations, coordinates (positions) and momenta appear completely symmetrically.

The reader might feel that an edifice (a general theory of dynamics) has been erected on a very tiny base (the harmonic oscillator). That is indeed true. But everything in this section can be developed quite rigorously from fundamental principles of physics. A good accessible reference is \cite{Goldstein1950}. As a last point, it can be shown that the equation of motion for a function $F$ of the dynamical variables can be written

\begin{equation}\label{eq: HamiltonEq}
{dF\over dt}=\{F,H\}+{\partial F\over\partial t}.
\end{equation}

The second term ${\partial F\slash\partial t}$, vanishes if there is no explicit time dependence for $F$. This is often the case.

%**********************************************
\subsection{Quantization}
Hamilton's equations for classical dynamics offers a very natural starting point for quantization. The rules are

\begin{enumerate}
\item Replace classical dynamical variables $A$ with the corresponding quantum operators $\hat A$ 
$$
A\longrightarrow \hat A
$$
\item Replace the Poisson brackets $\{\cdot ,\cdot\}$ with commutators $[\cdot ,\cdot]$ 
$$
\{A,B\}\longrightarrow{1\over i\hbar}\lbrack\hat A,\hat B\rbrack.
$$
\end{enumerate}

Upon these replacements, the equations of motion for the operators become

\begin{equation}\label{eq: HeisenbergEq}
{d\hat F\over dt}={1\over i\hbar}\lbrack \hat F,\hat H\rbrack.
\end{equation}

This form of quantum dynamics bears no obvious relation to the Schr\"odinger equation. The states does not even appear here. But in fact, there is a close relation and it will be explained in chapter 5. Suffice it here to say the present form of dynamics, which goes under the name {\itshape Heisenberg picture}, the time-development of the system is carried by the operators while the states are time-independent. That is the reason why they don't appear explicitly. In the Schr\"odinger equation approach, called the {\itshape Schr\"odinger picture}, the states carry the time-development while the operators are time-independent.

%**********************************************
\subsection{Dirac notation, a case of abstraction}
The Dirac notation is a very useful way of formulating quantum mechanics in that it allows quantum mechanical systems to be treated in a uniform way by abstracting away from particularities. In fact, a computer scientist might want to regard the Dirac formalism as providing an interface, specifying what properties and methods a system should support without entering into details on how to implement them.

Suppose a certain system is described by a wave function $\psi(x)$ expanded as in (\ref{eq: BasisExpansion})

$$
\psi(x)=\sum_{n=0}^\infty \alpha_nu_n(x).
$$

Now, the basis functions $u_n$ belong to a certain class of functions, all sharing common properties. Often it is just those properties that are important, not the explicit $x$-space representation. Letting $n$ label these properties, we can write the expansion in the form

$$
|\psi\rangle=\sum_{n=0}^\infty \alpha_n|n\rangle,
$$

where the {\it ket}-notation $|\cdot\rangle $ was introduced by Dirac to denote abstract quantum states.

Instead of having physical quantities represented as explicit differential operators acting on the wave functions, those quantities are now represented as abstract operators acting on the label $n$.

%**********************************************
\subsection{Summary of the classical harmonic oscillator}
Returning now to the harmonic oscillator, we can write the dynamical equations compactly using the general formalism just developed. The Hamiltonian is 

$$
H={p^2 \over 2m}+{kx^2\over 2}.
$$

With this particular form for the Hamiltonian, equations (\ref{eq: ParticlesDynEqx}) and (\ref{eq: ParticlesDynEqp})yield

$$
{dx\over dt}=\{x,{p^2 \over 2m}+{kx^2\over 2}\}=\{x,{p^2 \over 2m}\}={p\over m}
$$
$$
{dp\over dt}=\{p,{p^2 \over 2m}+{kx^2\over 2}\}=\{p,{kx^2\over 2}\}=-kx.
$$

Together these equations give 

$$
m{d^2x\over dt^2}=-kx$$

which is Newton's equation for a harmonic oscillator. To see this, note that the first equation gives $p=mv$ and insert this formula for $p$ in the second equation.

%**********************************************
\subsection{Creation and annihilation operators}

The groundwork is now laid for treating the harmonic oscillator using creation and annihilation operators. First, the classical Hamiltonian is rewritten in a suggestive way

$$
H={1\over 2}\big({p^2\over m}+kx^2\big)={1\over 2}{\sqrt{k\over m}}\big({p^2\over\sqrt{km}}+ \sqrt{km}\;x^2\big).
$$

Introducing $\omega = \sqrt{k\slash m}$, the Hamiltonian $H$ becomes\footnote{$\omega$ is related to the classical frequency of oscillations $f$ through $\omega = 2\pi f$. A useful formula is $m\omega=\sqrt{km}$}.

\begin{equation}\label{eq: HarmOscHamx2p2} 
H={1\over 2}\omega \big({p^2\over m\omega}+m\omega x^2\big).
\end{equation}

In anticipation of quantum mechanics, introduce $\hbar$ and do a further rewriting

$$
H={1\over 2}\hbar\omega\big({p^2\over m\omega\hbar}+{m\omega\over\hbar} x^2\big)=
$$
$$
{1\over 2}\hbar\omega\big({p\over\sqrt{m\omega\hbar}}-i\sqrt{m\omega\over\hbar}x\big)\big({p\over\sqrt{m\omega\hbar}}+i\sqrt{m\omega\over\hbar}x\big)=
$$
$$
{1\over 2}\hbar\omega z\bar{z},
$$
or for short

\begin{equation}\label{eq: zzBarFormHarmOscHam}
H={1\over 2}\hbar\omega z\bar{z}.
\end{equation}

The intuition behind this rewriting is that $H$ is a quadratic form, and therefore it should be possible to write it as a square. However, since $x$ and $p$ will become operators, care must be exercised regarding the order in which they appear in the product $z\bar{z}$. From now on, $x$ and $p$ are treated as operators, but we will drop the 'hat' notation, writing $A$ for $\hat A$.

In order to capitalize on the possibility to write $H$ as a square, define {\itshape creation} and {\itshape annihilation} operators $a$ and $a^{\dagger}$

\begin{eqnarray}
a={1\over \sqrt{2\hbar m\omega}}(p-im\omega x)\label{eq: Annihilation}
\\
a^{\dagger}={1\over \sqrt{2\hbar m\omega}}(p+im\omega x)\label{eq: Creation}.
\end{eqnarray}

The reason for giving them these, somewhat esoteric, names will become clear subsequently. Comparing these definitions with (\ref{eq: zzBarFormHarmOscHam}) suggests taking $z=\sqrt{2}a$ and $\bar z =\sqrt{2} a^{\dagger}$. However, since there is no reason to choose a particular ordering of the operators, a symmetric ordering will be used. Thus the Hamiltonian is written

$$
H={1\over 2}\hbar\omega(aa^{\dagger} + a^{\dagger}a).
\eqno{\addtocounter{equation}{1}\theequation}
$$

Inserting (\ref{eq: Annihilation}) and (\ref{eq: Creation}), and performing some careful algebra, yields

$$
H={1\over 2}\hbar\omega\big({1\over\sqrt{2\hbar m\omega}}\big)^2\Big(\big(p-im\omega x\big)\big(p+im\omega x\big)+
$$
$$
\big(p+im\omega x\big)\big(p-im\omega x\big)\Big)=
$$
$$
{1\over 4m}(p^2+im\omega px-im\omega xp+m^2\omega^2x^2+p^2-im\omega px+im\omega xp+m^2\omega^2x^2)=
$$
$$
{1\over 2m}(p^2+m^2\omega^2 x^2),
$$

which is the same formula as (\ref{eq: HarmOscHamx2p2}) slightly rearranged.

Note that the operator combinations $xp$ and $px$ cancel in the above calculation. They would not have done that, had not a symmetrical ordering been chosen.

So far, not very much has been achieved. In order to proceed, some properties of the creation and annihilation operators must be derived. In quantum mechanics, the commutators between operators are always important because much of the properties of a system are encoded into the commutators. We therefore calculate the commutator $\lbrack a,a^{\dagger}\rbrack$ using the definitions (\ref{eq: Annihilation}) and (\ref{eq: Creation}) and the basic commutators (\ref{eq: xpCommutator2})-(\ref{eq: ppCommutator})

$$
\lbrack a,a^{\dagger}\rbrack={1\over 2\hbar}(-i\lbrack x,p\rbrack + i\lbrack p,x \rbrack)=
$$
$$
{1\over 2\hbar}(-i(i\hbar)+i(-i\hbar))=1.
$$

This implies that $aa^{\dagger}=a^{\dagger}a+1$ and the Hamiltonian can be written as

\begin{equation}\label{eq: HarmOscHam}
H=\hbar\omega(a^{\dagger}a+{1\over 2}).
\end{equation}

The commutation relations for the creation and annihilation operators can now be summarized

\begin{eqnarray}
\lbrack a,a^{\dagger}\rbrack=1\label{eq: aaDagger}
\\
\lbrack a,a\rbrack=\lbrack a^{\dagger},a^{\dagger}\rbrack=0.\label{aaaDaggeraDagger}
\end{eqnarray}

So far no reference has been made to the states of the harmonic oscillator. It is time to introduce them now. Referring back to equation (\ref{eq: HarmOscHamx2p2}) we see that $H$ is a positive definite operator (the energy is positive classically), and therefore, on physical grounds, there must be a state with lowest energy. Denote this {\itshape ground state} with $|0\rangle$. In computing the energy for this state, we must know the effect of the creation and annihilation operators acting on it. We will choose

\begin{equation}\label{eq: GroundStateDefHarmonicOscillator}
a|0\rangle =0.
\end{equation}

The intuition behind this choice is that the ground state, being the lowest energy state, must be annihilated by the annihilation operator, but ultimately it is justified by the results that follow. The energy of the ground state can now be computed

$$
H|0\rangle =\hbar\omega(a^{\dagger}a+{1\over 2})|0\rangle ={\hbar\omega\over 2}|0\rangle.
$$

If there is a ground state, there ought to be {\itshape excited} states, i.e states with higher energy. As the terminology suggests, the next excited state above the ground state is created by the creation operator acting on the ground state. Denoting this state with $|1\rangle$, let us define tentatively

$$
|1\rangle=a^{\dagger}|0\rangle.
$$

Now, there is a consistency requirement on these equations. Since $\lbrack a,a^{\dagger}\rbrack=1$ it must be the case that 

$$
\lbrack a,a^{\dagger}\rbrack|0\rangle=1|0\rangle = |0\rangle.
$$

But this can now be checked explicitly

$$
|0\rangle=\lbrack a,a^{\dagger}\rbrack|0\rangle=aa^{\dagger}|0\rangle-a^{\dagger}a|0\rangle=a|1\rangle.
$$

Thus it must be the case that

$$
a|1\rangle=|0\rangle,
$$
the interpretation of which is that the first excited state is destroyed, or annihilated, by $a$.

Clearly, it must be possible to generalize this and construct an hierarchy of excited states by letting the creation operator act on the ground state repeatedly. The excited state $|n\rangle$ ought to be the ground state acted on by n creation operators, or writing a recursive definition

\begin{equation}\label{eq: StateNormalization1}
a^{\dagger}|n\rangle=\xi(n)|n+1\rangle\mbox{,   for   } n\geq 0,
\end{equation}
where $\xi(n)$ is an as yet undetermined normalization. This equation agrees with the previous formula for $|1\rangle$ when $n=0$ if we demand $\xi(0)=1$. 

On the other hand, acting with the annihilation operator on the state $|n\rangle$ should yield the state $|n-1\rangle$

\begin{equation}\label{eq: StateNormalization2}
a|n\rangle=\eta(n)|n-1\rangle.
\end{equation}

A careful analysis yields the coefficients $\xi(n)=\sqrt{n+1}$ and $\eta(n)=\sqrt{n}$, so that we have

\begin{eqnarray}
a^{\dagger}|n\rangle=\sqrt{n+1}\;|n+1\rangle\label{eq: RaiseState}
\\
a|n\rangle=\sqrt{n}\;|n-1\rangle.\label{eq: LowerState}
\end{eqnarray}

Equation (\ref{eq: RaiseState}) record the action of $a^{\dagger}$ as a {\it raising} operator, its action on a state is to increase the quantum number $n$ by 1. Likewise, Equation (\ref{eq: LowerState}) record the action of $a$ as a {\it lowering} operator, its action on a state is to decrease the quantum number $n$ by 1.

The name creation operator originates in the quantum theory of the electromagnetic field. The frequency modes of an electromagnetic field can be described by harmonic oscillators. In a quantum description of the electromagnetic field, a mode with frequency $f$ is corresponds to a harmonic oscillator with $\omega=2\pi f$. The intensity of the field corresponds to the number of photons in the mode, and the number of photons is precisely the quantum  number $n$ of the harmonic oscillator. As will be shown below, the energy in the mode is $\hbar\omega(n+{1\over 2})$. In this context, the action of the creation operator $a^{\dagger}$ is to create a new photon in the frequency mode. Correspondingly, the action of the {\it lowering} or annihilation operator $a$ is to decrease the quantum number $n$ or annihilate a photon in the mode.

Next, it follows that $a^{\dagger}a$ is a {\itshape number operator}, counting the excitation level of the state. Using the equations (\ref{eq: RaiseState}) and (\ref{eq: LowerState}) we get

$$
a^{\dagger}a|n\rangle=a^{\dagger}\sqrt n|n-1\rangle=\sqrt na^{\dagger}|n-1\rangle=\sqrt n\sqrt n|n\rangle=n|n\rangle
$$

In these calculations we are using the fact that numbers commutes with operators, i.e. the order can be interchanged freely.

Sometimes the number operator is denoted by $N$, writing $N=a^{\dagger}a$. Not surprisingly, the states $|n\rangle$ are {\itshape eigenstates} of the number operator, or

\begin{equation}\label{eq: NumberOperator}
N|n\rangle=n|n\rangle.
\end{equation}

The states $|n\rangle$ are furthermore eigenstates of the Hamiltonian since it can be written in terms of the number operator as $H=\hbar\omega(N+{1\over 2})$,

\begin{equation}\label{eq: HamiltonianEgenvalueEquation}
H|n\rangle=\hbar\omega(N+{1\over 2})|n\rangle=\hbar\omega(n+{1\over 2})|n\rangle.
\end{equation}

From this equation we can read of the energy spectrum of the linear harmonic oscillator,

\begin{equation}\label{eq: HarmOscSpectrum}
E_n=\hbar\omega(n+{1\over 2}).
\end{equation}

We have the following set of formulas, summarizing this algebraic treatment of the harmonic oscillator

$$
\cases{\lbrack a,a^{\dagger}\rbrack=1\cr
a|0\rangle=0\cr
a^{\dagger}|n\rangle=\sqrt{n+1}\,|n+1\rangle\cr
a|n\rangle=\sqrt{n}\,|n-1\rangle\cr
a^{\dagger}a|n\rangle=n|n\rangle}.
$$

Admittedly there is quite a lot of handwaving going into this 'derivation' of the harmonic oscillator spectrum, but this has more to do with the pedestrian approach of this section, than to the method as such. This algebraic approach to solving the harmonic oscillator can be made more rigorous. But it is clearly more abstract than the wave-function approach. The reader might wonder what is the actual content of an abstract equation such as $a^{\dagger}|n\rangle=\sqrt{n+1}\,|n\rangle$.

One, computer science oriented, way of thinking of equations like these is to regard them purely syntactically. Then, wherever we see the combination of symbols $a^{\dagger}|n\rangle$ we are allowed to replace them by $\sqrt{n+1}\,|n\rangle$. Of course, quite a few more rules are needed in order to 'calculate' syntactically with this model, but in principle, the whole theory can be phrased entirely abstractly in terms of formal syntactic rules. Semantics can be added to the model by interpreting the operators and the states. On such interpret ion is in terms of configuration space, derivatives and wave functions. Another one is in terms of (infinite dimensional) matrices and vectors. This can also be regarded as making a distinction between interface and implementation. Seen in this way, the equation $a^{\dagger}|n\rangle=\sqrt{n+1}\,|n\rangle$ belongs in the interface, being effectively a specification of a functionality to be provided by the implementation. The implementation then, could be in terms of wave functions or in terms of matrices, or perhaps in terms of some other for the purpose suitable (mathematical) constructs. In the next section on angular momentum and spin, we will see a concrete example of this.

%*************
\subsubsection{Derivation of the normalization conditions}
This section is somewhat technical, and do involve certain concept not yet discussed. The purpose is to derive the normalization coefficients $\xi(n)$ and $\eta(n)$. The reader might want to skip it for now and return after reading chapter 5.

The coefficients $\xi$ and $\eta$ are subject to some consistency conditions. First, since $[a,a^{\dagger}]=1$, we have

$$
[a,a^{\dagger}]|n\rangle=|n\rangle \Rightarrow 
$$
\begin{equation}\label{eq: FirstCondition}
\xi(n)\eta(n+1)-\xi(n-1)\eta(n)=1.
\end{equation}

Furthermore, the states $|n\rangle$ are subject to a orthonormality condition, analogous to (\ref{eq: OrtoNormGen})

$$
\langle n|m\rangle=\delta_{nm},
$$
and in particular

$$
\langle n|n\rangle=1.
$$

The question arises, what is $\langle n|$? A detailed explanation of this will be given i chapter 5. Here we can think of $\langle n|$ as a form of conjugate to $|n\rangle$. As regards the equation (\ref{eq: StateNormalization1}), this conjugation, denoted by a dagger $\dagger$, works as follows

$$
(a^{\dagger}|n\rangle)^{\dagger}=(\xi(n)|n+1\rangle)^{\dagger}\Rightarrow
$$
\begin{equation}\label{eq: StateNormalization1Conjugated}
\langle n|a=\langle n+1|\xi(n)^*,
\end{equation}
and as regards equation (\ref{eq: StateNormalization2})

$$
(a|n\rangle)^{\dagger}=(\eta(n)|n-1\rangle)^{\dagger}\Rightarrow
$$
\begin{equation}\label{eq: StateNormalization2Conjugated}
\langle n|a^{\dagger}=\langle n-1|\eta(n)^*
\end{equation}

Enforcing the condition $\langle n|n\rangle=1$ on the state $|n+1\rangle$, that is $\langle n+1|n+1\rangle=1$ and using (\ref{eq: StateNormalization1})  and (\ref{eq: StateNormalization2}) as well as (\ref{eq: StateNormalization1Conjugated}) and (\ref{eq: StateNormalization2Conjugated}) yields 

$$
\langle n+1|n+1\rangle={1\over \xi(n)\xi(n)^*}\langle n|aa^{\dagger}|n\rangle=
$$
$$
{1\over \xi(n)\xi(n)^*}\langle n|[a,a^{\dagger}]+a^{\dagger}a|n\rangle={1\over \xi(n)\xi(n)^*}\langle n|1+\eta(n)\eta(n)^*|n\rangle=
$$
$$
{1\over \xi(n)\xi(n)^*}\big(1+\eta(n)\eta(n)^*\big),
$$
where the common rewriting trick $aa^{\dagger}=[a,a^{\dagger}]+a^{\dagger}a$ has been used.

Concluding, we get 

$$
{1\over \xi(n)\xi(n)^*}\big(1+\eta(n)\eta(n)^*\big)=1,
$$
or more succinctly

\begin{equation}\label{eq: SecondCondition}
\xi(n)\xi(n)^* - \eta(n)\eta(n)^*=1
\end{equation}

Before trying to solve equations (\ref{eq: FirstCondition}) and (\ref{eq: SecondCondition}) we will make the simplifying assumption that $\xi$ and $\eta$ are real. This assumption can be justified after the fact. So equation (\ref{eq: SecondCondition}) becomes

\begin{equation}\label{eq: SecondConditionReal}
\xi(n)^2 - \eta(n)^2=1.
\end{equation}

The solution to (\ref{eq: FirstCondition}) and (\ref{eq: SecondConditionReal}) can now be constructed in an inductive way. Starting from $a|0\rangle=0$ which implies 

$$
\eta(0)=0
$$
we find that (\ref{eq: FirstCondition}) implies

$$
\xi(0)=1,
$$
which in its turn using (\ref{eq: SecondConditionReal}) implies

$$
\eta(1)=1.
$$

Going on in this way, using equations (\ref{eq: FirstCondition}) and (\ref{eq: SecondConditionReal}), we get the sequence of equations
\begin{eqnarray*}
\xi(1)=\sqrt 2
\\
\eta(2)=\sqrt 2
\\
\xi(2)=\sqrt 3
\\
\eta(3)=\sqrt 3
\\
\vdots
\\
\eta(n)=\sqrt n,
\end{eqnarray*}
or in general 

$$
\xi(n)=\sqrt{n+1}
$$
$$
\eta(n)=\sqrt n.
$$

It is easy to prove these equations using mathematical induction on equations (\ref{eq: FirstCondition}) and (\ref{eq: SecondConditionReal}).

A computer scientist cannot fail to register how closely this construction of the state space of the harmonic oscillator runs to the construction of the natural numbers as an inductive set. There are however important differences. This issue will be explored elsewhere.

%ANGULAR MOMENTUM AND SPIN
%*************************
\section{Angular momentum and spin}
In classical physics, angular momentum is a quantity related to rotation. So for example, a particle of mass $m$, rotating with the velocity $v$ in a circle at a distance $r$ from a center, has an angular momentum $L$ given by

$$
L={mvr}.
$$
This, however, is an oversimplification. Since the circle of rotation lies in a certain plane, and the velocity $v$ is a vector (the direction of which is changing as the particle moves in the circle), it turns out that angular momentum must be described by a vector quantity $\mathbf L$. In the simple case of a particle rotating with constant velocity $v$ in a circle, the direction of the vector $\mathbf L$ coincides with a vector normal to the plane of the circle. 

The angular momentum, being a vector, can be written in terms of its components $(L_x,L_y,L_z)$ in a rectangular coordinate system. Classically, all of the components of $\mathbf L$ can be determined to arbitrary precision.

When angular momentum for atomic quantum systems such as the hydrogen atom was studied, it turned out that the situation was radically different. Since atomic systems contain rotating components such as electrons, angular momentum played a central role in the development of quantum mechanics. Without going into either the history of the subject,\footnote{A good reference treating the history of atomic, nuclear and elementary particle physics is\cite{Pais1986}.} or the detailed theory, let us just record the basic facts.

In quantum mechanics the components of the angular momentum become hermitean operators $(\hat L_x,\hat L_y,\hat L_z)$. Having said that, we will at once drop the 'hats' over the operators. The order of application of these operators on a quantum state matters. This is recorded in the commutation relations

\begin{equation}\label{eq: AngularMomentumCommutationRelations1}
[L_x,L_y]=i\hbar L_z,\;\; [L_y,L_z]=i\hbar L_x,\;\; [L_z,L_x]=i\hbar L_y.
\end{equation}

The physical consequence of this is that not all three components of $\mathbf L$  are measurable simultaneously. Instead one normally considers the length square of $\mathbf L$

\begin{equation}\label{eq: AngularMomentumSquared}
\mathbf L^2=L_x^2+L_y^2+L_z^2.
\end{equation}

This quantity does commute with all components of $\mathbf L$ as can be shown by simply carrying out the commutator algebra, using (\ref{eq: AngularMomentumCommutationRelations1}). Therefore, in order to specify simultaneously measurable quantities for a rotating system, one normally chooses $\mathbf L^{2}$ and one component of $\mathbf L$, the standard choice being the $z$-component $L_z$.

This is all very abstract, and in a concrete case, as for example when treating the electron in a hydrogen atom, these operators are represented as differential operators acting on the configuration space wave-function of the electron. Such a concrete analysis shows that the angular momentum states can be characterized by two quantum numbers $l$ and $m$ related to the eigenvalues of the operators $\mathbf L^{2}$ and $L_z$. This involves quite a lot of long-winded calculations, and we will not perform them here. They can be found in any textbook on quantum mechanics. The type of calculations are similar to those that we reviewed in the case of the particle in a potential box, essentially solving the Scr\" odinger equation in three dimensions of space. 

Instead we will work in a more abstract way. The constant $\hbar$ setting the scale of quantum phenomena, plays no significant role in the following algebraic treatment of angular momentum, so we will rescale it to 1, or to put it differently, we will choose units of measurement where $\hbar=1.$\footnote{This is standard practice in quantum mechanics. It is actually possible to reinstate $\hbar$ in the formulas at a later stage via so called dimensional analysis. Another way to look at this is to rescale the operators, or absorb factors of $\surd\hbar$. This is actually what we did in the preceding section when defining the creation and annihilation operators.} 

The angular momentum eigenstates will be denoted by $|l,m\rangle$ where $m$ is the eigenvalue corresponding to the $z$-component of $\mathbf L$, i.e.

\begin{equation}\label{eq: LzEigenvalueEquation}
L_z|l,m\rangle=m|l,m\rangle.
\end{equation}

Since angular momentum is described by two commuting operators, it makes sense to label the eigenstates with two labels. As said, $l$ is a quantum number in some way related to the length of the angular momentum vector $\mathbf L$, but the exact  correspondence is left open at the moment. We will work with a fixed value for $l$.

The states are orthonormal

\begin{equation}\label{eq: AngularMomentumStatesOrthonrmality}
\langle l,m|l,n\rangle=\delta_{mn}.
\end{equation}

Some further properties of this representation will now be derived. It will become apparent that this can be done in rather close analogy with the way in which the harmonic oscillator was treated. In that case the operators $x$ and $p$ was replaced by the creation and annihilation operators. The creation operator increases the quantum number $n$ (acting like a kind of successor) whereas the annihilation operator, decreases the quantum number (like a predecessor). The first step will be to introduce new operators $L_+$ and $L_-$ with properties similar to creation and annihilation operators with the difference that now we will get a finite spectrum.

\begin{eqnarray}
L_+=L_x+iL_y\label{eq: RaisingOperator}
\\
L_-=L_x-IL_y\label{eq: LoweringOperator}.
\end{eqnarray}

Next, the commutation relations (\ref{eq: AngularMomentumCommutationRelations1}), are rewritten in terms of these operators and $L_z$

\begin{equation}\label{eq: AngularMomentumCommutationRelations2}
[L_z,L_+]=L_+,\;\; [L_z,L_-]=-L_-,\;\; [L_+,L_-]=2L_z.
\end{equation}

To understand the action of the operator $L_+$, perform the following calculation

$$
L_z\big(L_+|l,m\rangle\big)=\big([L_z,L_+]+L_+L_z\big)|l,m\rangle=
$$
$$
\big(L_++mL_+\big)|l,m\rangle=(m+1)L_+|l,m\rangle.
$$
 
This is a calculation \footnote{Note the use of the standard trick, $AB=[A,B]+BA$, to take advantage of the commutation relations for a pair of operators $A$ and $B$.} of the $L_z$-eigenvalue for the state $L_+|l,m\rangle$ and it shows that as compared to the state $|l,m\rangle$, the state $L_+|l,m\rangle$ has $L_+|l,m\rangle$ eigenvalue $m+1$, i.e. the eigenvalue is one unit larger. In the same way, it can be shown that the state $L_-|l,m\rangle$ has $L_z$-eigenvalue $m-1$ as compared to the state $|l,m\rangle$. For this reason, the operators $L_+$ and $L_-$ are called {\it raising} and {\it lowering} operators respectively. It now remains to calculate the spectrum of eigenstates for $L_z$ and $\mathbf L^2$.

The intuition is the following. We have a physical system with a certain angular momentum, somehow parameterized by the quantum number $l$. The quantum number $m$ corresponds to the $z$-component. Obviously, if the total angular momentum has a finite value, the $z$-component must also be bounded. Even classically, no component of $\mathbf L$ can be larger than $\mathbf L$ itself. Then, thinking of $L_+$ as an operator that increases $z$-component of the angular momentum, it makes sense to postulate the existence of a state with the highest possible value for $m$, denoted by $|l,l\rangle$, that is annihilated by $L_+$, or

\begin{equation}\label{eq: AngularMomentumHighestWeightState}
L_+|l,l\rangle=0.
\end{equation}

This equation plays a similar role to that of equation (\ref{eq: GroundStateDefHarmonicOscillator}) for the harmonic oscillator as providing a base case for building the spectrum of states. The state $|l,l\rangle$ is called the {\it highest weight state}.
 
Acting on the state $|l,l\rangle$ with the lowering operator $L_-$ should yield the state $|l,l-1\rangle$, having one unit lower $m$-value. It is perhaps tempting then to assume that $L_-|l,l\rangle=|l,l-1\rangle$, but that is wrong. There are normalization factors to take into account. Instead, a detailed analysis will show

\begin{equation}
L_-|l,m\rangle=\zeta(m)|l,m-1\rangle
\end{equation}
as well as
\begin{equation}
L_+|l,m\rangle=\zeta(m+1)|l,m+1\rangle
\end{equation}

with normalization factors $\zeta(m)$. These factors have the following form

\begin{equation}\label{eq: AngularMomentumNormalization}
\zeta(m)=\sqrt{l(l+1)-m(m-1)}.
\end{equation}

Applying $L_-$ repeatedly, we must at some stage arrive at a state with lowest possible value for the $L_z$-eigenvalue. A careful analysis shows that this lowest weight is $-m$. So for a given $l$, the spectrum of the operator $L_z$ is \linebreak $\{-l,-l+1,\ldots,l-1,l\}$ yielding in total $2l+1$ states. Now, the eigenvalues of the operator $\mathbf L^2$ can also be calculated. First rewrite $\mathbf L^2$ as

\begin{equation}\label{eq: L2}
\mathbf L^2={1\over 2}\big(L_+L_-+L_-L_+\big)+L_z^2,
\end{equation}
using the definitions (\ref{eq: RaisingOperator}) and (\ref{eq: LoweringOperator}). Calculating the eigenvalues of $\mathbf L^2$ is now a simple matter of applying $\mathbf L^2$ to a state $|l,m\rangle$ and using the appropriate equations. The result is

\begin{equation}\label{eq: L2Eigenvalue}
\mathbf L^2|l,m\rangle=l(l+1)|l,m\rangle,
\end{equation}
Thus the eigenvalue of $\mathbf L^2$ in a state $|l,m\rangle$ is $l(l+1)$. 

%*************
\subsubsection{Derivation of the normalization conditions}
In this section, we will prove that the normalization coefficients has the form given by equation (\ref{eq: AngularMomentumNormalization}). We will do it in the form of a proof by induction. To simplify, we will assume them to be real, i.e. $\zeta^*=\zeta$. This is in fact a choice  that is always possible to make.

\subsubsection{Proposition}
The following recursive equations hold for $k\geq 0$

\begin{eqnarray}
L_-|l,l-k\rangle=\zeta(l-k)|l,l-k-1\rangle\label{eq: L-gen}
\\
L_+|l,l-k-1\rangle=\zeta(l-k)|l,l-k\rangle\label{eq: L+gen}
\\
\zeta(l-k)^2-\zeta(l-k+1)^2=2(l-k)\label{eq: N2gen}
\end{eqnarray}
where the first equation is just a definition and need not be proved.

\subsubsection{Proof}
For the base case put k=0. Then the equations read

\begin{eqnarray}
L_-|l,l\rangle=\zeta(l)|l,l-1\rangle\label{eq: L-0}
\\
L_+|l,l-1\rangle=\zeta(l)|l,l\rangle\label{eq: L+0}
\\
\zeta(l)^2=2l\label{eq: N20}
\end{eqnarray}

The last equation follows from the definition (\ref{eq: AngularMomentumHighestWeightState}) of the highest weight state
$$
L_+|l,l\rangle=0,
$$
i.e. there is no coefficient corresponding to $l+1$, or $\zeta(l+1)=0$.

The state $L_-|l,l\rangle$ must be normalized, so that
$$
\langle l,l|L_+L_-|l,l\rangle=\zeta(l)^*\zeta(l)\langle l,l-1|l,l-1\rangle=\zeta(l)^*\zeta(l)=\zeta(l)^2
$$
On the other hand, using the angular momentum algebra
$$
\langle l,l|L_+L_-|l,l\rangle=\langle l,l|L_z|l,l\rangle=2l\langle l,l|L_z|l,l\rangle=2l,
$$
so that 
$$
\zeta(l)^2=2l.
$$

This proves (\ref{eq: N20}).

Next performing some further algebra
$$
L_+|l,l-1\rangle={1\over \zeta(l)}L_+L_-|l,l\rangle={1\over \zeta(l)}[L_+,L_-]|l,l\rangle=
$$
$$
{2\over \zeta(l)}L_z|l,l\rangle={2l\over \zeta(l)}|l,l\rangle={\zeta(l)}|l,l\rangle
$$
This proves (\ref{eq: L+0}). 

For the induction step, assume that the equations (\ref{eq: L+gen}) (\ref{eq: N2gen}) are true for a certain $k$. First we consider \ref{eq: N2gen}) for $k+1$. Calculate the norm of the state $L_-|l,l-k-1\rangle$ using equation (\ref{eq: L-gen}), which as pointed out, is just a definition, we get

$$
\langle l,l-k-1|L_+L_-|l,l-k-1\rangle=\zeta(l-k-1)^*\zeta(l-k-1)\langle l,l-k-2|l,l-k-2\rangle=
$$
$$
\zeta(l-k-1)^2.
$$
On the other hand, using the angular momentum algebra and the induction hypothesis

$$
\langle l,l-k-1|L_+L_-|l,l-k-1\rangle=\langle l,l-k-1|[L_+L_-]+L_-L_+|l,l-k-1\rangle=
$$
$$
\langle l,l-k-1|2L_z+L_-L_+|l,l-k-1\rangle=2(l-k-1)+\zeta(l-k)^*\zeta(l-k)=
$$
$$
2(l-k-1)+\zeta(l-k)^2.
$$
Combining the last two equations yields
$$
\zeta(l-k-1)^2=2(l-k-1)+\zeta(l-k)^2,
$$
which is precisely (\ref{eq: N2gen}) with $k+1$ instead of $k$. 

Next we consider (\ref{eq: L+gen}) for $k+1$. Using the operator algebra and the induction hypothesis yields

$$
L_+|l,l-k-2\rangle={1\over \zeta(l-k-1)}L_+L_-|l,l-k-1\rangle=
$$
$$
{1\over \zeta(l-k-1)}\big([L_+,L_-]+L_-L_+\big)|l,l-k-1\rangle=
$$
$$
{1\over \zeta(l-k-1)}\big(2L_z+L_-L_+\big)|l,l-k-1\rangle=
$$
$$
{1\over \zeta(l-k-1)}\big(2(l-k-1)+\zeta(l-k)^2\big)|l,l-k-1\rangle=
$$
$$
{\zeta(l-k-1)^2\over \zeta(l-k-1)}|l,l-k-1\rangle=\zeta(l-k-1)|l,l-k-1\rangle.
$$

The recursive equations (\ref{eq: N2gen}) can be solved. The result is

\begin{equation}
\zeta(l-k)=\sqrt{(k+1)(2l-k)}.
\end{equation}

There is an upper limit for $k$. Just as there is a state with highest value for $m$, namely $|l,l\rangle$, there must be a state with lowest value for $m$. This occurs when $k=2l$ in the above equation, corresponding to 

$$
L_-|l,-l\rangle=0,
$$

which is (\ref{eq: L-gen}) for $k=2l$. So we learn that the spectrum for the operator $L_z$ ranges in unit steps from a maximum weight of $l$ to a minimum weight of $-l$. Furthermore, since $k$ is an integer, and $2l-k$ must be zero for some k, the possible values of $l$ are $\{0,1/2,1,3/2,2,\ldots\}$.

Finally, putting $l-k=m$ we arrive at (\ref{eq: AngularMomentumNormalization}).

%********
%SPIN 1/2
\subsection{Spin 1/2}
In quantum physics, there is a distinction made between angular momentum and {\it spin}. Angular momentum refers to, just as in classical physics, to rotational motion in space. Spin, on the other hand, refers to an intrinsic property of a particle. Though it can be thought of as some kind of rotation around an internal axis of the particle, just as one would do for a spinning classical body, there is actually no need for, and no experimental basis for such a picture. Spin is best regarded as an intrinsic quantity, along with other intrinsic quantities such as mass, electric charge et cetera.

The simplest non-trivial, and for quantum computation, most important example, is the case of spin 1/2, or $l={1\over 2}$. An example of a particle having spin 1/2 is the electron.

The abstract theory of angular momentum given in the preceding section, can be made concrete by representing the states $|l,m\rangle$ and operators $L$ by vectors and matrices. Thinking of the preceding abstract theory as an interface, we will now provide an implementation. We will do it in the case $l={1\over 2}$ as that will be needed subsequently when discussing quantum computation.

%**********
\subsubsection{Implementation of spin one half}
With $l={1\over 2}$, the spectrum of m is $\{{1\over 2},-{1\over 2}\}$ with the highest weight state $|{1\over 2},{1\over 2}\rangle$ and the lowest weight state $|{1\over 2},-{1\over 2}\rangle$. This notation is to heavy, and customarily one uses a leaner syntax $|{1\over 2}\rangle$ and $|-{1\over 2}\rangle$. These two states are implemented by 2-dimensional vectors unit vectors as

\begin{eqnarray}\label{eq: SpinOneHalfBasis}
|{1\over 2}\rangle=\pmatrix{1\cr 0}
\\
|-{1\over 2}\rangle=\pmatrix{0\cr 1}.
\end{eqnarray}

The components of the angular momentum operators (spin operators) are given by
\begin{eqnarray}\label{eq: SpinOperators1}
L_x={1\over 2}\pmatrix{0& 1\cr 1& 0}
\\
L_y={1\over 2}\pmatrix{0& -i\cr i& 0}
\\
L_z={1\over 2}\pmatrix{1& 0\cr 0& -1},
\end{eqnarray}

But we are more interested in $L_+$ and $L_-$ which are easily calculated from $L_x$ and $L_-$

\begin{eqnarray}\label{eq: SpinOperators2}
L_+=\pmatrix{0& 1\cr 0& 0}
\\
L_-=\pmatrix{0& 0\cr 1& 0}.
\end{eqnarray}

Employing this concrete realizations, almost trivial matrix calculations yield

\begin{eqnarray}\label{eq: SpinCalculations}
L_+|{1\over 2}\rangle=\pmatrix{0& 1\cr 0& 0}\pmatrix{1\cr 0}=0
\\
L_+|-{1\over 2}\rangle=\pmatrix{0& 1\cr 0& 0}\pmatrix{0\cr 1}=\pmatrix{1\cr 0}=|{1\over 2}\rangle
\\
L_-|{1\over 2}\rangle=\pmatrix{0& 0\cr 1& 0}\pmatrix{1\cr 0}=\pmatrix{0\cr 1}=|-{1\over 2}\rangle
\\
L_-|-{1\over 2}\rangle=\pmatrix{0& 0\cr 1& 0}\pmatrix{0\cr 1}=0,
\end{eqnarray}
thus implementing the raising and lowering operations of $L_+$ and $L_-$. Comparing to the previous section, it can also be checked that the normalizations $\zeta({1\over2})$ and $\zeta(-{1\over2})$ come out right.

Next we have to check the action of $L_z$

\begin{eqnarray}\label{eq: SpinCalculations2}
L_z|{1\over 2}\rangle={1\over 2}\pmatrix{1& 0\cr 0& -1}\pmatrix{1\cr 0}={1\over 2}\pmatrix{1\cr 0}={1\over 2}|{1\over 2}\rangle
\\
L_z|-{1\over 2}\rangle={1\over 2}\pmatrix{1& 0\cr 0& -1}\pmatrix{0\cr 1}=-{1\over 2}\pmatrix{0\cr 1}=-{1\over 2}|-{1\over 2}\rangle.
\end{eqnarray}

Thus we have a concrete realization of the angular momentum theory of the preceding section in the particular case of $l={1\over 2}$.

%**********
\subsubsection{Connection to quantum computation}
Spin 1/2 is an example of a two-state quantum system. In quantum computation such systems are referred to as {\it qubits}. In that context, a slightly different notation is used. Instead of $|{1\over 2}\rangle$ and $|-{1\over 2}\rangle$, the states are denoted by $|0\rangle$ and  $|1\rangle$ respectively, and of course, they are to be thought of as quantum versions of the classical bit values 0 and 1.

%**********
\subsubsection{Pauli matrices}
In the context of describing spin 1/2 is customary to express the angular momentum operators in terms of the {\it Pauli matrices}. As is seen below, it is really a trivial rescaling by a factor 1/2 involved, but theoretical physicists seem to be fond of shuffling factors of 1/2 around.\footnote{As well as factors of 2, $\surd 2$, $\pi$ et cetera!}

\begin{eqnarray}\label{eq: PauliMatrices}
\sigma_x=\pmatrix{0& 1\cr 1& 0}
\\
\sigma_y=\pmatrix{0& -i\cr i& 0}
\\
\sigma_z=\pmatrix{1& 0\cr 0& -1}.
\end{eqnarray}

We now leave the theory of angular momentum at this stage.

\chapter{General quantum theory }
In this chapter, quantum theory will be outlined in an abstract and formal way suitable for discussing quantum computational models and quantum complexity theory. The concepts of the preceding chapter will reappear but in a much more formal setting. A modern general reference to quantum mechanics suitable for studies into quantum computation is the book by A. Peres \cite{Peres1993}.

%STATE SPACES
%************
\section{State spaces}
The state spaces of quantum mechanics are modeled on Hilbert spaces. These spaces can be either finite dimensional or infinite dimensional. The infinite dimensional case requires a rather elaborate mathematical treatment if one wants to be stringent. For quantum computation it suffices to consider finite dimensional Hilbert spaces, as any real quantum computer would have a finite size in terms of number of states. 

In the first section, the theory of vector spaces will reviewed. The intuition behind the theory of vector spaces is the familiar real vectors of ordinary three-dimensional space. Thus if ${\bf v}_1$ and ${\bf v}_2$ are two vectors in the space, so is $\alpha_1{\bf v}_1 +\alpha_2{\bf v}_2$, where $\alpha_1$ and $\alpha_2$ are real numbers. These notions will be made precise below.

A concise and readable physics-style reference to vector spaces is C. Isham's book \cite{Isham1989}.

%*****************************
\subsection{Vector spaces}
Let ${\bf V}$ denote an n-dimensional vector space over the complex numbers $\bf{C}$. This means that for any vectors ${\bf x_1}$, ${\bf x_2}$ and ${\bf x}$ in ${\bf V}$ and any complex numbers $\alpha_1$, $\alpha_2$ and $\alpha$ the following equations expressing linearity hold

\begin{eqnarray}
\alpha({\bf x}_1+{\bf x}_2)=\alpha {\bf x}_1+\alpha {\bf x}_2
\\
(\alpha_1+\alpha_2){\bf x}=\alpha_1 {\bf x}+\alpha_2 {\bf x}
\\
\alpha_1(\alpha_2){\bf x}=(\alpha_1\alpha_2){\bf x}
\\
1{\bf x}={\bf x}
\\
0{\bf x}={\bf 0}
\end{eqnarray}

The space $\bf{V}$ can be thought of as $n$-tuples of complex numbers arranged as column vectors so that we can write ${\bf V}={\bf C}^n$.\footnote {A more correct way to express this is: there is an isomorphism $i: \bf{V}\rightarrow {\bf C}^n$.} An {\itshape inner product} (or a {\itshape scalar product}) over $\bf{V}$ is defined as a complex valued function $\langle {\bf x},{\bf y}\rangle$ defined on $\bf{V}\times \bf{V}$, subject to the conditions

\begin{eqnarray}
\langle {\bf x},{\bf x}\rangle\geq 0\mbox{  and  }\langle {\bf x},{\bf x}\rangle = 0\mbox{  if and only if  }{\bf x}=0
\label{eq: InnerProduct1}
\\
\langle{\bf x},\alpha_1{\bf y}_1+\alpha_2{\bf y}_2\rangle = \alpha_1\langle {\bf x},{\bf y}_1\rangle+\alpha_2\langle {\bf x},{\bf y}_2\rangle
\label{eq: InnerProduct2}
\\
\langle {\bf x},{\bf y}\rangle = {\langle {\bf y},{\bf x}\rangle}^*
\label{eq: InnerProduct3}
\end{eqnarray}
for all ${\bf x}, {\bf y}, {\bf y}_1,{\bf y}_2 \in \bf{V}$ and all $\alpha_1,\alpha_2 \in \bf{C}$. The complex conjugate of a complex number $\alpha$ is denoted by $\alpha^*$. Note that $\langle {\bf x},{\bf x}\rangle$ is a real number by condition (c).

Two vectors ${\bf x}$ and ${\bf y}$ are said to {\itshape orthogonal} if $\langle {\bf x},{\bf y}\rangle = 0$.

A vector space supplied with an inner product is called an {\itshape inner-product vector space}.

It can be shown that the inner product satisfies the {\it Schwarz inequality} 

\begin{equation}\label{eq: ScharwzEquality}
|\langle {\bf x},{\bf y}\rangle |\leq\sqrt{\langle {\bf x},{\bf x}\rangle}\sqrt{\langle {\bf y},{\bf y}\rangle}.
\end{equation}

%******************
\subsubsection{Norm}
From the inner product, a {\itshape norm} can be defined by $\|{\bf x}\|=\sqrt{\langle {\bf x},{\bf x}\rangle}$. From this definition and the properties of the inner product, it follows that

\begin{eqnarray}
\|\alpha {\bf x}\|=|\alpha|\|{\bf x}\|\mbox{  for all complex numbers  }\alpha
\\
\|{\bf x}\|\geq 0\mbox{  with equality only if  }\|{\bf x}\|=0.
\\
\|{\bf x}+{\bf y}\|\leq\|{\bf x}\|+\|{\bf y}\|,
\end{eqnarray}
ensuring that $\|{\bf x}\|$ satisfies all requirements for a proper norm on a vector space. The last property, (c), is the {\itshape triangle inequality}, which can be shown from the Schwarz inequality. 

A vector ${\bf x}$ is said to be {\itshape normalized} if $\|{\bf x}\|=1$.

\subsubsection{Quantum states and rays}
Quantum states are represented by normalized vectors ${\Psi}$. Actually, if the normalized states ${\Psi}$ and ${\Psi}'$ are related by the equation ${\Psi}'=\alpha{\Psi}$ where $|\alpha|=1$, then $\|{\Psi}\|=\|{\Psi}'\|=1$, and they are thought of as representing the same state. This is refereed to as saying that the two vectors belong to the same {\itshape ray} in the space. Thus it is more correct to say that quantum states are represented by equivalence classes of states.

%***
\subsubsection{Linear independence and basis sets of vectors}
A finite set of vectors $\{{\bf x}_1,{\bf x}_2,\ldots {\bf x}_k\}$ is {\itshape linearly dependent} it there exists some set of complex numbers $\{\alpha_1,\alpha_2,\ldots \alpha_k\}$ (not all of them zero) such that 

$$
\sum_{i=1}^{k}\alpha_i {\bf x}_i=0
$$.

If there is no such set of complex numbers, then the set of vectors is said to be {\itshape linearly independent}. This gives a precise way of defining the dimension of a finite dimensional vector space. A vector space is $n$-dimensional if it contains a subset of $n$ linearly independent vectors but no subset of $n+1$ linearly independent vectors.

A subset $\{{\bf e}_1,{\bf e}_2,\ldots {\bf e}_n\}$ of an $n$-dimensional vector space $\bf V$ is a {\itshape basis} for $\bf V$ if any vector ${\bf x}$ in $\bf V$ can be expanded as 

\begin{equation}\label{eq: BasisExpansion}
{\bf x}=\sum_{i=1}^{n}\alpha_i {\bf e}_i.
\end{equation}

The numbers $\alpha_i$ are called {\itshape expansion coeffients}. They are unique. This follows from the linear independence of the basis vectors.

%***
\subsubsection{Orthonormal basis vectors}
A pair of vectors ${\bf x}$ and ${\bf y}$ are said to {\itshape orthonormal pair} if they are both normalized and orthogonal to each other. An orthonormal basis for an $n$-dimensional vector space is a set of basis vectors $\{{\bf e}_1,{\bf e}_2,\ldots {\bf e}_n\}$, all of which are normalized and in which every pair is orthogonal. Or more concisely

\begin{equation}\label{eq: BasisOrtonormality}
\langle {\bf e}_i,{\bf e}_j\rangle = \delta_{ij} \mbox{ for  } i,j=1,2,\ldots,n,
\end{equation}
where $\delta_{ij}$ is the {\it Kronecker delta} which is equal to 1 for $i=j$ and 0 otherwise.

It can be shown that every finite-dimensional vector space has an orthonormal basis. A constructive proof of this is given by the inductive {\itshape Gram-Schmidt} procedure. Suppose $\{{\bf v}_1,{\bf v}_2,\ldots {\bf v}_n\}$ is a basis for the vector space under consideration. For the base case, define

\begin{equation}\label{eq: GramSchmidtBase}
{\bf e}_1=\frac{{\bf v}_1}{\|{\bf v}_1\|}.
\end{equation}

Then define inductively for $1\leq k\leq n-1$

\begin{equation}\label{eq: GramSchmidtInduct}
{\bf e}_{k+1}=\frac{{\bf v}_k-\sum_{i=1}^k\langle {\bf e}_i,{\bf v}_{k+1}\rangle{\bf e}_i}{\|{\bf v}_k-\sum_{i=1}^k\langle {\bf e}_i,{\bf v}_{k+1}\rangle{\bf e}_i\|}.
\end{equation}

It is straightforward to verify by direct calculation that this yields an orthonormal basis $\{{\bf e}_1,{\bf e}_2,\ldots {\bf e}_n\}$.

Orthonormal basis vectors are useful as it is possible to compute the expansion coefficients explicitly. Suppose we have a vector ${\bf x}$ with  (unknown) expansion

$$
{\bf x}=\sum_{i=1}^{n}\alpha_i {\bf e}_i.
$$

Taking inner-product on both sides of the equation with an arbitrary basis vector ${\bf e}_j$ yields the following short calculation

$$
\langle {\bf e}_j,{\bf x}\rangle =\langle {\bf e}_j,\sum_{i=1}^{n}\alpha_i {\bf e}_i\rangle =\sum_{i=1}^{n}\langle{\bf e}_j,\alpha_i {\bf e}_i\rangle
$$
$$
=\sum_{i=1}^{n}\alpha_i\langle{\bf e}_j,{\bf e}_i\rangle=\sum_{i=1}^{n}\alpha_i\delta_{ij}=\alpha_i.
$$

Thus we get the simple formula for the expansion coefficients

\begin{equation}\label{eq: ExpCoeffFormula}
\alpha_i=\langle {\bf e}_j,{\bf x}\rangle.
\end{equation}

%***
\subsubsection{Dual vector space}
To each vector ${\bf x}$ in ${\bf V}$ there is a {\itshape dual} vector ${\bf x}^\dagger$. If ${\bf x}$ is represented as a column vector, then ${\bf x}^\dagger$ is represented by a row vector whose elements are the complex conjugates of the elements of ${\bf x}$. In this representation, the inner product between two vectors ${\bf x}$ and ${\bf y}$ can be defined as 

\begin{equation}\label{eq: InnerProductDefExplicit}
\langle {\bf x},{\bf y}\rangle = x_1^*y_1+x_2^*y_2+ \ldots + x_n^*y_n.
\end{equation}

This form of the inner product satisfies all the conditions (\ref{eq: InnerProduct1}), (\ref{eq: InnerProduct2}) and (\ref{eq: InnerProduct3}).

Another, more abstract point of view, is to consider ${\bf x}^\dagger$ as linear functional from ${\bf V}$ to ${\bf C}$ defined by 
${\bf x}^\dagger({\bf y})=\langle {\bf x},{\bf y}\rangle$. However, in practical calculations the explicit vector representation is useful.

%********************************
\subsection{Hilbert spaces}
In order for a general vector space ${\bf V}$ to be a Hilbert space, two requirements must be meat: (i) there must exist a norm defined in terms of an inner product, and (ii) the set of vectors must be complete with respect to the norm. Completeness means that all Cauchy sequences of vectors converges to vectors in the space \cite{Simmons1963}. This last point is somewhat elaborated in the next section. 

\subsubsection{Completeness}
A Hilbert space $\cal H$ is complete in the sense that if $\{{\bf x}_n\}$ is a sequence in $\cal H$ with $\lim_{n,m\rightarrow \infty}\|{\bf x}_n-{\bf x}_m\|=0$ then there is an ${\bf x}$ in $\cal H$ to which the sequence $\{{\bf x}_n\}$ converges, i.e $\lim_{n\rightarrow \infty}\|{\bf x}_n\|={\bf x}$. Thus, in a Hilbert space, every Cauchy sequence is convergent. (The converse, that every convergent sequence is a Cauchy sequence, is true in every normed vector space).

For infinite-dimensional vector spaces, this condition is not automatically true. In certain cases an intricate process of completing the space with all limits of Cauchy sequences can be preformed (in much the same way as the set of rational numbers are completed to form the set of real numbers). The requirement of completeness is needed in order that the usual tools of analysis; performing limits, differentiating et cetera, can be used. 

Finite dimensional inner-product vector spaces over the complex numbers are Hilbert spaces. This is follows from the fact that the complex numbers are complete with respect to their usual absolute value norm, and this is enough to ensure completeness of finite-dimensional vector spaces over the complex numbers. No process of completion is needed in this case.

In the context of quantum computation, this is actually more than we want, since completeness in this sense means that there are vectors in the Hilbert space whose components are non-computable numbers. Non-computable functions could therefore be hidden within the specification of the quantum computer. We will return to this point later on. 

%********************************
\subsection{Dirac notation}
After this general introduction to vector spaces, we will now change the notation somewhat in order to be in conformity with mainstream quantum theory.

Dirac invented a notational system that is very useful both for formal treatment of quantum mechanics and explicit calculations. A vector in the vector space is denoted by a {\itshape ket} $|\;\rangle$. Inside the ket, one places any symbol or symbols that characterizes the state, so for example the vector ${\bf x}$ can be denoted by $|x\rangle$. A generic quantum state is often denoted by $|\psi\rangle$. 

This system is very versatile. If we are working with a specific orthonormal basis for a certain vector space, the basis vectors can be denoted concisely by $|i\rangle$, i.e. we just record the index within the ket symbol. The expansion of a vector $|x\rangle$ in terms of the basis $|i\rangle$ is written as

\begin{equation}\label{eq: DiracExpansion}
|x\rangle=\sum_{i=1}^{n}\alpha_i|i\rangle,
\end{equation}
in analogy to equation (\ref{eq: BasisExpansion}).

In this notational system, the dual to $|x\rangle$ is denoted by $\langle x|$ and we have $|x\rangle ^\dagger = \langle x |$. A vector $\langle\psi|$ in the dual space is called a {\itshape bra}. The inner product is written $\langle x,y\rangle =\langle x|y\rangle$. Thus, the orthonormality requirement for a basis becomes

\begin{equation}\label{eq: DiracOrthoNormality}
\langle i|j\rangle=\delta_{ij},
\end{equation}
and the formula (\ref{eq: ExpCoeffFormula}) for the expansion coefficients becomes

\begin{equation}\label{eq: DiracExpCoeffFormula}
\alpha_i=\langle i|x\rangle.
\end{equation}

If the ket vector $|x\rangle$ is represented concretely by a column vector, the corresponding dual bra vector is represented by a row vector, the elements of which are complex conjugates of the elements of the ket vector, or

\begin{equation}\label{eq: ConjugationExplicit}
|x\rangle=\pmatrix{\alpha_1\cr \alpha_2\cr\vdots\cr \alpha_n} \Rightarrow \langle x|=(\alpha_1^*,\alpha_2^*,\ldots ,\alpha_n^*).
\end{equation}

In this notation we have to decide how to treat the equation $\langle x,y\rangle = {\langle y,x\rangle}^*$ of the definition of the inner product. Referring to equation (\ref{eq: InnerProductDefExplicit}) we see that taking the complex conjugate we get

$$
\langle x,y\rangle^* = (x_1^*y_1+x_2^*y_2+ \ldots + x_n^*y_n)^*
$$
$$
=y_1^*x_1+y_2^*x_2+ \ldots + y_n^*x_n=\langle y,x\rangle
$$
making the identification natural

\begin{equation}\label{eq: BraKetConjugation}
\langle x|y\rangle^*=\langle y|x\rangle
\end{equation}
where

\begin{equation}\label{eq: BraAndKetConjugation}
\langle y|=|y\rangle^{\dagger} \mbox{ and } |x\rangle=\langle x|^{\dagger}.
\end{equation}

From now on, the Dirac notation will be used almost exclusively. The terms state, vector, bra and ket will be used interchangeably in the following. No confusion can arise. 

The Dirac system of notation captures a deep property of quantum mechanics, namely a that a physical system can be represented in many different ways. The explicit representation that we use carries very little significance and can be chosen for convenience.

%********************************
\subsection{Tensor products}
Using {\itshape tensor products} new (larger) vector spaces can be built out of existing vector spaces. This is a common practice in quantum mechanics where for example the joint states $|\psi_{12}\rangle$ of two independent systems can be built from the states of the constituents $|\psi_{1}\rangle$ and $|\psi_{2}\rangle$. 

To make this notion precise, let ${\bf V}$ and ${\bf W}$ be vector spaces of dimension $n$ and $m$, and let $|v\rangle$ and $|w\rangle$ denote generic vectors in these spaces respectively. The tensor product ${\bf V}\otimes{\bf W}$ is defined as the $nm$-dimensional vector space consisting of all linear combinations of 'tensor products' of vectors $|v\rangle\otimes|w\rangle$. This is an abstract product, which by definition satisfies 

\begin{eqnarray}
\alpha(|v\rangle\otimes|w\rangle)=(\alpha|v\rangle)\otimes|w\rangle=|v\rangle\otimes(\alpha|w\rangle)
\label{eq: TensorProduct1}
\\
(|v_1\rangle+|v_2\rangle)\otimes|w\rangle=|v_1\rangle\otimes|w\rangle+|v_2\rangle\otimes|w\rangle
\label{eq: TensorProduct2}
\\
|v\rangle\otimes(|w_1\rangle+|w_2\rangle)=|v\rangle\otimes|w_1\rangle+|v\rangle|w_2\rangle.
\label{eq: TensorProduct3}
\end{eqnarray}

These conditions on $|v\rangle\otimes|w\rangle$ suffices to prove that ${\bf V}\otimes{\bf W}$ is indeed a linear vector space, i.e, the conditions (1.1) are satisfied.

For an example of a concrete realization of the tensor product, consider two vectors $|x\rangle$ and $|y\rangle$

$$
|x\rangle=\pmatrix{\alpha_1\cr \alpha_2} \mbox{ and } |y\rangle=\pmatrix{\beta_1\cr \beta_2}
$$
in a 2-dimensional vector space. The tensor product of the vectors becomes

$$
|x\rangle\otimes|y\rangle=\pmatrix{\alpha_1\beta_1\cr\alpha_1\beta_2\cr\alpha_2\beta_1\cr\alpha_2\beta_2}.
$$

%OPERATORS AND DYNAMICAL VARIABLES
%*********************************
\section{Operators and dynamical variables}
The intuition behind (abstract) quantum mechanics is that if vectors in a Hilbert space are used to describe states of a physical system, then it is natural to use linear operators acting on the space to describe the dynamics, i.e changes of state. A linear operator acting on a state yields a new state. In the following section we will make the basis for this intuition exact. The dynamical variables of classical physics, like position, momentum and energy, will be mapped to linear operators in quantum mechanics.
 
%********************************
\subsection{Linear operators}
A {\itshape linear operator} $A$ on a vector space ${\bf V}$ is a linear map from ${\bf V}$ to itself, that is $A:{\bf V}\rightarrow{\bf V}$. The image $A(|x\rangle)$ of a vector $|x\rangle$ is written $A|x\rangle$. The linearity requirement is expressed as 

\begin{equation}\label{eq: LinearityReq}
A(\alpha_1|x_1\rangle + \alpha_2|x_2\rangle)=\alpha_1A|x_1\rangle + \alpha_2A|x_2\rangle
\end{equation}
for complex numbers $\alpha_1,\alpha_2$ and vectors $|x_1\rangle,|x_1\rangle$.

Two special linear operators are the identity operator $I_{\bf V}$ and zero operator $0$, with properties

\begin{eqnarray}
I_{\bf V}|x\rangle=|x\rangle
\\
0|x\rangle=0.
\end{eqnarray}

The subscript ${\bf V}$ on the identity operator is often dropped when no confusion can arise. The symbol $0$ is somewhat overloaded, denoting ordinary complex number 0, the zero vector, and the zero operator. This causes no confusion in practice though. Note however, that $|0\rangle$ does not denote the zero vector! 

An operator $A$ is {\itshape invertible} if there exists an inverse operator $A^{-1}$ with the property

\begin{equation}\label{eq: InverseOpDef}
A^{-1}A=AA^{-1}=I.
\end{equation}

%***
\subsubsection{Matrix representation}
Linear operators on a vector space can be represented by matrices. \linebreak Let $\{|1\rangle,|2\rangle\ldots,|n\rangle\}$ be a basis (not necessarily orthonormal) for the vector space. Any vector $|x\rangle$ can be expanded as in (\ref{eq: DiracExpansion}),

$$
|x\rangle=\sum_{i=1}^n\alpha_i|i\rangle.
$$

Applying the linear operator $A$ on both sides of the equation and using linearity yields

$$
A|x\rangle=\sum_{i=1}^n\alpha_iA|i\rangle.
$$

For each $i$, $A|i\rangle$ is a vector in the vector space, and since $\{|1\rangle,|2\rangle\ldots,|n\rangle\}$ is a basis, there must exist complex numbers $A_{ji}$ for $i,j=1,2,\ldots,n$ such that $A|i\rangle$ can be expanded

\begin{equation}\label{eq: OperatorOnBaseVector}
A|i\rangle=\sum_{j=1}^nA_{ji}|j\rangle.
\end{equation}

Inserting this into the previous equation gives

$$
A|x\rangle=\sum_{i=1}^n\sum_{j=1}^n\alpha_iA_{ji}|j\rangle=\sum_{j=1}^n(\sum_{i=1}^nA_{ji}\alpha_i)|j\rangle.
$$

Representing vectors concretely as column vectors of expansion coefficients as in equation (1.12) we see that the action of the linear operator $A$ on components of a vector $|x\rangle$ can be written as a transformation equation

\begin{equation}\label{eq: OperatorTransformation}
\pmatrix{\alpha_1\cr \alpha_2\cr\vdots\cr \alpha_n}\rightarrow\pmatrix{A_{11}&A_{12}&\cdots&A_{1n}\cr
	A_{21}&A_{22}&\cdots&A_{2n}\cr
	\vdots&\vdots& &\vdots\cr
	A_{n1}&A_{n2}&\cdots&A_{nn}\cr}\pmatrix{\alpha_1\cr \alpha_2\cr\vdots\cr \alpha_n}.
\end{equation}

Already at this stage it is clear that the theory of transformations on vector spaces offers the potential for setting up a model of computation. Letting the state of the computer be represented by the vector $|x\rangle$, linear operators induces transitions between states of the computer. The details must of course be elaborated, which is the subject of subsequent chapters.

If the vector space has an inner product and the basis is orthonormal there is a convenient way to calculate the matrix elements $A_{ji}$ of the linear operator. Taking the inner product with a dual basis vector $\langle k|$ on both sides of equation (\ref{eq: OperatorOnBaseVector}) and using orthonormality of the basis vectors gives

$$
\langle k|A|i\rangle=\langle k|\sum_{j=1}^nA_{ji}|j\rangle=\sum_{j=1}^nA_{ji}\langle k|j\rangle
$$
$$
=\sum_{j=1}^nA_{ji}\delta_{kj}=A_{ki},
$$
that is 

\begin{equation}\label{eq: OperatorMatrixElements}
A_{ki}=\langle k|A|i\rangle.
\end{equation}

This is a representation for the matrix elements that is very often used in quantum mechanics.

Note that the matrix representation of a certain linear operator on a vector space depends on the basis used, different bases gives different matrix representations, and consequently the matrix elements given by (\ref{eq: OperatorMatrixElements}) are different.

Also note that, as we are only considering operators mapping ${\bf V}$ to ${\bf V}$, the matrices representing the operators are $n\times n$ matrices.

\subsection{Outer products}
%********************************
Let $|x\rangle$ and $|y\rangle$ be two vectors in a vector space. By the {\itshape outer product} between these vectors we mean $|x\rangle\langle y|$. This can be considered to define a linear operator on the vector space as is seen from the following formal calculation

$$
(|x\rangle\langle y|)|z\rangle=|x\rangle(\langle y|z\rangle)=\langle y|z\rangle|x\rangle.
\eqno{\addtocounter{equation}{1}\theequation}
$$

Thus the vector $|z\rangle$ is mapped to the vector $\mu|x\rangle$ where $\mu$ is the complex number $\langle y|z\rangle$.

The usefulness of this concept becomes clear when it is applied to orthonormal basis vectors. Let $|i\rangle$ be an orthonormal basis for a vector space $V$. We have the expansion (\ref{eq: DiracExpansion}) of an arbitrary vector $|x\rangle$

$$
|x\rangle=\sum_{i=1}^n\alpha_i|i\rangle
$$
and the equation (\ref{eq: DiracExpCoeffFormula}) for the expansion coefficients
$$
\alpha_i=\langle i|x\rangle.
$$

Then consider the operator
\begin{equation}\label{eq: BasisOuterProduct}
\sum_{i=1}^n|i\rangle\langle i|.
\end{equation}
 
Letting it act on the vector $|x\rangle$ yields

$$
(\sum_{i=1}^n|i\rangle\langle i|)|x\rangle=\sum_{i=1}^n|i\rangle\langle i|x\rangle
$$

$$
=\sum_{i=1}^n\langle i|x\rangle|i\rangle=\sum_{i=1}^n\alpha_i|i\rangle=|x\rangle.
$$

But this equation is true for any vector $|x\rangle$ so we can identify the operator in equation (\ref{eq: BasisOuterProduct}) with the identity operator, or

\begin{equation}\label{eq: IdentityAsBasisOuterProduct}
\sum_{i=1}^n|i\rangle\langle i|= I.
\end{equation}

This is known as the {\itshape completeness relation}. 

The reader might worry about the ambiguities in the notation when writing expressions such as $|x\rangle\langle y|z\rangle$. It is not clear whether this should be read as the operator $|x\rangle\langle y|$ acting on the state $|z\rangle$ or the number $\langle y|z\rangle$ multiplying the state $|x\rangle$. However, there is no ambiguity and the expression can be read in either way. It denotes a certain state which can be calculated either as $(|x\rangle\langle y|)|z\rangle$ or $(\langle y|z\rangle)|x\rangle$. This is one aspect of the strength and versatility of the Dirac notation.

%********************************
\subsection{Projectors}
An important class of operators are the {\it projectors}. These are operators that project a state onto a subspace of the Hilbert space. Suppose we have an $n$-dimensional Hilbert space with an orthonormal basis  $\{|i\rangle\}_{i=1}^n$ and let $\{(k)\}$ denote a $k$-dimensional subset of the basis vectors. Then consider the operators

\begin{equation}\label{eq: ProjectionOperator}
P_{(k)}=\sum_{(k)}|i\rangle\langle i|.
\end{equation}

Taking $\{(k)\}=\{|i\rangle\}_{i=1}^n$ this is just the identity operator $I$. 

Next consider an arbitrary state $|x\rangle=\sum_{j=1}^n\alpha_j|j\rangle$ and let $P_{(k)}$ act on this state

$$
P_{(k)}|x\rangle=\sum_{(k)}|i\rangle\langle i|\big(\sum_{j=1}^n\alpha_j|j\rangle\big)=\sum_{(k)}\sum_{j=1}^n\alpha_j|i\rangle\langle i|j\rangle=\sum_{(k)}\alpha_i|i\rangle
$$
effectively restricting the summation to the subset ${(k)}$. Thus $P_{(k)}$ projects the state onto the substate spanned by the subset of basis vectors ${(k)}$.

An important property of projection operators is that acting twice with the same projector have no further action on the state. This is almost trivial, as can be seen by acting on more time with $P_{(k)}$ on the projected state $\sum_{(k)}\alpha_i|i\rangle$. Thus we find the general operator equation for projectors

\begin{equation}\label{eq: Idempotency}
P_{(k)}P_{(k)}=P_{(k)}.
\end{equation}

An important subclass of projectors are the projectors $P_i$ onto the basis states themselves. These are given simply by

\begin{equation}\label{eq: BasisProjector}
P_i=|i\rangle\langle i|
\end{equation}
satisfying the equation

\begin{equation}\label{eq: BasisIdempotency}
P_iP_j=P_i\delta_{ij}.
\end{equation}

%********************************
\subsection{Adjoints}
The adjoint $A^{\dagger}$ of a complex matrix $A$ is the matrix obtained by transposing and complex conjugating the matrix elements

\begin{equation}\label{eq: MatrixAdjoint}
(A_{ij})^{\dagger}=A^*_{ji}.
\end{equation}

In order to define the abstract notion of adjoint operators, the action of an operator on a bra vector has to be defined. This is a point where some confusion might arise as to how employ the notational system.

As argued in [Dirac], the inner product of a bra vector $\langle y|$ with a ket vector $A|x\rangle$ is a complex number that depends linearly on $|x\rangle$, therefore it can likewise be considered as the inner product of $|x\rangle$ with some, as yet undefined, bra vector. This bra vector depends linearly on $\langle y|$, so it can be considered as the result of applying a linear operator to $\langle y|$. Since this linear operator is uniquely determined by the original linear operator $A$ it can be considered to be the same operator. 

Then, choosing the convention of writing the action of the linear operator $A$ on $\langle y|$ as $\langle y|A$, i.e with the operator to the right of the bra, we get the two ways of writing the inner product discussed above

$$
\langle y|(A|x\rangle) \mbox{ or } (\langle y|A)|x\rangle.
$$
But from the linearity, this 'product' is clearly associative, and we can it write simply as

$$
\langle y|A|x\rangle
$$
where the operator can be considered to act either to the right or to left. The correctness of this is also born out by writing out the product concretely as matrices and row and column vectors.

The {\itshape adjoint} of the operator $A$ is defined as that operator $A^{\dagger}$, which acting on an arbitrary bra vector $\langle x|$, yields the same vector as the dual to the vector $A|x\rangle$, or in formulas

\begin{equation}\label{eq: AdjointDef}
\langle x|A^{\dagger}=(A|x\rangle)^{\dagger}.
\end{equation}
Again, this definition can be justified using linearity.

From (\ref{eq: AdjointDef}) follows the important property 

\begin{equation}\label{eq: AdjointDefAlternative}
\langle y|A^{\dagger}|x\rangle=\langle x|A|y\rangle^*.
\end{equation}

This equation could in fact be used as an alternative definition of the adjoint. Let us derive it since the short calculation illustrates the workings of the formalism. Start with the right hand side, taking the complex conjugate of $\langle x|A|y\rangle$

$$
\langle x|A|y\rangle^*=\big(\langle x|(A|y\rangle)\big)^*=\big(A|y\rangle\big)^{\dagger}\langle x|^{\dagger}=\langle y|A^{\dagger}|x\rangle
$$
where in the first step parenthesis are introduced to emphasizes which parts of the expressions are grouped together, next equations (\ref{eq: BraKetConjugation}) and (\ref{eq: BraAndKetConjugation}) are used, and finally the definition of the adjoint (\ref{eq: AdjointDef}) is employed.

Note that in terms of matrices, the adjoint is the same as the conjugate-transpose, or $A^{\dagger}=(A^*)^T$

%***
\subsubsection{Hermitean and unitary operators}
Of special interest in quantum mechanics are hermitean and unitary operators. They play the roles of representing observable quantities and generators of transformations respectively.

An operator $A$ is said to be {\itshape hermitean} or {\itshape self-adjoint} if 
$$
A^{\dagger}=A.
\eqno{\addtocounter{equation}{1}\theequation}
$$
An operator $U$ is said to be {\itshape unitary} if 
$$
U^{\dagger}=U^{-1}.
\eqno{\addtocounter{equation}{1}\theequation}
$$
Hermitean operators corresponds to observable physical quantities. Unitary operators corresponds to transformations of states.

%********************************
\subsection{Composition of operators}
Since a linear operator acting on a state is again a state, composition of operators is naturally defined as

$$
(AB)|x\rangle=A(B|x\rangle)=AB|x\rangle.
\eqno{\addtocounter{equation}{1}\theequation}
$$
Composition is associative
$$
A(BC)=(AB)C=ABC
\eqno{\addtocounter{equation}{1}\theequation}
$$ 
so there is no need to use parentheses. 

The 'product' of two operators $A$ and $B$ can be concretely realized in terms of ordinary matrix multiplication given matrix representations of the operators (in the same basis). As in matrix multiplication, the product is in general not commutative. The non-commutativity of quantum mechanical operators is an important property of quantum mechanics and leads to the celebrated uncertainty relations connecting results of measurements of non-commuting observables. But we will come to this is due time.

%***
\subsubsection{Commutators}
Certain combinations of operators often occur in quantum mechanics. One is the {\itshape commutator} between two operators $A$ and $B$. It is defined as

\begin{equation}\label{eq: CommutatorDef}
\lbrack A,B \rbrack=AB-BA.
\end{equation}

Since, in general, operators don't commute, the commutator is in general non-zero. Sets of hermitean operators that do commute among themselves are especially important in that they can represent sets of physical quantities that can be measured simultaneously.

%TRANSFORMATIONS AND SYMMETRIES
%******************************
\section{Transformations and symmetries}
Unitary operators effect symmetry transformations of quantum states and quantum operators. Symmetries are transformations of states that do not affect observable quantities, i.e they do not change the results of measurements. 

%***
\subsubsection{A first look at measurement}
A measurement always results in a number. It is a fundamental property of quantum mechanics that the states themselves are not observable or measurable. Essentially the only way to get numbers out of quantum mechanics is by taking inner products of states. Since the states are described by vectors it is reasonable to suspect that physical quantities are described by linear operators. Thus the result of a measurement is in some way related to inner products of the form $\langle\phi|A|\phi\rangle $. Such inner products are often called {\itshape diagonal matrix elements} in analogy to (\ref{eq: OperatorMatrixElements}). Furthermore, if $A$ is an hermitean operator, $\langle\phi|A|\phi\rangle $ is a real number which is interpreted as the expectation value for the quantity represented by $A$. The theory of measurement will developed in section 5.6 after some more terminology is introduced. 

%***
\subsubsection{Symmetry transformations}
In order to study symmetry transformations, suppose $|\psi\rangle$ and $|\phi\rangle$ are two quantum states. Acting on these states with the unitary operator $U$ we get the transformed states $U|\psi\rangle$ and $U|\phi\rangle$. It is customary to write transformations as

\begin{equation}\label{eq: SymmetryTransKet}
|\psi\rangle\rightarrow|\psi'\rangle=U|\psi\rangle.
\end{equation}

Likewise, the transformation of a bra vector is

\begin{equation}\label{eq: SymmetryTransBra}
\langle\phi|\rightarrow\langle\phi'|=\langle\phi|U^{\dagger}.
\end{equation}

That this is the correct form of a transformation of a bra vector follows from the equation $(U|\phi\rangle)^\dagger=\langle\phi|U^\dagger$ applied to the transformation of a ket vector.

With $U$ a unitary operator, the inner product between the states $|\phi\rangle$ and $|\psi\rangle$ is unaffected by this transformation. We get

$$
\langle\phi|\psi\rangle\rightarrow\langle\phi|U^\dagger U|\psi\rangle=\langle\phi|U^{-1} U|\psi\rangle=\langle\phi|\psi\rangle.
$$

The corresponding form for a transformation of a linear operator can be derived by demanding the matrix element $\langle\phi|A|\psi\rangle$ to be invariant under a transformation. We know how to transform states of the form $|\psi\rangle$. Consider transforming states of the form $A|\psi\rangle$, i.e states acted on by a linear operator $A$. The transformed state is $UA|\psi\rangle$, which can be expanded as

$$
UA|\psi\rangle=UA(U^{-1}U)|\psi\rangle=(UAU^{-1})U|\psi\rangle.
$$
In this way we separate the transformation of the state $A|\psi\rangle$ into a transformation of the state $|\psi\rangle$ and the operator $A$. Thus it is natural to define a transformation of a linear operator $A$ as 

\begin{equation}\label{eq: OperatorTrans}
A\rightarrow A'=UAU^{-1}=UAU^{\dagger}.
\end{equation}

That this is a reasonable definition is born out by calculating the transformation of the matrix element $\langle\phi|A|\psi\rangle$

$$
\langle\phi|A|\psi\rangle\rightarrow(\langle\phi|U^{\dagger})UAU^{\dagger}(U|\psi\rangle)
$$
$$
=\langle\phi|(U^{\dagger}U)A(U^{\dagger}U)|\psi\rangle)=\langle\phi|A|\psi\rangle),
$$
which shows the invariance of the matrix element under the transformation.

So although states and operators are affected by symmetry transformations, measurable quantities are not; this is the essence of symmetry in quantum mechanics.

%EIGENVECTORS AND EIGENVALUES
%****************************
\section{Eigenvectors and eigenvalues}
An {\itshape eigenvector} of a linear operator $A$ on a vector space is a non-zero vector $|x\rangle$ such that

\begin{equation}\label{eq: EigenValueEq1}
A|x\rangle\ = \lambda |x\rangle
\end{equation}
where the {\itshape eigenvalue} $\lambda$ is a complex number. Introducing the identity operator on the right hand side of the equation, it can be rewritten as a proper matrix equation

\begin{equation}\label{eq: EigenValueEq2}
(A-\lambda I)|x\rangle=0.
\end{equation}

From the theory of linear equations it follows that this equation has no non-zero solutions $|x\rangle$ unless the determinant of the matrix $A-\lambda I$ is zero. If the determinant is zero, then the vector $|x\rangle$ is identically zero since the equation is homogenous (the right hand side being zero). So, in order to get non-trivial eigenvectors, we require

\begin{equation}\label{eq: SecularEq}
\det(A-\lambda I)=0.
\end{equation}

This equation, called the {\itshape secular equation}, is an $n$-th degree equation for the unknown $\lambda$ and thus it has $n$ complex, not necessarily distinct, roots. These are the eigenvalues of the operator $A$. Once the eigenvalues are known, the corresponding eigenvectors can be calculated. Several distinct, linearly independent, eigenvectors might correspond to one and the same eigenvalue. In that case the eigenvalue is said to be {\itshape degenerate}. The degree of degeneracy is equal to the number of distinct linearly independent eigenvectors corresponding to the degenerate eigenvector.

%***
\subsubsection{Diagonalization}
A matrix is said to be diagonal if it has non-zero elements only on the diagonal. Using outer products of basis vectors $|i\rangle$ a diagonal operator can be written as

\begin{equation}\label{eq: OperatorDiagonal}
A=\sum_i\lambda_i|i\rangle\langle i|.
\end{equation}

That this actually represents a diagonal matrix is clear from using equation (\ref{eq: OperatorMatrixElements}) to compute the matrix elements. We get

$$
\langle k|A|j\rangle=\sum_i\lambda_i\langle k|i\rangle\langle i|j\rangle=\lambda_k\delta_{kj}.
$$

That is, only the diagonal elements are non-zero, and equal to the numbers $\lambda_i$. 

It would be natural to identify the $\lambda_i$ with the eigenvalues of the operator. Indeed, a diagonal operator trivially satisfies the eigenvalue equation (1.35) with eigenvectors equal to the basis vectors $|i\rangle$. So the question arises, when is an operator diagonalizable?

%********************************
\subsection{Spectral decomposition}
Here we will just state an important theorem that allows us to use diagonal representations for certain classes of operators.

An operator $A$ on a vector space is said to be {\it normal} if $A^{\dagger}A=AA^{\dagger}$. It follows immediately that a Hermitean operator is normal. Also, any unitary operator $U$ is also normal. This follows from the simple calculation

$$
U^{\dagger}U=U^{-1}U=I=UU^{-1}=UU^{\dagger}.
$$

%***
\subsubsection{Spectral decomposition theorem}
Let $A$ be a normal operator on a vector space ${\bf V}$. Then $A$ can be diagonalized with respect to some orthonormal basis for ${\bf V}$. Conversely, any diagonalizable operator is normal.

This means that the eigenvalue equation (\ref{eq: EigenValueEq1}) can be solved and the operator can be represented explicitly as in equation (\ref{eq: OperatorDiagonal}). To be definite, 

$$
A=\sum_i\lambda_i|i\rangle\langle i|
$$
where $\lambda_i$ are the eigenvalues of $A$, $|i\rangle$ is an orthonormal basis and each $|i\rangle$ is an eigenvector of $A$. This equation can also be trivially rewritten in terms of projectors $P_i$

\begin{equation}\label{eq: OperatorDiagonalProjector}
A=\sum_i\lambda_iP_i.
\end{equation}

In particular, hermitean and unitary operators can be diagonalized. Proofs of the spectral decomposition theorem can be found in \cite{Simmons1963} and \cite{NielsenChuang2000}.

%***
\subsubsection{Diagonalization using unitary transformations}
%***
We saw in section 5.3 that symmetry transformations are effected by unitary operators. Choosing a suitable unitary operator, a normal operator can be transformed into a diagonal form. Let $A$ be a normal operator. We want to find a unitary operator $D$ that transforms $A$ into diagonal form. Explicitly as in (\ref{eq: OperatorTrans})

\begin{equation}\label{eq: OTD}
A'=DAD^{-1}.
\end{equation}
where we demand that $A'$ is diagonal. In order to be concrete, we represent the operators by the corresponding matrices, so that $A'_{kl}=A_k'\delta_{kl}$. Then multiplying the equation (\ref{eq: OTD}) by $D$ from the left gives $DA=A'D$. Writing this last equation in terms of matrices yields a short calculation (note the subtle changes of indices)

$$
\sum_mD_{km}A_{ml}=\sum_mA'_{km}D_{ml}=\sum_mA'_k\delta_{km}D_{ml}=A'_kD_{kl}=\sum_mA'_k\delta_{ml}D_{km},
$$
or

$$
\sum_mD_{km}(A_{ml}-A'_k\delta_{ml}).
$$

Now, fixing the index $k$, we have $n$ homogeneous equations for the transformation matrix elements $D_{km}$. These equations have non-trivial solution if and only if the determinant of the coefficients vanish, i.e. that the determinant of the matrix $A_{ml}-A'_k\delta_{ml}$ is zero,
$$
det(A_{ml}-A'_k\delta_{ml})=0.
$$

Thus we get back the secular equation (\ref{eq: SecularEq}) explicitly. Solving this equation, we get the $n$, not necessarily distinct, eigenvalues of the original matrix $A$. Note that the eigenvalues does not depend on the diagonalizing matrix $D$.

\subsubsection{Diagonalization of hermitean operators}
Suppose A is a hermitean operator. Then it is diagonalizable and can be written as in equation (\ref{eq: OperatorDiagonal}), 

$$
A=\sum_i\lambda_i|i\rangle\langle i|.
$$
Taking the hermitean conjugate, we get

$$
A^{\dagger}=\sum_i(\lambda_i|i\rangle\langle i|)^{\dagger}=\sum_i\lambda_i^*|i\rangle\langle i|,
$$
since, obviously $(|i\rangle\langle i|)^{\dagger}=|i\rangle\langle i|$ for each $i$. But $A=A^{\dagger}$ so that we must have 

$$
\sum_i\lambda_i^*|i\rangle\langle i|=\sum_i\lambda_i|i\rangle\langle i|.
$$
This is only possible if all eigenvalues are real numbers, or $\lambda_i^{\dagger}=\lambda_i$.

Thus, hermitean operators have real eigenvalues. And conversely, if an operator have all eigenvalues real, then it is hermitean.

%***
\subsubsection{Simultaneous diagonalization theorem}
Suppose two operators $A$ and $B$ are diagonal in the same basis. Then it is easily shown that they commute. This follows since the product of two diagonal matrices is itself diagonal, and the elements on the diagonal is simply the product of the diagonal elements of $A$ and $B$.

$$
\pmatrix{A_{11}&0&\cdots&0\cr
	0&A_{22}&\cdots&0\cr
	\vdots&\vdots& &\vdots\cr
	0&0&\cdots&A_{nn}\cr}\pmatrix{B_{11}&0&\cdots&0\cr
		0&B_{22}&\cdots&0\cr
		\vdots&\vdots& &\vdots\cr
		0&0&\cdots&B_{nn}\cr}=
$$
$$\pmatrix{A_{11}B_{11}&0&\cdots&0\cr
			0&A_{22}B_{22}&\cdots&0\cr
			\vdots&\vdots& &\vdots\cr
			0&0&\cdots&A_{nn}B_{nn}\cr}
$$

The converse is also true, if two operators commute, then they are simultaneously diagonalizable in the same basis. For a proof, see \cite{NielsenChuang2000}.

%**********************************************
\section{Quantum dynamics}
The time evolution, or dynamics, of a closed quantum system can be described in two related ways. A system is closed if there is no interaction with the rest of the world. In practice, this might not be a realistic assumption. In principle it is not possible to isolate one piece of the world from the rest, there are always interactions between system and environment. The assumption is that either this interaction can be arbitrarily weak or controlled. The usefulness of the closedness assumption is that all of the systems dynamics is encoded in the Hamiltonian. 

%***
\subsubsection{Schr\"odinger equation}
Traditionally, the dynamics is described by the {\it Schr\"odinger equation}. This is a (first order) differential equation in the time variable $t$, equating the time derivative of the state to the action of the Hamiltonian operator on the state

\begin{equation}\label{eq: ScrodingerEq}
i\hbar{d|\psi\rangle\over dt} = H|\psi\rangle,
\end{equation}
where $\hbar$ is a fundamental physical constant setting the scale of quantum phenomena.\footnote{Its value is  $6.626\cdot10^{-34}$ Js} In theoretical contexts its value is often taken to be 1. This amounts to a choice of measuring units, where the 'natural' scale of phenomena is taken to be submicroscopic.

The form of the Hamiltonian depends on the representation chosen for the states. In the configuration space representation presented in chapter 4, the states are wave functions, and the Hamiltonian is a partial differential operator. In quantum computation contexts, $H$ will be a matrix acting on superpositions of computational basis states. 

Finding the proper Hamiltonian for a physical system is in general a difficult problem. It is not generally considered as a question within quantum mechanics itself, since as we have pointed out, quantum mechanics is just a framework for formulating physical theories. However this last point might very well change as our understanding of fundamental physics develops.\footnote{It might be the case that a would be "theory of everything" comes packaged with quantum mechanics as an inseparable part.}

Being a hermitean operator, $H$ can be diagonalized. Since the Hamiltonian is physically related to the energy of the system, the corresponding eigenvalues and eigenstates are referred to as {\it energy eigenvalues} and {\it energy eigenstates}. Naming the eigenvalues with $E_n$ and eigenstates $|n\rangle$, we have the spectral decomposition

\begin{equation}\label{eq: HamiltonianExpansion}
H=\sum_nE_n|n\rangle\langle n|.
\end{equation}

The lowest energy eigenvalue is the {\it ground state energy} and the corresponding eigenstate is simply the{\it ground state}. It is interesting to insert the expansion (\ref{eq: HamiltonianExpansion}) into the Schr\"odinger equation. A short calculation yields the following equation holding for an arbitrary energy state

$$
i\hbar{d|n\rangle\over dt}=E_n|n\rangle,
$$
the solution of which is 

$$
|n\rangle=\exp(-iE_nt/\hbar)|n\rangle.
$$

So, in a certain sense, the time dependence for energy eigenstates are trivial, it is just an overall oscillating phase factor. It does play a role in superpositions though, where different eigenstates oscillates with different frequencies. The frequency of oscillation $\omega$ is defined by $\omega=E/\hbar$ so that the energy is often written as $E_n=\hbar\omega_n$.

%*************************************
\subsubsection{Unitary transformation}
The time development of a quantum system between two times $t_1$ and $t_2$ can also be described by a unitary transformation. Let $|\psi(t_1)\rangle$ and $|\psi(t_2)\rangle$ be the state at the two times respectively. Then

\begin{equation}\label{eq: UnitaryEvolution}
|\psi(t_2)\rangle=U(t_1,t_2)|\psi(t_1)\rangle
\end{equation}
where $U(t_1,t_2)$ is a unitary operator that depends only on the two times $t_1$ and $t_2$, i.e. there is no other time dependence in $U$.

The two ways of prescribing the dynamics of the quantum state can be related by formally solving the Schr\"odinger equation. Provided we grant us the privilege to formally exponentiate the Hamiltonian operator, a solution to equation (\ref{eq: ScrodingerEq}) can be written

\begin{equation}\label{eq: SchrodingerEqSolution}
|\psi(t)\rangle=\exp(-iHt/\hbar)|\psi(0)\rangle.
\end{equation}

The intuition here is that the state starts in the state $|\psi(0)\rangle$ at an initial time $t=0$ and develops into $|\psi(t)\rangle$ at time $t$. Let us check this by a short calculation

$$
i\hbar{d\over dt}|\psi(t)\rangle=\big(i\hbar{d\over dt}\exp(-iHt/\hbar)\big)|\psi(0)\rangle=
$$

$$
(i\hbar)(-iH/\hbar)\exp(-iHt/\hbar)|\psi(0)\rangle=H|\psi(t)\rangle
$$
Clearly, this calculation presupposes that the Hamiltonian has no explicit time dependence.

Equation (\ref{eq: SchrodingerEqSolution}) can also be written somewhat more generally as

\begin{equation}\label{eq: SchrodingerEqSolutionGen}
|\psi(t_2)\rangle=\exp(-iH(t_2-t_1)/\hbar)|\psi(t_1)\rangle,
\end{equation}
expressing time development from time $t_1$ to time $t_2$. Now comparing this formal solution to the Schr\"odinger equation we see the connection between the two ways of prescribing the dynamics of the system. The unitary operator $U$ should be equated to the exponential of the Hamiltonian, or 

$$
U(t_1,t_2)=\exp(-iH(t_2-t_1)/\hbar).
$$

%******************************************
\subsubsection{Unitarity and reversibility}
The unitarity of time evolution leads immediately to the reversibility of the dynamics. Since $U(t_1,t_2)$ is invertible we can recover the 'initial' state $|\psi(t_1)\rangle$ from the 'final' state $|\psi(t_2)\rangle$ by multiplying the equation (\ref{eq: UnitaryEvolution}) by $U(t_1,t_2)^{\dagger}$

$$
U(t_1,t_2)^{\dagger}|\psi(t_2)\rangle=U(t_1,t_2)^{\dagger}U(t_1,t_2)|\psi(t_1)\rangle=|\psi(t_1)\rangle.
$$

There is one more issue that must be clarified in the context of quantum dynamics, and that is the different so called 'pictures'. In quantum dynamics, there is a choice as to where the time dependence resides. One choice is to let the states carry all the time dependence and letting the operators be time independent. This is the {\it Schr\"odinger picture}. Another choice is to have the states themselves be time independent and letting all time dependence be carried by the operators. This is the {\it Heisenberg picture}. There are also intermediate pictures, where the time dependence is split in a well defined way between states and operators. One such picture is the {\it interaction picture}, which is useful in calculations.
%*********************************************
\subsection{Schr\"odinger picture}
Of the Schr\"odinger picture there is not much more to be said. In fact, it is the Schr\"odinger picture that we have implicitly used in the preceeding sections. The Hamiltonian and all other operators are time independent and the time dependence is carried by the states. Formally solving the Schr\"odinger equation as in (\ref{eq: SchrodingerEqSolution}) makes this explicit.

%**********************************************
\subsection{Heisenberg picture}
The transition to the Heisenberg picture is interesting and we will carry it through in some detail. First let $|\psi_S(t)\rangle$ and $|\phi_S(t)\rangle$ be two quantum states where the index $S$ is used to indicate that these are taken in the Schr\"odinger picture. Later in the discussion we will introduce the corresponding Heisenberg picture states $|\psi_H\rangle=|\psi_S(0)\rangle$ and $|\phi_H\rangle=|\phi_S(0)\rangle$. Next consider an operator $A_S$. In order to transform all time dependence from the states to the operator, we will consider the time derivate of the matrix element of $A_S$ with $|\psi_S(t)\rangle$ and $|\phi_S(t)\rangle$. Thus we perform the calculation

$$
{d\over dt}\langle\psi_S(t)|A_S|\phi_S(t)\rangle=
$$
$$
\big({d\over dt}\langle\psi_S(t)|\big)A_S|\phi_S(t)\rangle+\langle\psi_S(t)|A_S{d\over dt}(|\phi_S(t)\rangle)=
$$
$$
{1\over i\hbar}\langle\psi_S(t)|(A_SH-HA_S)|\phi_S(t)\rangle={1\over i\hbar}\langle\psi_S(t)|[A_S,H]|\phi_S(t)\rangle,
$$
where the Schr\"odinger equation has been used. 

Next we use the formal solution (\ref{eq: SchrodingerEqSolution}) to the Schr\"odinger equation

\begin{eqnarray}
\langle\psi_S(t)|=\exp(iHt/\hbar)\langle\psi_S(0)|\label{eq: SchoKet}
\\
|\phi_S(t)\rangle=\exp(-iHt/\hbar)|\phi_S(0)\rangle\label{eq: SchoBra}.
\end{eqnarray}

Substituting these expressions into the last step of the calculation gives 

$$
{1\over i\hbar}\langle\psi_S(0)|[\exp(iHt/\hbar)A_S\exp(-iHt/\hbar),H]|\phi_S(0)\rangle,
$$
where we have used the fact that $H$ commutes with $\exp(\pm iHt/\hbar)$.

This is the proper place to define the Heisenberg picture operator $A_H$

\begin{equation}\label{eq: HeisenbergPictureOperator}
A_H=\exp(iHt/\hbar)A_S\exp(-iHt/\hbar).
\end{equation}

Thus the result of this calculation is
$$
{d\over dt}\langle\psi_S(t)|A_S|\phi_S(t)\rangle={1\over i\hbar}\langle\psi_H|[A_H,H]|\phi_H\rangle.
$$

Furthermore, making the substitutions (\ref{eq: SchoKet}), (\ref{eq: SchoBra}) and (\ref{eq: HeisenbergPictureOperator}) in the left hand side also, yields

$$
{d\over dt}\langle\psi_H|A_H|\phi_H\rangle={1\over i\hbar}\langle\psi_H|[A_H,H]|\phi_H\rangle.
$$

Now, since the states $|\psi\rangle$ and $|\phi\rangle$ are arbitrary, this equation must be valid for the operators

\begin{equation}\label{eq: HeisenbergEqMotionOp}
{dA_H\over dt}={1\over i\hbar}[A_H,H].
\end{equation}

The discussion above shows that the time dependence can be transformed from the states to the operators. The dynamical equation (\ref{eq: HeisenbergEqMotionOp}) can however be more easily derived directly from equation (\ref{eq: HeisenbergPictureOperator}) by direct differentiation. 

As a last point, we can now make contact with the discussion in chapter 4 where we discussed quantization of classical systems. The dynamical equation in the Heisenberg picture is actually identical to equation (\ref{eq: HeisenbergEq}) of chapter 4.

%**********************************************
\section{Quantum measurement}
Measurement is the process of getting numbers out of quantum systems. It is perhaps the most non-intuitive aspect of quantum mechanics. It has also been (and still is) an area of controversy and discussion related to the interpretation of quantum mechanics. In classical physics it is in principle always possible to find out all properties of a state to any desired degree of accuracy by making appropriate measurements. Not so in quantum mechanics where the state itself is not measurable. The only information we can get out of the system is certain numbers (corresponding to the eigenvalues of operators) with certain probabilities predicted by the theory. We will not review the vast literature about quantum measurement here (which would an herculean task) but rather present modern main stream measurement theory.\footnote{However it might be that the proper understanding of the interpretation and measurement issues in quantum mechanics could have bearings on computation theory.}

Some intuition can be obtained from thinking about how things are generally done in quantum mechanics. States are transformed by applying (unitary) operators to them. Especially, dynamics is expressed by applying the unitary time development operator $U(t_2,t_1)$ to the state. Another clue comes from the eigenvalue equation, where applying an hermitean operator to a state extracts a real number. Therefore it makes sense to define quantum measurement in terms of applying an operator to a state. 

%*************************************
\subsection{Projective measurement}
The context of projective measurement is the following. Suppose that the system under consideration is in an (unknown) state $|\psi\rangle$. The object is to measure a certain physical quantity $O$ say. This quantity is represented by an hermitean operator $\cal O$. Then $\cal O$ can be diagonalized and be written 
\begin{equation}\label{eq: ProjectiveMeasurment}
{\cal O}=\sum_{i=1}^n\lambda_iP_i
\end{equation}
in terms of the projectors $P_i$. Now the only (real) numbers that are present in this context are the eigenvalues $\lambda_i$ so we might suspect that the outcomes of the measurement must be related to these numbers. Measurement is then defined by the following postulate

Measuring the operator $\cal O$ in the state $|\psi\rangle$ gives the result $\lambda_i$ with probability 
\begin{equation}\label{eq: MeasurmentProbabilities1}
p(\lambda_i)=\langle\psi|P_i|\psi\rangle.
\end{equation}

If the outcome of the measurement was $\lambda_i$, the state of the system immediately after the measurement is

\begin{equation}\label{eq: NewState1}
{|P_i|\psi\rangle}\over{\sqrt{p(\lambda_i)}}
\end{equation}

Without loss of generality we can think of the state in terms of an expansion in the eigenstates of the operator $\cal O$

$$
|\psi\rangle=\sum_{j=1}^n\alpha_j|j\rangle.
$$

Note, however, that the coefficients in this expansion are unknown, unless we have deliberately {\it prepared} the system in a certain superposition. Now the probability can be calculated explicitly
$$
p(\lambda_i)=(\sum_{k=1}^n\alpha_k^*\langle k|)P_i(\sum_{j=1}^n\alpha_j|j\rangle)=(\sum_{k=1}^n\alpha_k^*\langle k|)\alpha_i|i\rangle=
$$
$$
\alpha_i^*\alpha_i=|\alpha_i|^2.
$$
The state of the system after the measurement becomes
$$
{{|P_i|\psi\rangle}\over{\sqrt{p(\lambda_i)}}}={{\alpha_i|i\rangle}\over{\sqrt{|\alpha_i|^2}}}
$$

%*************************************
\subsection{General measurement}

It is possible to define a slightly more general concept of quantum measurement. Instead of having an observable with a spectral resolution as in equation (\ref{eq: ProjectiveMeasurment}), it is sufficient to have a set of measurement operators $\{M_i\}$ acting on the state space of the system. The index $i$ refers to the outcome of the measurement. These operators are subject to a completeness requirement

\begin{equation}\label{eq:}
\sum_i M_i^{\dagger}M_i=I
\end{equation}

expressing the fact that the sum of probabilities must be 1. This follows since the probability for outcome $k$ is defined as

\begin{equation}\label{eq: MeasurmentProbabilities2}
p(k)=\langle\psi|M_k^{\dagger}M_k|\psi\rangle.
\end{equation}

In this case, the state of the system after the measurement is

\begin{equation}\label{eq: NewState2}
{M_k|\psi\rangle}\over{\sqrt{p(k)}}.
\end{equation}

It is thus easy to see that projective measurements are a special case of general measurements. To go from a general measurement to a projective measurement, we demand that the measurement operators $M_i$ are orthogonal projectors, that is they are hermitean and satisfy $M_iM_j=\delta_{ij}M_i$.

%*************************************
\subsection{POVM measurement}
Yet another special case of general measurements is the so called POVM\footnote{Positive Operator Valued Measurement} measurements. The idea is the following. If one is only interested in the probabilities, and not the resulting states, it is enough to know the combination $M_k^{\dagger}M_k$, the operators $M_k$ themselves are not needed. If we define new operators $E_k=M_k^{\dagger}M_k$, then $E_k$ is a positive operator such that $\sum_i E_i=I$ and the probabilities are given by $p(k)=\langle\psi|E_k|\psi\rangle$. 

Turning this argument around, a POVM measurement is defined by any set of operators $\{E_i\}$ such that each operator $E_i$ is positive and the completeness relation $\sum_iE_i=I$ holds. The probability of outcome $k$ becomes $p(k)=\langle\psi|E_k|\psi\rangle$. Of course, not having the "square roots" of the $E_i$, the new states cannot be computed.

\chapter{Abstract quantum computation: The circuit model}
In this chapter, the theory of quantum computation will be outlined in an abstract way, without recourse to physical considerations. This is in analogy with how classical computation theory is generally developed, relying on models of computation which abstracts away from the details of actual physical computing machinery. We will first look at the quantum Turing machine model, but will then turn to the quantum circuit model. 

The quantum Turing Machine (QTM) is a quantum generalization of the classical Turing machine, and as such, can be considered to be  a model of a programmable quantum computer. It is not a practical model for either algorithm construction or actual physical implementation. The QTM model is, however, useful in discussing complexity theory, and has been used extensively in that context \cite{BennetBernsteinBrassardVazirani97}.

The quantum circuit model (QCM) is easier to work with as regards algorithm construction, and is closer to a physical realization.

The models are claimed to be equivalent, but there seems to be a few unclear points, mainly having to do with subtleties as regards the QTM's. This is an area of active research. For that reason, QTM's will not be treated here. 

The original references are \cite{Deutsch1985} and \cite{Deutsch1989}.

%QUANTUM ALPHABETS AND LANGUAGES
%*******************************
\section{Quantum alphabets, strings  and languages}\label{sec: QuantumAlphabets}
As a starting point, let us see how far a quantum generalization of the concepts of alphabets and languages will carry us. An alphabet is a finite, non-empty set of symbols. Here we will use such an alphabet to label the quantum states of a system. This is an abstract labeling and the realization in terms of a concrete physical system will, as said, not concern us here. Suppose we have an alphabet $\Sigma=\{S_1,S_2,\ldots,S_n\}$, then the corresponding set of quantum states are

\begin{equation}\label{eq: QuantumAlphabet}
\Sigma_Q=\{|S_1\rangle,|S_2\rangle,\ldots,|S_n\rangle\}.
\end{equation}

These states are taken to span an orthonormal basis, i.e.

\begin{equation}\label{eq: QuantumAlphabetOrthonorm}
\langle S_i|S_j\rangle=\delta_{ij}.
\end{equation}

It seems reasonable to introduce the term {\it quantum alphabet} for such sets of states, though the term is not in common use.

Next, the classical concept of a string of length $L$, "$S_{{i_1}}S_{{i_2}}\cdots S_{{i_L}}$", is generalized to the corresponding composite quantum state 

\begin{equation}\label{eq: QuantStringCompBase}
|S_{{i_1}}\rangle |S_{{i_2}}\rangle\cdots |S_{{i_L}}\rangle=|S_{{i_1}}S_{{i_2}}\cdots S_{{i_L}}\rangle.
\end{equation}

This notation is a shorthand for the direct product notation 

$$
|S_{{i_1}}\rangle\otimes|S_{{i_2}}\rangle\otimes\cdots \otimes|S_{{i_L}}\rangle.
$$

The orthonormality condition (\ref{eq: QuantumAlphabetOrthonorm}) generalizes to
\begin{equation}\label{eq: QuanAlphOrthonormGen}
\langle {\cal S}|{\cal T}\rangle=\delta_{{\cal S}{\cal T}},
\end{equation}
where the notation ${\cal S}$ is used for the string "$S_{{i_1}}S_{{i_2}}\cdots S_{{i_L}}$". If the string length needs to be recorded we could write ${\cal S}_L$.

It is clear that the symbols $S_i$ only serve as labels, each of them ranging over the set $\Sigma$. Just as in the classical case, the actual symbols used play no role. The main difference as compared to the classical case is the possibility to consider linear superpositions of the states ${|\cal S}\rangle$. Thus the states of $\Sigma_Q$ span a $n^L$\,-dimensional Hilbert space isomorphic to the complex vector space $({\mathbf C^n})^{\otimes L}$. 

Imposing a lexicographic ordering $lex: \{{\cal S}_L\}\rightarrow N$, we have the general state $|\psi\rangle_L$

\begin{equation}\label{eq: QuantumString}
|\psi\rangle_L=\sum_{lex({\cal S}_L)}\alpha_{lex({\cal S}_L)} |{\cal S}_L\rangle,
\end{equation}
which could be denoted a {\it quantum string} of length $L$. Here $|{\cal S}_L\rangle$ play the role of basis states.

Normalization of the states $|\psi\rangle_L$, i.e. the demand that $\langle\psi|\psi\rangle=1$ leads to the usual restriction on the coefficients 
\begin{equation}\label{eq: QuantumStringNormalization}
\sum_{lex({\cal S}_L)}|\alpha_{lex({\cal S}_L)}|^2=1.
\end{equation}

The states considered so far have a fixed string length and are therefore finite in number. This is the normal situation when describing discrete quantum systems. In order to support Turing-like quantum computation, we have to consider arbitrary length strings. This is because fixed string length implies a finite size memory. Machines with a finite size memory are not really Turing machines, they are {\it finite state machines} and as such do not support universal computation. So, just as in the classical case, the tape must be potentially infinite.

A potentially infinite tape in the quantum case implies a potentially infinite dimensional Hilbert space. When a new tape square is added (or activated) during the computation, the tape Hilbert space goes from $n^L$-dimensional to $n^{(L+1)}$-dimensional.

Classically this situation is described by considering the set of all strings $\Sigma^*$ over the alphabet $\Sigma$. The set $\Sigma^*$ thus contains all strings that could be written on the potentially infinite tape. In the quantum case, the state $|\psi\rangle_L$ in \ref{eq: QuantumString} contains all length $L$ strings, provided all the coefficients are nonzero. Thus the analogue of the classical set $\Sigma^*$ ought to be $\{|\psi\rangle_L\}_{L=0}^\infty$, which we will denote as $\Sigma_Q^*$, the set of all quantum strings. It is the set of all superpositions of states $|{\cal S}_L\rangle$ for all string lengths $L$.

Classically, a language is a subset of $\Sigma^*$, or equivalently, an element of the power set ${\cal P}(\Sigma^*)$. By analogy, we define a {\it quantum language} as a subset of the set of all quantum states, i.e. as an element of the power set ${\cal P}(\Sigma_Q^*)={\cal P}(\{|\psi\rangle_L\}_{L=0}^\infty)$. 

%***
\subsubsection{A leaner notation}
The notation introduced so far, being a generalization of classical notions, is a bit to heavy handed. Utilizing the economy of the Dirac notation, we write $|i\rangle$ instead of $|S_i\rangle$, just keeping the indices. In the case of an alphabet with, say $n$ symbols, $i$ can be thought of as ranging over the numbers $1,2,\ldots,n$. Thus the state of equation (\ref{eq: QuantStringCompBase}) is written $|i_1i_2\cdots i_L\rangle$. A general state is then

\begin{equation}\label{eq: QubitCompBase}
|\psi\rangle_L=\sum\alpha_{i_1i_2\cdots i_L}|i_1i_2\cdots i_L\rangle.
\end{equation}

In conclusion then, $|i\rangle$ spans a $n$-dimensional Hilbert space isomorphic to ${\mathbf C}^n$. Likewise, $|i_1i_2\cdots i_L\rangle$ spans a $n^L$-dimensional Hilbert space isomorphic to $({\mathbf C}^n)^{\otimes L}$. A general state in this space is given by \ref{eq: QubitCompBase}.

%***
\subsubsection{Qubits}
A special case of this construction is the qubit and the qubit string corresponding to the alphabet $\{0,1\}$. Let $x$ denote a single bit. A classical $n$\,-bit string is then "$x_1x_2\ldots x_n$". A single qubit is denoted by $| x \rangle$ and a multi qubit state by $|x_1x_2\ldots x_n\rangle$. These states are called {\itshape computational basis states} or the {\it classical basis} \cite{Mermin2002}. Measurements on the quantum computer are often thought of as being made in this basis and it can therefore be considered as the connection to classical I/O-streams.

A convenient notation is often used for the computational basis states. Using the binary number representation for natural numbers, we can denote the state 

\begin{equation}\label{eq: CompBaseExpansion}
| x_1x_2\ldots x_n\rangle = \;| x_1\cdot 2^{n-1}+x_2\cdot2^{n-2}+\ldots x_{n-1}\cdot 2+x_n\rangle_n.
\end{equation}

The subscript $n$ is needed to indicate the number of qubits, but is often suppressed. An example of this notation is $|1001\rangle=|9\rangle_4$.

A general state of the quantum computer is a complex linear superposition of the basis states

\begin{equation}\label{eq: LinSupBaseState}
\sum_{i=0}^{2^n-1}\alpha_i| i\rangle_n,
\end{equation}
and the coefficients are again subject to the restriction

\begin{equation}\label{eq: AmplNormQubit}
\sum_{i=0}^{2^n-1}|\alpha_i|^2=1.
\end{equation}

The Hilbert space of one qubit is isomorphic to the complex vector space $\mathbf C^2$, and the $n\,$-qubit Hilbert space is isomorphic to $(\mathbf C^2)^{\otimes n}$. Thus, an $n\,$-bit quantum state carries exponentially more information than an n-bit classical string. Classically, it is not possible to linearly combine different bit strings.\footnote{The classical case could be seen as corresponding to all the coefficients except one being zero, the non-zero one being equal to 1.}

We also record the sometimes convenient explicit representation of the basis states in terms of $2^n\,$-dimensional basis vectors, for example

$$
|5\rangle_3=\pmatrix{0\cr0\cr0\cr0\cr0\cr1\cr0\cr0}.
$$

\section{The circuit model of quantum computation}\label{sec: CircuitModel}
The quantum Turing machine model is not practical when it comes to actual algorithm construction, and just as in the classical case, it is a theoretical construct far from real computer design. The model is difficult to work with since the state of the computer is a superposition of not just the data on the tape, but also the head position and the internal configuration. This leads to tricky questions about the halting of the machine, as different branches of the computation may take different number of steps to complete their respective computations. \cite{HaltingProblem}. 

The quantum circuit model is a generalization of the classical circuit model. Instead of bits, qubits are transmitted in the wires. The classical logic gates are replaced by quantum gates represented by unitary operators. In this model, the state of the computer is a superposition of the data only. The actual wiring of the circuit and the number of gates applied to the data are treated classically. 

The model is formulated in terms of unitary computation matrices, that given arbitrary $n$\,-qubit input vectors, produce the desired $n$\,-qubit output vectors. Algorithm construction amounts to composing such matrices out of simpler, primitive matrices acting on just a few qubits at a time. There are two important questions; finding a universal set of building blocks, i.e. programming primitives, and finding methods for efficient algorithm construction.

The circuit model is, however, subject to certain limitations. Its classical counterpart is the reversible logic circuit described in chapter 2. A logic circuit computes a fixed function for a given range of input, say $n$ bits. If the output is required for data beyond this range, the circuit must be extended to $m>n$ input bits. In principle, an algorithm is needed for this, or put differently, an algorithm is needed to generate the uniform circuit family $\{C_n\}_{n=1}^\infty$ computing the required function for arbitrary length input. The task of assembling the uniform circuit family cannot be performed by another circuit \cite{Deutsch1989}. Therefore, in itself, the circuit model is not a complete computational model, and the same is true for quantum circuits.

Therefore, we must refine the notion of algorithm construction in the circuit model, to providing a uniform circuit family for the problem at hand. 

\begin{figure}[h]
\epsfbox{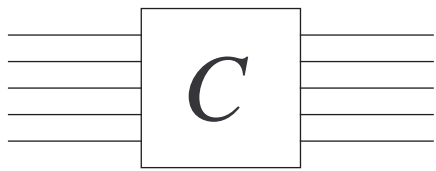}
\caption{A general circuit.}
\label{fig: GeneralCircuit}
\end{figure}

Since each wire carry a two-state quantum bit, an $n\,$-qubit circuit $C_n$ performs a unitary operation represented by a $2^n\times2^n$ unitary matrix $U_{C_n}$.

If all the input wires are used for data, then a given circuit performs one and the same algorithm on the data. Thus, the program is hardwired into the circuit, and in this sense, the circuit is not a general purpose computer. But nothing prevents us from considering some of the inputs as supplying a program, or rather an instruction, to be carried out on the rest of the input, which is then the data proper.\footnote{Of course, on a certain level of abstraction one need not make any distinction between data and program.} Universality in the context of the circuit model will be discussed below.

%***************************
\subsection{Gates and wires}\label{subsec: GatesWires}
An abstract quantum circuit is built out of wires and gates.\footnote{We leave open the physical implementation of the wires. They should not be thought of as classical wires, but rather in terms of some unspecified interaction between gate outputs and inputs, or by gates sharing qubits.} The wires carry the qubits between the gates, from outputs to inputs. The qubit processing takes place in the gates. There is to be no feedback wires. The number of output wires and input wires are equal for individual gates as well as for the complete circuit. 

\begin{figure}[h]
\epsfbox{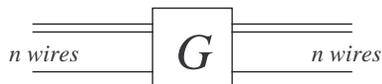}
\caption{A generic quantum gate.}
\label{fig: GeneralGate}
\end{figure}

A generic quantum gate $G$ performs the unitary operation $|\psi_{out}\rangle=U_G|\psi_{in}\rangle$. The $N\times N$ unitary matrix $U_{G_n}$ representing an $n$-qubit gate with $N=2^n$ belong to the Lie group $U(N)$.

%****************************
\subsection{General notation}\label{subsec: GeneralNotation}
A general one-qubit unitary gate is represented by a $2\times 2$ unitary matrix

\begin{equation}\label{eq: UnitaryMatrix2by2}
U=\pmatrix{u_{00}&u_{01}\cr u_{10}&u_{11}},
\end{equation}
and likewise, $n\,$-qubit gates are represented by $2^n\times2^n$ matrices where the matrix elements are denoted by $u_{ij}$ with indices ranging from 0 to $2^n-1$.\footnote{Other index ranges are sometimes convenient.} For fixed index $i$, $u_{ij}$ are row vectors, and for fixed index $j$, $u_{ij}$ are column vectors. In either view, unitarity for the matrix is equivalent to orthonormality of these row and column vectors respectively. This is a property of unitary matrices that is often useful precisely when deciding unitarity.

These matrices are realizations of unitary operators $U$ in some orthonormal basis, most often in the computational basis. Graphically they are represented by gates or circuit elements.\footnote{A few words on terminology; a {\it gate} ({\it circuit element}) is represented by a {\it unitary operator} which is realized as a {\it unitary matrix} in some basis. The same holds for the complete circuit, itself built out of gates.}

An important class of $2^{m+1}\times2^{m+1}$ matrices are the {\it controlled} gates $\Lambda_m(U)$ defined by

\begin{equation}\label{eq: ControlledGate}
\Lambda_m(U)(|x_1,\ldots,x_m,y\rangle=\cases{u_{y0}|x_1,\ldots,x_m,0\rangle+u_{y1}|x_1,\ldots,x_m,1\rangle\cr\quad\mbox{ if } \wedge^m_{k=1}x_k=1\cr\cr |x_1,\ldots,x_m,y\rangle\cr\quad\mbox{ if } \wedge^m_{k=1}x_k=0}
\end{equation}
or in a different notation where $x_1x_2\cdots x_n$ denotes the product of the bits $x_1,x_2,\ldots ,x_n$

$$
\Lambda_m(U)|x_1,\ldots,x_m,y\rangle=|x_1,\ldots,x_m,y\rangle U^{x_1x_2\cdots x_n}|y\rangle,
$$
or, explicitly, in block-matrix form

$$
\pmatrix{I_{2^m}&0&0\cr 0&u_{00}&u_{01}\cr 0&u_{10}&u_{11}},
$$
where $I_{2^m}$ denotes the $2^m\times 2^m$ identity matrix. The operator $\Lambda_m(U)$ applies the operation $U$ to the $(m+1)\,$-th qubit conditioned on the first $m$ qubits all being equal to 1, otherwise nothing is done. The controlled operations are the quantum analogs of the selection primitive of classical computation.

Another notation in common use for controlled gates, is $C^m(G)$. The gate $G$ need not be a single qubit gate, though in most cases it is.

\begin{figure}[h]
\epsfbox{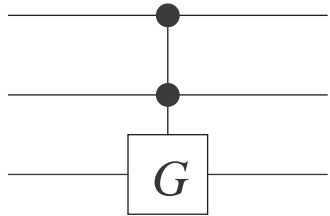}
\caption{A $\Lambda_2(G)$ gate.}
\label{fig: ControlledControlledSingleGate}
\end{figure}

Figure \ref{fig: ControlledControlledSingleGate} shows the diagrammatic representation of a $\Lambda_2(G)$ gate. Note that conditioning on the control bit being 1 is denoted by an filled circle on the corresponding wire.

\begin{figure}[h]
\epsfbox{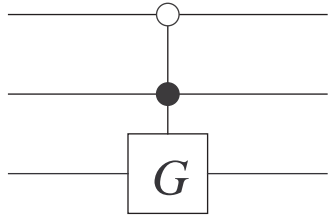}
\caption{A generalized $\Lambda_2(G)$ gate with conditioning on 0 and 1.}
\label{fig: ControlledAntiControlledSingleGate}
\end{figure}

There is nothing special about conditioning on 1. It is sometimes convenient to condition on 0, or on combinations of 0 and 1 for different control qubits. No special notation will be introduced for this case, but I will refer to it as a generalized $\Lambda_n(G)$, and an example is given in figure \ref{fig: ControlledAntiControlledSingleGate} to exemplify the concept.

%********************************************
\subsection{Special discrete one-qubit gates}\label{subsec: SpecDiscrete1QubitGates}
We first list a set of simple 1-qubit gates. In section 4.3.1, the spin-1/2 Pauli matrices were introduced. With a change of notation they are

\begin{equation}\label{eq: PauliMatricesQCNotation}
X=\pmatrix{0& 1\cr 1& 0},\quad Y=\pmatrix{0& -i\cr i& 0},\quad Z=\pmatrix{1& 0\cr 0& -1}.
\end{equation}

For the algebraic identities and commutation relations satisfied by these matrices, refer back to section 4.3.1.

Apart from these gates, the Hadamard gate $H$, the phase gate $S$, and the $\pi/8$\,-gate $T$ are given by the matrices

\begin{equation}\label{eq: HadamardPhasePi8}
H={1\over\sqrt{2}}\pmatrix{1& 1\cr 1& -1},\quad S=\pmatrix{1& 0\cr 0& i},\quad T=\pmatrix{1& 0\cr 0& \exp(i\pi/4)}.
\end{equation}

These single qubit gates are important, as they can be used together with the CNOT-gate to give universal sets of discrete quantum gates.

When simplifying circuits, the following identities are useful

\begin{equation}\label{eq: HPauliHIdentities}
HXH=Z,\quad HYH=-Y,\quad  HZH=X.
\end{equation}

The Hadamard gate can be used to produce equally weighted superpositions as the following simple example shows

\begin{eqnarray}\label{eq: HadamardSuperposition}
H|0\rangle={1\over\sqrt 2}(|0\rangle+|1\rangle),
\\
H|1\rangle={1\over\sqrt 2}(|0\rangle-|1\rangle).
\end{eqnarray}

%****************************************
\subsection{One-qubit rotation operators}\label{subsec: 1QubitRotationOperators}
By formally exponentiating the Pauli matrices one obtains a set of continuous rotation operators

\begin{eqnarray}
R_x(\theta)=e^{-i\theta X/2}=\cos{\theta\over 2}I-i\sin{\theta\over 2}X=\pmatrix{\cos{\theta\over 2}& -i\sin{\theta\over 2}\cr -i\sin{\theta\over 2}&\cos{\theta\over 2}},\label{eq: RotationOperatorX}
\\
R_y(\theta)=e^{-i\theta Y/2}=\cos{\theta\over 2}I-i\sin{\theta\over 2}Y=\pmatrix{\cos{\theta\over 2}& -\sin{\theta\over 2}\cr \sin{\theta\over 2}&\cos{\theta\over 2}},\label{eq: RotationOperatorY}
\\
R_z(\theta)=e^{-i\theta Z/2}=\cos{\theta\over 2}I-i\sin{\theta\over 2}Z=\pmatrix{e^{-i\theta/2}& 0\cr 0 & e^{i\theta/2}}\label{eq: RotationOperatorZ}.
\end{eqnarray}

The matrices in the right hand sides of these equations can be derived using the following general formula for functions of Pauli matrices

\begin{equation}\label{eq: FunctionsOfPaulimatrices}
f(\theta\,\overline n\cdot\overline{\sigma})={f(\theta)+f(-\theta)\over 2}I+{f(\theta)-f(-\theta)\over 2}\,\overline n\cdot\overline{\sigma},
\end{equation}
where $\hat n = (n_x,n_y,n_z)$ is a three dimensional normal vector and $\overline{\sigma}=(\sigma_x,\sigma_y,\sigma_z)=(X,Y,Z)$.

From the definition of the rotation operators, it is clear that the following equations hold

\begin{equation}\label{eq: SumFormulaRotationOp}
R_i(\alpha)+R_i(\beta)=R_i(\alpha +\beta)\quad \mbox{   where   } i\in\{x,y,z\}.
\end{equation}

Some more useful identities are

\begin{eqnarray}
XR_x(\theta)X=R_x(\theta)\label{eq: XRotX},
\\
XR_y(\theta)X=R_y(-\theta)\label{eq: YRotY},
\\
XR_z(\theta)X=R_z(-\theta)\label{eq: ZRotZ}.
\end{eqnarray}

Obviously, there are lots of such simple identities for single qubit gates.

%*******************************************************
\subsection{The rotation operators and the Bloch sphere}\label{subsec: Bloch}
\begin{figure}[h]
\epsfbox{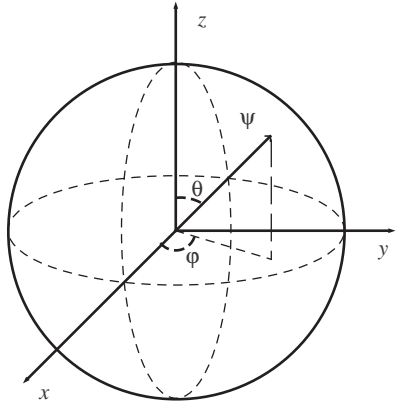}
\caption{The Bloch sphere.}
\label{fig: BlochSphere}
\end{figure}

The rotation operators can indeed be interpreted as rotations around the $x$, $y$ and $z$ axes respectively, in a three dimensional space. In such an interpretation, a general qubit $|\psi\rangle$ can be represented as

\begin{equation}\label{eq: ThetaPhiQubit}
|\psi\rangle=\cos{\theta\over 2}|0\rangle+e^{i\phi}\sin{\theta\over 2}|1\rangle, 
\end{equation}
with angles $\theta$ and $\phi$ defined as in figure \ref{fig: BlochSphere}.

The basis vectors $|0\rangle$ and $|1\rangle$ are represented by $(0,0,1)$ and $(0,0,-1)$ (corresponding to $\theta=0,\phi=0$ and $\theta=\pi,\phi=0$), respectively.

Note that this representation (\ref{eq: ThetaPhiQubit}), of the qubit follows from the general form 
$$
|\psi\rangle=\alpha|0\rangle+\beta|1\rangle
$$ 
by writing $\alpha=e^{i\gamma}\cos{\theta\over 2}$ and $\beta=e^{i\gamma}e^{i\phi}\sin{\theta\over 2}$ so that 

$$
|\psi\rangle=e^{i\gamma}(\cos{\theta\over 2}|0\rangle+e^{i\phi}\sin{\theta\over 2}|1\rangle)
$$
and then dropping the overall unphysical phase factor $e^{i\gamma}$. Thus the number of real physical degrees of freedom of a qubit is 2.

A general rotation by an angle $\alpha$ about an axis $\hat n=(n_x,n_y,n_z)$ is given by the operator $R_{\hat n}(\alpha)$

\begin{equation}\label{eq: GeneralRotation}
R_{\hat n}(\alpha)=\exp(-i\alpha{\hat n}\cdot{\overline\sigma}/2)=\cos\Big({\alpha\over 2}\Big)I-i\sin\Big({\alpha\over 2}\Big)(n_xX+n_yY+n_zZ).
\end{equation}

%**********************************************
\subsection{Single qubit phase-shift operators}\label{subsec: SingleQubitPhaseShiftOperators}
The following phase-shift operators are sometimes useful

\begin{equation}\label{eq: PhaseShift}
P(\alpha)=\pmatrix{e^{i\alpha}&0\cr 0&e^{i\alpha}}=e^{i\alpha}I,
\end{equation}
\begin{equation}\label{eq: PhaseShift}
E(\alpha)=\pmatrix{1&0\cr 0&e^{i\alpha}}=P(\alpha/2)R_z(\alpha).
\end{equation}

%**********************************************
\subsection{Some special controlled operations}\label{subsec: SpecialControlledOperations}
Two of the most important and useful controlled operations are the two-qubit CNOT\,-gate and the three-qubit Toffoli\,-gate. 

%***
\subsubsection{Controlled NOT}
The CNOT-gate is an example of a two-qubit controlled operation. It also goes under the name (quantum) XOR. Its matrix representation in the computational basis is

\begin{equation}\label{eq: CNOT}
{\rm CNOT}=\Lambda_1(X)=\pmatrix{1&0&0&0\cr0&1&0&0\cr0&0&0&1\cr0&0&1&0}
\end{equation}

\begin{figure}[h]
\epsfbox{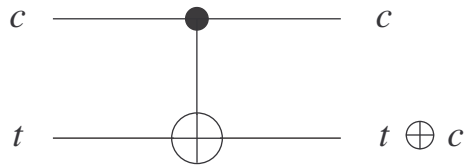}
\caption{Controlled NOT gate.}
\label{fig: ControlledNOT}
\end{figure}

The CNOT-gate performs a NOT (i.e. an $X$) operation on the target bit $t$ conditioned on the control bit $c$ being 1.\footnote{Inevitably, language tends to get sloppy, and "bit" will be used when it is really the qubit basis states that are refereed to.}

%***
\subsubsection{Toffoli gate}
The Toffoli gate is $\Lambda_2(X)$. As noted in chapter 2, it is universal for classical reversible computation. 

\begin{figure}[h]
\epsfbox{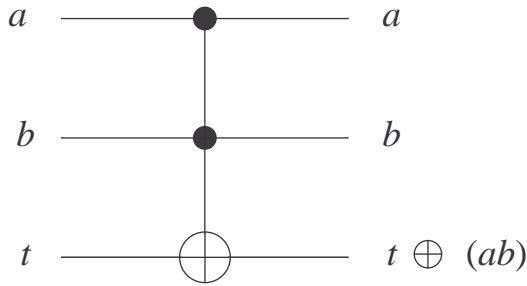}
\caption{The Toffoli gate.}
\label{fig: ToffoliGate}
\end{figure}

%**************************************
\subsection{Some practical "machinery"}\label{subsec: SpecialMachinery}
The relation between a quantum circuit and the corresponding unitary matrix is not entirely straightforward to figure out. For the benefit of the reader we here record some "practical tools of the trade" often left implicit in the literature.

%***
\subsubsection{Gates acting independently}
Let $G$ denote a general single qubit gate. Then the action of $G$ on the $j$\,-th qubit on a multi-qubit quantum state is denoted by $G(j)$ and defined by

\begin{equation}\label{eq: GateOnJthQubit}
G(j)|x_1x_2\cdots x_j\cdots x_n\rangle=|x_1x_2\cdots x_{j-1}\rangle\otimes \big(G(j)|x_j\rangle\big)\otimes|x_{j+1}\cdots x_n\rangle.
\end{equation}

The generalization to several single qubit gates acting on different qubits in the register is straightforward. In particular, specializing to the case of two single qubit gates acting independently on a two-qubit state, we have

\begin{equation}\label{eq: TwoGatesOnTwoQubits}
A(1)\otimes B(2)|x_1x_2\rangle=A(1)B(2)|x_1x_2\rangle=\big (A(1)|x_1\rangle\big )\otimes \big (B(2)|x_2\rangle\big ). 
\end{equation}

\begin{figure}[h]
\epsfbox{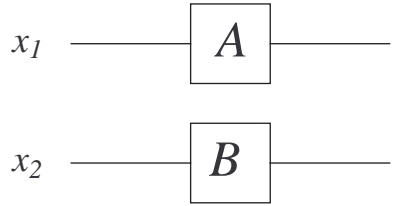}
\caption{Two gates acting independently.}
\label{fig: GatesActingIndependtly}
\end{figure}

In order to work out this equation explicitly, let the states be given by

\begin{equation}\label{eq: TwoGatesOnTwoQubitsEx1}
|x_1\rangle=\alpha_1|0\rangle+\beta_1|1\rangle=\pmatrix{\alpha_1\cr \beta_1}, \quad |x_2\rangle=\alpha_2|0\rangle+\beta_2|1\rangle=\pmatrix{\alpha_2\cr \beta_2},
\end{equation}
so that the composite state is

\begin{equation}\label{eq: TwoGatesOnTwoQubitsEx2}
|x_1x_2\rangle=\pmatrix{\alpha_1\alpha_2\cr \alpha_1\beta_2\cr \beta_1\alpha_2\cr \beta_1\beta_2}.
\end{equation}

Next, let the operators be

\begin{equation}\label{eq: TwoGatesOnTwoQubitsEx3}
A=\pmatrix{a_{11}&a_{12}\cr a_{21}&a_{22}}, \quad B=\pmatrix{b_{11}&b_{12}\cr b_{21}&b_{22}}.
\end{equation}

The right hand side of equation (\ref{eq: TwoGatesOnTwoQubits}) becomes

$$
\pmatrix{a_{11}&a_{12}\cr a_{21}&a_{22}}\pmatrix{\alpha_1\cr \beta_1}\otimes \pmatrix{b_{11}&b_{12}\cr b_{21}&b_{22}}\pmatrix{\alpha_2\cr \beta_2}=
$$
\begin{equation}\label{eq: TwoGatesOnTwoQubitsEx4}
\pmatrix{(a_{11}\alpha_1+a_{12}\beta_1)(b_{11}\alpha_2+b_{12}\beta_2)\cr
(a_{11}\alpha_1+a_{12}\beta_1)(b_{21}\alpha_2+b_{22}\beta_2)\cr
(a_{21}\alpha_1+a_{22}\beta_1)(b_{11}\alpha_2+b_{12}\beta_2)\cr (a_{21}\alpha_1+a_{22}\beta_1)(b_{21}\alpha_2+b_{22}\beta_2)}.
\end{equation}

The operator product in the left hand side of (\ref{eq: TwoGatesOnTwoQubits}) must be defined so that it reproduce this expression. In analogy with the definition of the $\otimes$ product of vectors in (\ref{eq: TwoGatesOnTwoQubitsEx2}), it is natural to define

$$
A\otimes B=\pmatrix{a_{11}B&a_{12}B\cr a_{21}B&a_{22}B}=
$$
\begin{equation}\label{eq: TwoGatesOnTwoQubitsEx5}
\pmatrix{a_{11}b_{11}&a_{11}b_{12}&a_{12}b_{11}&a_{12}b_{12}\cr
a_{11}b_{21}&a_{11}b_{22}&a_{12}b_{21}&a_{12}b_{22}\cr
a_{21}b_{11}&a_{21}b_{12}&a_{22}b_{11}&a_{22}b_{12}\cr
a_{21}b_{21}&a_{21}b_{22}&a_{22}b_{21}&a_{22}b_{22}}.
\end{equation}

Then the left and side of (\ref{eq: TwoGatesOnTwoQubits}) becomes

\begin{equation}\label{eq: TwoGatesOnTwoQubitsEx6}
\pmatrix{a_{11}b_{11}&a_{11}b_{12}&a_{12}b_{11}&a_{12}b_{12}\cr
a_{11}b_{21}&a_{11}b_{22}&a_{12}b_{21}&a_{12}b_{22}\cr
a_{21}b_{11}&a_{21}b_{12}&a_{22}b_{11}&a_{22}b_{12}\cr
a_{21}b_{21}&a_{21}b_{22}&a_{22}b_{21}&a_{22}b_{22}}\pmatrix{\alpha_1\alpha_2\cr \alpha_1\beta_2\cr \beta_1\alpha_2\cr \beta_1\beta_2}
\end{equation}

Multiplying out (\ref{eq: TwoGatesOnTwoQubitsEx4}) and (\ref{eq: TwoGatesOnTwoQubitsEx6}) and doing some careful bookkeeping, shows the equality of these two expressions.

Two special cases are useful to note in order to gain some intuition. When $A=I$, the unity matrix, we get in block diagonal form

\begin{equation}\label{eq: }
I\otimes B=\pmatrix{B&0\cr 0&B}.
\end{equation}

On the other hand, when $B=I$

\begin{equation}\label{eq: }
A\otimes I=\pmatrix{a_{11}&0&a_{12}&0\cr 0&a_{11}&0&a_{12}\cr a_{21}&0&a_{22}&0\cr 0&a_{21}&0&a_{22}}.
\end{equation}

Note that general controlled operations cannot be achieved by direct products of operators, i.e. by operators acting independently. This can be seen by inspecting the explicit product $A\otimes B$ in \ref{eq: TwoGatesOnTwoQubitsEx5}.

%***
\subsubsection{Explicit controlled operations}
\begin{figure}[h]
\epsfbox{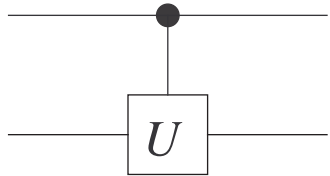}
\caption{Conditioning on the first qubit.}
\label{fig: ContrUFirstBit}
\end{figure}

In the notation $\Lambda_m(U)$, the target qubit is supposed to be the qubit with lowest place value in the binary number $x_1\ldots x_my$, that is $y$. However, the target can be any of the qubits. In that case some care must be exercised when writing out an explicit matrix representation. A simple example will illustrate this. Let $U$ be the $2\times 2$ unitary matrix 
$$
\pmatrix{a&b\cr c&d}.
$$
Then the controlled operation $\Lambda_1(U)$ of figure \ref{fig: ContrUFirstBit} 

has the explicit form

$$
\pmatrix{1&0&0&0\cr
0&1&0&0\cr
0&0&a&b\cr
0&0&c&d}.
$$

\begin{figure}[h]
\epsfbox{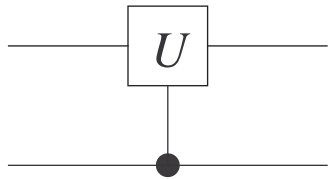}
\caption{Conditioning on the second qubit.}
\label{fig: ContrUSecondBit}
\end{figure}

On the other hand, conditioning on the second qubit as in figure \ref{fig: ContrUSecondBit}

we get the explicit matrix representation
$$
\pmatrix{1&0&0&0\cr
0&a&0&b\cr
0&0&1&0\cr
0&c&0&d}.
$$

From the form of the matrix it is clear that it acts non-trivially on the computational basis states $|01\rangle$ and $|11\rangle$, i.e. conditioned on the second qubit being 1.

%***
\subsubsection{A note and directions in $n$-qubit space}
A qubit is two dimensional, therefore an $n$-qubit state is $2^n$-dimensional. Often, one is interested in lower dimensional subspaces of the full space. It is convenient to use the term {\it level} to point out a particular direction in this space.  For instance, consider a 3-qubit computational basis state $|s\rangle$ with binary expansion $|s_1s_2s_3\rangle$ and explicit vector representation 
$$
{\mathbf e}_s=\pmatrix{e_0\cr e_1\cr\vdots\cr e_s\cr\vdots\cr\cr e_7},
$$
where all $e_i=0$ except $e_s=1$. Thus, a certain computational basis state $|s\rangle$ points out direction ${\mathbf e}_s$ in the full Hilbert space. Such a direction is sometimes called a level. Lower dimensional unitary matrixes can act on subspaces of levels of the full space.

%***
\subsubsection{Order of multiplication}
\begin{figure}[h]
\epsfbox{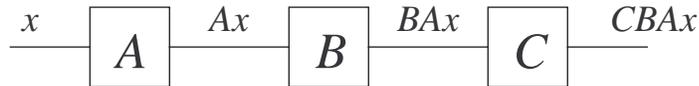}
\caption{Order of multiplication for consecutive gates.}
\label{fig: OrderOfMultiplication}
\end{figure}

When consecutive gates are applied to a qubit as in figure (\ref{fig: OrderOfMultiplication}), the gates are applied from left to right. This is clear, since following the circuit from left to right, the gate $A$ is first applied to the qubit $|x\rangle$, then $B$ and finally $C$, resulting in the state $CBA|x\rangle$.

%************************
\subsection{Entanglement}\label{subsec: Entanglement}
It is important to realize that the quantum states entering and leaving a gate might be entangled. Entanglement is a property of composite quantum systems. 

A state of a composite quantum state is {\it entangled} when the state cannot be decomposed as product of states of the component systems. In fact, a general $n$-qubit state 
$$|\psi\rangle_n=\sum\alpha_{b_1b_2\cdots b_n}|b_1b_2\cdots b_n\rangle
$$ 
as in equation \ref{eq: QubitCompBase}, is said to be {\it decomposable} when it can be expressed in terms of a tensor product $$|x_1x_2\cdots x_n\rangle=|x_1\rangle\otimes|x_2\rangle\otimes\cdots \otimes|x_n\rangle$$
of component qubit states $|x_i\rangle=\alpha_i|0\rangle+\beta_i|1\rangle$. A state is {\it entangled} if and only if it is non-decomposable.

The concept is easy to illustrate in the case of a two qubit system. The computational basis of such a system is

$$
|00\rangle, \quad |01\rangle, \quad |10\rangle, \quad |11\rangle.
$$

Consider then the state

\begin{equation}\label{eq: EntangledQubits}
\alpha|00\rangle + \beta|11\rangle.
\end{equation}

By studying the coefficients in the expansion of the product of two qubits

$$
\big (\alpha_1|0\rangle+\beta_1|1\rangle \big )\big (\alpha_2|0\rangle+\beta_2|1\rangle \big )=
$$
\begin{equation}\label{eq: }
\alpha_1\alpha_2|00\rangle+\alpha_1\beta_2|01\rangle+\beta_1\alpha_2|10\rangle+\beta_1\beta_2|11\rangle
\end{equation}

it is clear that there is no way to choose the coefficients $\alpha_1,\alpha_2,\beta_1$ and $\beta_2$ so that $\alpha_1\alpha_2\not=0$, $\alpha_1\beta_2=0$, $\beta_1\alpha_2=0$ and $\beta_1\beta_2\not=0$ simultaneously in order to reproduce the entangled state (\ref{eq: EntangledQubits}) as a product of single qubit states.

The entangled state (\ref{eq: EntangledQubits}) can be produced by the application of the CNOT-gate to a product state. The following figure illustrates an example.

\begin{figure}[h]
\epsfbox{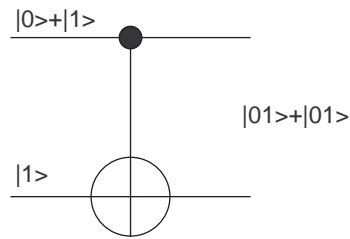}
\caption{Entangling a two-qubit product state using CNOT.}
\label{fig: Entanglement}
\end{figure}

Working with un-normalized states for simplicity, we have the incoming product state

$$
\big(|0\rangle+|1\rangle\big)|1\rangle=|01\rangle+|11\rangle=\pmatrix{0\cr 1\cr 0\cr 0}+\pmatrix{0\cr 0\cr 0\cr 1}=\pmatrix{0\cr 1\cr 0\cr 1}.
$$
Applying the CNOT gate yields

$$
\pmatrix{1&0&0&0\cr 0&1&0&0\cr 0&0&0&1\cr 0&0&1&0}\pmatrix{0\cr 1\cr 0\cr 1}=\pmatrix{0\cr 1\cr 1\cr 0}=\pmatrix{0\cr 1\cr 0\cr 0}+\pmatrix{0\cr 0\cr 1\cr 0}=|01\rangle+|10\rangle.
$$

It is interesting and important to note that the CNOT gate cannot be written as a $\otimes$\,- product of two single qubit gates. This is clear from studying the matrix elements of the product $A\otimes B$ in equation (\ref{eq: TwoGatesOnTwoQubitsEx5}). In fact, there is no way to produce entanglement using only single qubit gates. This observation will be put in context in section X.X.X on universal quantum gates.

Mathematically, entanglement is quite trivial, but the concept is far from trivial from a physical point of view, and has been a subject of discussion since the mid nineteen thirties. We will return briefly to this discussion in the last chapter.

%*********************************************
\subsection{Some important gate constructions}\label{subsec: GateConstructions}
We need to able to build complicated gates out of simpler ones. These simple gates are, apart from the special discrete one-qubit gates, also general one-qubit gates $U$ and CNOT gates. We will call these {\it basic} or {\it elementary} operations. Here follows a few useful constructions.

%***
\subsubsection{Decomposition of a single qubit gate into Z and Y rotations}
Every unitary $2\times 2$ matrix $U$ can be expressed as

\begin{equation}\label{eq: ZYDecomposition}
U=e^{i\alpha}R_z(\beta)R_y(\gamma)R_z(\delta),
\end{equation}
in terms of the rotation operators \ref{eq: RotationOperatorY} and \ref{eq: RotationOperatorZ}, and where the parameters $\alpha,\beta,\gamma$ and $\delta$ are real.

\subsubsection{Proof}
First note that the unitarity constraint $UU^{\dagger}=1$ on a general $2\times 2$ complex matrix $U$ reduces the number of real parameters from 8 to 4. Furthermore, a matrix is unitary if and only if its row vectors and column vectors are orthonormal. Therefore, every unitary $2\times 2$ matrix can be expressed in terms of four real parameters $\alpha,\beta,\gamma$ and $\delta$ as

\begin{equation}\label{eq: 2X2Unitary}
\pmatrix{e^{i(\alpha-\beta/2-\delta/2)}\cos{\gamma\over 2}&-e^{i(\alpha-\beta/2+\delta/2)}\sin{\gamma\over 2}\cr
e^{i(\alpha+\beta/2-\delta/2)}\sin{\gamma\over 2}&e^{i(\alpha+\beta/2+\delta/2)}\cos{\gamma\over 2}}.
\end{equation}

For example, the column vectors are orthogonal, since

$$
\big(e^{i(\alpha-\beta/2-\delta/2)}\cos{\gamma\over 2})^*\big(-e^{i(\alpha-\beta/2+\delta/2)}\sin{\gamma\over 2}\big)+
$$
$$
\big(e^{i(\alpha+\beta/2-\delta/2)}\sin{\gamma\over 2})^*(e^{i(\alpha+\beta/2+\delta/2)}\cos{\gamma\over 2})=
$$
$$
-e^\delta\cos{\gamma\over 2}\sin{\gamma\over 2}+e^\delta\sin{\gamma\over 2}\cos{\gamma\over 2}=0.
$$
The rest of the conditions on the matrix \ref{eq: 2X2Unitary} can be checked similarly.

Multiplying out equation \ref{eq: ZYDecomposition}, yields exactly the matrix in \ref{eq: 2X2Unitary}. 

%***
\subsubsection{Decomposition of a general $\Lambda_1(U)$ gate}
The above result allows a decomposition of general $\Lambda_1(U)$ gate in terms of single qubit gates and CNOT's. 

Let $U$ be single qubit unitary gate. Then there exist single qubit operators $A,B$ and $C$ such that

\begin{eqnarray}
ABC=I\label{eq: ABCI}
\\
e^{i\alpha}AXBXC=U\label{eq: AXBXC}.
\end{eqnarray}

%***
\subsubsection{Proof}
In terms of the rotation operators (\ref{eq: RotationOperatorX}), (\ref{eq: RotationOperatorY}) and (\ref{eq: RotationOperatorZ}), put 

$$
A=R_z(\beta)R_y(\gamma/2),
$$
$$
B=R_y(-\gamma/2)R_z(-(\delta +\beta)/2),
$$
$$
C=R_z((\delta -\beta)/2).
$$
Then
$$
ABC=R_z(\beta)R_y(\gamma/2)R_y(-\gamma/2)R_z(-(\delta +\beta)/2)R_z((\delta -\beta)/2)=I.
$$
Next, using the identities \ref{eq: YRotY}, \ref{eq: ZRotZ} as well as $X^2=I$
$$
XBX=XR_y(-\gamma/2)XXR_z(-(\delta +\beta)/2)=R_y(\gamma/2)R_z((\delta +\beta)/2),
$$
so that
$$
AXBXC=R_z(\beta)R_y(\gamma/2)R_y(\gamma/2)R_z((\delta +\beta)/2)R_z((\delta -\beta)/2)=
$$
$$
R_x(\beta)R_y(\gamma)R_z(\delta).
$$

Thus $U$ can be decomposed as $U=e^{i\alpha}AXBXC$ with ABC=I. 

\begin{figure}[h]
\epsfbox{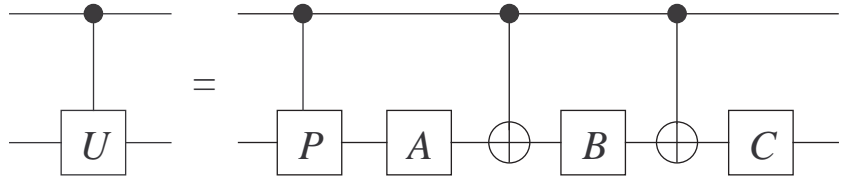}
\caption{Simulation of a controlled-U operation in terms of single qubit operations and CNOT's.}
\label{fig: ControlledUDecompositionV2}
\end{figure}

These equations allow a decomposition of a general controlled-$U$ operation as in figure \ref{fig: ControlledUDecompositionV2}. When the control qubit is $|0\rangle$, the operation $ABC$ is applied to the target qubit. On the other hand, when the control qubit is $|1\rangle$, the operation $P(\alpha)AXBXC$ is applied to the target qubit.

\begin{figure}[h]
\epsfbox{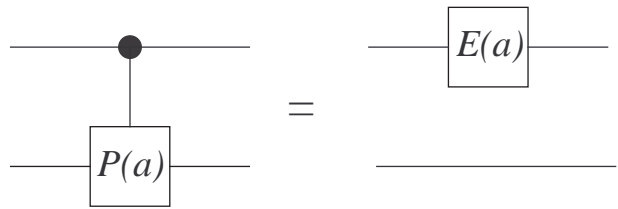}
\caption{Identity between $\Lambda_1(P(\alpha))$ and $E(\alpha)\otimes I$.}
\label{fig: ControlledPhaseIdentity}
\end{figure}

\begin{figure}[h]
\epsfbox{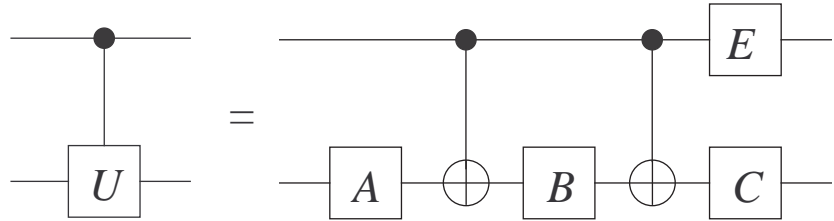}
\caption{Simplified decomposition of a controlled-U operation in terms of single qubit operations and CNOT's.}
\label{fig: ControlledUDecomposition}
\end{figure}

In order to obtain a somewhat more simple diagrammatic representation, the circuit identity of figure \ref{fig: ControlledPhaseIdentity} is useful. This identity can be derived using the methods of section 6.3.7. Using the identity, we get the circuit of figure \ref{fig: ControlledUDecomposition}.

\pagebreak

Thus, the operation $\Lambda_1(U)$ can be decomposed into six basic operations.

%***
\subsubsection{Decomposition of a general $\Lambda_2(U)$ gate}
\begin{figure}[h]
\epsfbox{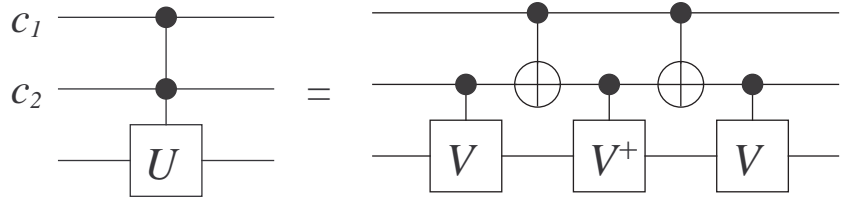}
\caption{Simulation of a $\Lambda_2(U)$ operation in terms of $\Lambda_1(V)$ operations and CNOT's.}
\label{fig: ContrContrUDecomp}
\end{figure}

For any unitary $2\times 2$ operation $U$, a $\Lambda_2(U)$ can be decomposed as in figure \ref{fig: ContrContrUDecomp} where $V$ is a matrix that satisfies $V^2=U$.

\pagebreak

%***
\subsubsection{Proof}
The gate construction is proved correct by an examination of the four cases of combinations of basis states for the two target qubits $|c_1\rangle$ and $|c_2\rangle$
\begin{itemize}
\item When $c_1=0$ and $c_2=0$, $I$ is applied to the target. 
\item When $c_1=0$ and $c_2=1$, $VV^{\dagger}=I$ is applied to the target.
\item When $c_1=1$ and $c_2=0$, $V^{\dagger}V=I$ is applied to the target. 
\item Finally, when $c_1=1$ and $c_2=1$, $VV=U$ is applied to the target.
\end{itemize}

The circuit identities developed so far allow for the decomposition of a general $\Lambda_2(U)$ gate in terms of 16 single qubit gates and CNOT gates. A naive counting based on figures \ref{fig: ControlledUDecomposition} and \ref{fig: ContrContrUDecomp} would imply the need to use 20 elementary gates, but a closer examination shows that a few single qubit gates can be merged and eliminated. The details are left to the reader.

%\pagebreak
%**********
\subsubsection{Decomposition of a general $\Lambda_{n-1}(U)$ gate}
The decomposition of $\Lambda_{2}(U)$ can be generalized to more than 2 control qubits. In fact, for any $2\times 2$ unitary operator $U$, the controlled operation $\Lambda_{n-1}(U)$ can be implemented using only ${\Theta}(n^2)$ elementary operations. There are several such constructions \cite{BBCDMSSSW1995,NielsenChuang2000}. The one below is from \cite{BBCDMSSSW1995}.

\begin{figure}[h]
\epsfbox{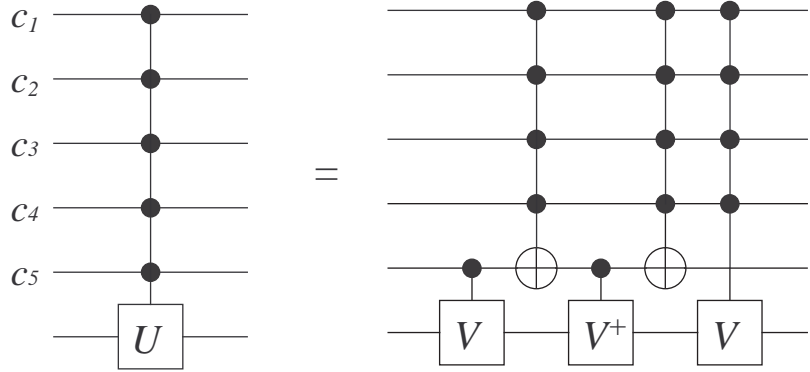}
\caption{Simulation of a $\Lambda_5(U)$ operation.}
\label{fig: ContrNminus1UDecomp}
\end{figure}

Consider the gate construction in figure \ref{fig: ContrNminus1UDecomp}, which is an example with $n=5$.

This is obviously a generalization of the decomposition of $\Lambda_2(U)$ and the proof is similar with $V$ an operator such that $V^2=1$.

To estimate the complexity of this decomposition, we need a construction of the generalized Toffoli gate $\Lambda_{n-2}(X)$. There are constructions of these gates of order ${\Theta(n)}$ in the number of Toffoli gates and elementary gates used. The reader is referred to \cite{BBCDMSSSW1995} for details, a paper that contains a wealth of gate constructions.

The overall complexity $C(n-1)$ of $\Lambda_{n-1}(U)$ can now be estimated. The cost of simulating the gates $\Lambda_1(V)$ and $\Lambda_1(V^{\dagger})$ is constant, and the complexity of $\Lambda_{n-2}(X)$ is ${\Theta(n)}$. Then, the complexity of $\Lambda_{n-2}(V)$ is $C(n-2)$ by applying the construction of figure \ref{fig: ContrNminus1UDecomp} recursively. This yields a recurrence equation for the cost $C(n)$
$$
C(n-1)=C(n-2)+\Theta(n).
$$
Since $(n-1)^2-(n-2)^2$ is in $\Theta(n)$, it is clear that $C(n)$ is in $\Theta(n^2)$.\footnote{This solution is a particular solution. A more careful treatment of this equation shows that the solution to the homogenous equation, which in general yields exponential behavior, in this case gives a solution $c1^n=c$, a constant.}

%***********
\subsection{Decomposing general two-level unitary operation on $n$-qubit states}\label{subsec: DecompGeneral2LevelUnitaries}
Let $U$ be a two-level unitary matrix acting on an $n$ qubit state. The two levels involved can be any two directions in the full space of the state.  Suppose the two directions are given by the computational basis states $|s\rangle=|s_1s_2\cdots s_n\rangle$ and $|t\rangle=|t_1t_2\cdots t_n\rangle$ respectively. These two directions can differ in between 1 and $n$ places. In order to write $U$ in terms of a single qubit operation, swap operations must be applied to yield two directions that differ in precisely one place corresponding to one particular qubit. The swap operations can be performed by generalized $\Lambda_{n-1}(X)$ operations. An example will clarify the situation. 

Consider a 3 qubit computer and a particular two level matrix $U$

$$
U=\pmatrix{1&0&0&0&0&0&0&0\cr
0&a&0&0&0&0&b&0\cr
0&0&1&0&0&0&0&0\cr
0&0&0&1&0&0&0&0\cr
0&0&0&0&1&0&0&0\cr
0&0&0&0&0&1&0&0\cr
0&c&0&0&0&0&d&0\cr
0&0&0&0&0&0&0&1\cr}.
$$

Denote by {\mathversion{bold} $u$} the two-dimensional matrix 

$$
U'= \pmatrix{a&b\cr c&d}.
$$

The matrix $U$ acts non-trivially on the computational basis states $|001\rangle$ and $|110\rangle$. Using so called Gray coding, the binary number 001 can be transformed by one-bit flips into 110 through the steps (they are not unique)
$$
001\rightarrow000\rightarrow010\rightarrow110
$$
The first two steps can be performed by the circuit

\begin{figure}[h]
\epsfbox{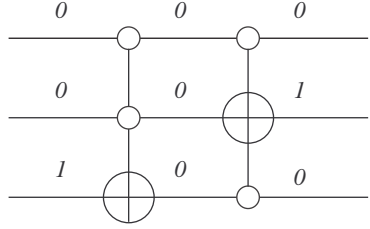}
\caption{Implementation of example Gray code transformation.}
\label{fig: GrayCode}
\end{figure}

The idea is to use generalized controlled NOT gates to transform $|001\rangle$ into $|101\rangle$ which differ in the first qubit, and then apply $U'$ to the first qubit conditioned on the second and third qubit, i.e. a generalized $\Lambda_2(U')$ . Finally, the swap operations are applied in reverse. The full circuit thus becomes as pictured in figure \ref{fig: GrayCode}.

\begin{figure}[h]
\epsfbox{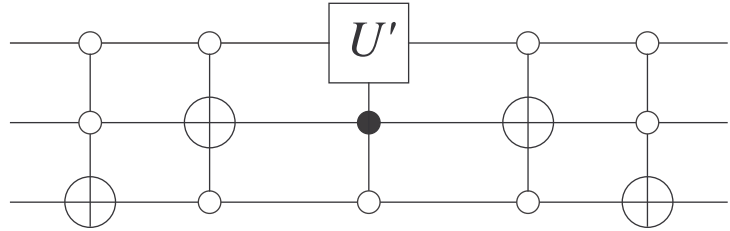}
\caption{Example of decomposition of two-level unitary operation on 3-qubit state.}
\label{fig: GrayCode}
\end{figure}

The general case is a straightforward generalization of this procedure. 

The number of gates needed to achieve this decomposition can now be calculated. First, at most $2(n-1)$ generalized CNOT operations $\Lambda_{n-1}(X)$ to swap the input state as described above, and then back again. Each such swap operation can be decomposed into ${\cal O}(n)$ single qubit and CNOT gates according to section \ref{subsec: ImplementationControlledNU}. The same holds for the $\Lambda_{n-1}(U)$ operation, yielding an overall ${\cal O}(n^2)$ complexity for this gate construction.

These gate constructions will be used in the following sections discussing universality for the quantum circuit model.

%*******************************************
\subsection{Universal sets of quantum gates}\label{subsec: UniversalQuantumGates}
In any computational model,\footnote{Except in the Gandy machine model \cite{Gandy1980}} there is a choice as to what constitutes the basic programming primitives. In the non-reversible classical circuit model, NAND-gates together with FANOUT, is a universal set of gates. In the reversible classical circuit model, the Toffoli gate is universal. Another choice is the Fredkin gate. Using either of these gates, any other logic operation as well as FANOUT can be performed. 

In the quantum circuit model, the situation is more complicated. The unitary quantum gates form a continuum and an $N\times N$ unitary matrix has $N(N-1)/2$ complex parameters $u_{ij}$. The possible sets of universal operations are therefore much richer than in the classical case where the set of gates is discrete and finite. However, it can be shown that an arbitrary $N\times N$ unitary matrix $U$, acting on $N$\,-dimensional vector space, can be expressed as a product lower-dimensional unitary matrices acting on subspaces of the vector space. In fact, $U$ can be decomposed entirely in terms of matrices acting non-trivially on just two-dimensional subspaces. For a proof see \cite{ReckEtAl1994} and \cite{NielsenChuang2000}. Such a construction is however in general not efficient in terms of the number of required two-dimensional matrices, as the number of matrices needed is ${\mathcal O}(N^2)$, or more precisely $2^{n-1}(2^n-1)$. Remember that in this context the dimension of the vector spaces are $N=2^n$.\footnote{To avoid confusion, keep in mind that the number of wires $n$ is equal to the number of qubits. But each qubit spans a 2-dimensional complex vector space, thus the linear dimension of the vectors and matrices is really $N=2^n$ and $N\times N$ respectively. This is explicit when using the computational basis states.} Furthermore, the matrices appearing in such a construction might not be possible to realize in a real physical system. What we need is a discrete set of standard gates out of which any unitary operation can be composed. In general we cannot expect such a construction to be exact, but rather obtained to within a certain approximation.

%**********************************************
\subsection{Exact and approximate universality}\label{subsec: ExactApproximateUniversality}
The universality results reported in the literature are somewhat bewildering. In order to navigate among them, two distinctions should be kept in mind. First, the distinction between exact universality and approximate universality. Secondly, the distinction between a finite set of standard gates and an infinite set of gates (parameterized by some variables). Note at once that a finite set of discrete standard gates can only be universal in the approximate sense, since the unitary matrices $U(N)$ is a continuous set, while circuits built using a discrete set of gates can only generate a countable set of matrices. 

Thus exact universality requires circuits with an infinite number of discrete gates, or circuits using a finite number of gates continously parameterized by a set of variables. 

For practical purposes, approximate universality is the important concept, since any real quantum computer must presumable be built from a set of discrete standard gates.

The two universality concepts, exact universality and approximate universality, are defined below.

%***
\subsubsection{Definition}
Let $\mathbf{G}$ be a set of gates with $r$ elements, 

\begin{equation}\label{eq: UniversalSetsOfGates}
\mathbf{G}=\{G_{1,n_1},\cdots ,G_{r,n_r}\},
\end{equation}
where $G_{j,n_j}$ is a gate that acts on $n_j$ qubits. The set is (approximately) universal if any unitary operator $U$ acting on $n$ input qubits can be decomposed into a finite product $\widetilde U$ of successive actions of gates $G_{j,n_j}$ so that the error $E(U,\widetilde U)$
$$
E(U,\widetilde U)\leq\epsilon
$$
where the error is defined as 
$$
E(U,V)={max\atop |\psi\rangle}||(U,V)|\psi\rangle||.
$$
The maximum is taken over the set of all normalized states in the state space of the computer.

In other words, there is a finite circuit (i.e. with a finite number of gates), approximating $U$, built from the gates in $\mathbf{G}$ \cite{GalindoMartin-Delgado2002}. 

The set is exactly universal if a finite circuit exactly reproduces an arbitrary operation $U_n$. In that case the set $\mathbf{G}$ must contain an infinite number of gates (in the sense referred to above). 

Otherwise, with a finite set $\mathbf{G}$, it would take (in general) an infinitely large circuit, which would contradict the requirement that a computer must operate by finite means.

%***
\subsubsection{Linear addition of errors}
With this definition of error it can be shown that errors add at most linearly \cite{NielsenChuang2000}. Suppose $V_1,V_2,\ldots ,V_m$ is a sequence of gates approximating another sequence of gates $U_1,U_2,\ldots ,U_m$, then

$$
E(U_mU_{m-1}\ldots U_1,V_mV_{m-1}\ldots V_1)\leq\sum_{j=1}^mE(U_j,V_j).
$$

%*****************************************************
\subsection{Exact universality of two-level unitaries}\label{subsec: ExactUniversality2LevelUnitaries}
Let us restate the general decomposition result mentioned above.

Let $U$ be an $N$-dimensional unitary matrix. Then it can be decomposed into a product of at most $N(N-1)/2$, 2-dimensional (two-level) unitary matrices \cite{ReckEtAl1994} acting on two-dimensional subspaces of the full $N$-dimensional space.

\subsubsection*{Outline of proof}

Let $U(N)$ be a general $N\times N$-dimensional unitary matrix. Let $T_{PQ}$ be $N\times N$ identity matrices with the elements $T_{pp}$, $T_{pq}$, $T_{qp}$ and $T_{qq}$ replaced with the corresponding elements from a certain $2\times 2$ unitary matrix $C$, i.e. 

$$
T_{pp}=C_{11},\quad T_{pq}=C_{12},\quad T_{qp}=C_{21},\quad T_{qq}=C_{22}\quad
$$
 
This matrix performs a unitary transformation on the $N$-dimensional state space, leaving an $N-2$-dimensional subspace unchanged. By a process analogous to Gauss-elimination, these matrices can be used to make all non-diagonal matrix elements of $U(N)$ zero by a judicious choice of parameters in the $C$-matrices. 

The matrix $U(N)$ is multiplied on the right by a succession of $T$-matrices,

$$
U(N)T_{N,N-1}T_{N,N-2}\cdots T_{N,1}=\pmatrix{U(N-1)&0\cr 0&e^{i\alpha_N}},
$$
where $R(N)=T_{N,N-1}T_{N,N-2}\cdots T_{N,1}$

Now this can be applied recursively to the matrix $U(N-1)$ down to $N=2$, when we have obtained a diagonal matrix with phase factors on the diagonal. Applying one final diagonal phase matric $D$ then yields

$$
U(N)R(N)R(N-1)\cdots R(2)D=I(N)
$$

and $U(N)$ is obtained as

$$
U(N)=D^{\dagger}R(2)^{\dagger}\cdots R(N-1)^{\dagger}R(N)^{\dagger}.
$$

The number of $T$-matrices used in this construction is ${N\choose 2}$.

Furthermore, there exist $N$-dimensional unitary matrices that cannot be decomposed into less than $N-1$ two-level unitaries. To see this, consider the case of a general diagonal unitary matrix.

%*******************************************
\subsection{Summary of universality results}\label{subsec: Summary}
One nice review of universal quantum gates can be found in \cite{DiVencenzo1992}.

%***
\subsubsection{The Deutsch gate}
In the first paper defining the quantum circuit model, Deutsch \cite{Deutsch1989} showed that there is a 3-qubit universal quantum gate. The gate is of the form $\Lambda_2(U)$ where $U=iR_x(\pi\alpha)$ and $\alpha$ is any irrational number. 

This gate is approximately universal. The proof will not be repeated here, but the result can be intuitively understood by noting that applying arbitrary integer powers $p$ of $R_x(\pi\alpha)$ to a qubit $|u\rangle$ yields

$$
R_x(\pi\alpha)^p|u\rangle=e^{-i{n\pi\alpha\over 2}\sigma_x}|u\rangle.
$$

Since $\alpha$ is irrational, the numbers $n\pi\alpha/2\;(\rm{mod}\; 2\pi)$ approximates arbitrarily well any number $\lambda$ in the interval $[0,2\pi[$. Therefore any operation $e^{-i\lambda\sigma_x}$ can be arbitrarily well approximated. 
Therefore, all gates of the form $\Lambda_2(ie^{-i\lambda\sigma_x})$ can be approximated. In particular, since the Toffoli gate corresponds to $\lambda=\pi/2$, it can be approximated.

%***
\subsubsection{Two-qubit universal gates}
Then DiVincenzo \cite{DiVincenzo1995} showed that the Deutsch gate could be realized by a set of one and two bit gates. This result was improved by Barenco \cite{Barenco1995} who showed that a single two-bit gate is universal. This gate depends on three angular parameters $\phi,\alpha$ and $\theta$. They are irrational with respect to each other and to $\pi$.

\begin{equation}\label{eq: BarencoGate}
A(\phi,\alpha,\theta)=\pmatrix{1&0&0&0\cr 
0&1&0&0\cr
0&0&e^{i\alpha}\cos\theta&-ie^{i(\alpha-\phi)}\sin\theta\cr 
0&0&e^{i(\alpha+\phi)}\sin\theta&e^{i\alpha}\cos\theta}
\end{equation}

%***
\subsubsection{Universality of almost any two-qubit gate}
Although the Deutsch gate and the Barenco gate are of a certain form, it turns out that there is nothing special about these particular gates. In \cite{DeutschBarencoEkert1995} and \cite{Lloyd1995} it was shown that almost any $n$\,-bit gate with $n\geq 2$ is universal. The question as to which gates are not universal was clarified in \cite{Brylinski2001}, a result which we will return to below.

%***
\subsubsection{Single qubit and CNOT universality}
Relying on the fact that the Deutsch gate $\Lambda_2(U)$  is universal, and the gate constructions of section \ref{subsec: GateConstructions}, where it was shown that any $\Lambda_2(U)$ gate can be decomposed in terms of single qubit and CNOT gates, it follows that single qubit and CNOT gates are universal for quantum computation. This is approximate universality, since the Deutsch gate is approximately universal. 

However it is possible to show that single qubit and CNOT gates are in fact exactly universal. This stronger results follows from the exact decomposition of a general $N$-dimensional unitary matrix in terms of 2-dimensional unitaries acting on two-dimensional subspaces. Relying on the construction in section \ref{subsec: DecompGeneral2LevelUnitaries}, this can be simulated with single-qubit gates and CNOT-gates.

%**********************************
\subsection{Discrete sets of gates}\label{subsec: DiscreteSetsGates}
Finally we arrive at a discussion on finite sets of practical one qubit gates together with CNOT gates. Relying on the previous discussion, it now suffices to effectively approximate general one-qubit gates with a discrete set of one-qubit gates. 

The {\it standard set} of gates  consists of the Hadamard gate $H$, the $\pi/8$ gate $T$ and the phase gate $S$ \cite {BMPRV1999}. The phase gate is not really needed for universality, but it is needed in order to perform the approximations {\it fault tolerantly}, an important topic that will be briefly mentioned below. 

The proof that the gates $H$, $\pi/8$ and $T$ are sufficient to approximate any single qubit gate, is quite complicated. It will not be repeated here, the reader is instead referred to \cite{NielsenChuang2000}. 

%***
\subsubsection*{Efficiency of the gate constructions}
In order to estimate the efficiency of this construction, consider an n-qubit circuit approximating a $U(2^n)$ operation. Suppose the circuit requires $f(n)$ gates, most of which need not be in the universal set of gates so that most of them must be approximated. Suppose furthermore that the total tolerance to errors is required to be $\epsilon$. Then each gate must be approximated to within an error $\delta=\epsilon/f(n)$ since the errors adds at most linearly. The overall efficiency is thus determined by the efficiency with which a single qubit gate can be approximated. Naively, one would suspect that roughly $\Theta(1/\delta)$ gates from the discrete set would be needed, resulting in an overall efficiency of $\Omega(f(n)/\delta)=\Omega(f(n)/(\epsilon/f(n))=\Omega(f(n)^2/\epsilon)$. 

However, according to the {\it Solovay-Kitaev} theorem (which unfortunately is too complicated to prove here) a general single qubit gate can be approximated within an error $\delta$ using ${\cal O}(\log ^c(1/\delta))$ gates from the discrete set. The constant $c$ is number approximately equal to 2. Therefore, the overall efficiency becomes ${\cal O}(f(n)\log ^c(f(n)/\delta))$, which is a polylogarithmic increase in the number of gates as compared to original circuit.

What about the function $f(n)$? An upper bound for this function can be calculated based on the gate constructions in sections \ref{subsec: DecompGeneral2LevelUnitaries} and \ref{subsec: ExactUniversality2LevelUnitaries}. From the decomposition of an arbitrary $U(2^n)$ matrix in terms of two-level unitaries (section \ref{subsec: ExactUniversality2LevelUnitaries}), we know that it takes $2^n(2^n-1)/2$ gates, i.e. ${\cal O}(4^n)$. Then, according to section \ref{subsec: DecompGeneral2LevelUnitaries}, each such two-level unitary gate (acting on the state space of $n$ qubits) can be implemented using ${\cal O}(n^2)$ single qubit and CNOT gates. Therefore, the function $f(n)$ is in ${\cal O}(n^24^n)$.

Putting everything together we see that the overall complexity is \break ${\cal O}(n^24^n{\log^c}(n^24^n/\delta))$, which is exponential in $n$. But the exponential behavior arises not from the approximation of single qubit gates by discrete single qubit gates, but from the complexity of breaking down general $U(2^n)$ matrices in terms of two-level matrices. Thus, fast quantum algorithms cannot rely on naive universality constructions.

%***************************
\subsection{General results}\label{subsec: GeneralResults}
The particular universality results reviewed to above are beautifully subsumed under a general theorem \cite{Brylinski2001}. In order to state the theorem, the notion of an {\it imprimitive} gate must be defined. 

A two-qubit gate $V$ is said to be {\it primitive} if it maps decomposable states into decomposable states, i.e. if $|x\rangle$ and $|y\rangle$ are qubits, then there are qubits $|u\rangle$ and $|v\rangle$ such that $V|x\rangle|y\rangle=|u\rangle|v\rangle$. $V$ is imprimitive if it is not primitive.

There is a simple condition to determine whether a gate is primitive or not. Let $P$ be an operator that swaps the two qubits in a product state, i.e. $P|x\rangle|y\rangle=|y\rangle|x\rangle$. Then it can be shown that $V$ is primitive if and only if $V$ can be expressed as $S\otimes T$ or as $(S\otimes T)P$ for some single qubit gates $S$ and $T$, so that $V$ acts as $V|x\rangle|y\rangle=S|x\rangle\otimes T|y\rangle$ or as $V|x\rangle|y\rangle=S|y\rangle\otimes T|x\rangle$. 

\subsubsection{Theorem}
Given a two-qubit gate $V$, the following conditions are equivalent
\begin{itemize}
\item the collection of all single qubit gates together with $V$ is approximately universal
\item the collection of all single qubit gates together with $V$ is exactly universal
\item V is imprimitive
\end{itemize}

Of course, a discrete set of single qubit gates can never yield exact universality.

\end{document}